\documentclass[letterpaper,twocolumn,english]{revtex4-2}
\usepackage[T1]{fontenc}
\usepackage[utf8]{inputenc}
\usepackage{color}
\usepackage{babel}
\usepackage{array}
\usepackage{float}
\usepackage{mathrsfs}
\usepackage{amsmath}
\usepackage{amssymb}
\usepackage{graphicx}
\usepackage{setspace}
\usepackage{wasysym}
\usepackage{xargs}[2008/03/08]
\usepackage[pdfusetitle,
 bookmarks=true,bookmarksnumbered=false,bookmarksopen=false,
 breaklinks=false,pdfborder={0 0 1},backref=false,colorlinks=true]
 {hyperref}
\hypersetup{
 pdfborderstyle=}

\makeatletter

\newcommand*\LyXZeroWidthSpace{\hspace{0pt}}

\providecommand{\tabularnewline}{\\}

\usepackage{babel}
\usepackage[justification=raggedright]{caption}

 {

      }

\usepackage{orcidlink}
\usepackage{ragged2e}
\DeclareCaptionJustification{justified}{\justifying}
\usepackage{colortbl}
\definecolor{lightpink}{RGB}{255,182,193} % 淡粉红色
\definecolor{lightorange}{RGB}{255,218,185} % 淡橙色
\definecolor{lightblue}{RGB}{255,255,224} % 淡黄色
\definecolor{lightgreen}{RGB}{144,238,144} % 淡绿色
\definecolor{lightyellow}{RGB}{187,240,253} % 淡蓝色
\definecolor{lightpurple}{RGB}{216,191,216} % 淡紫色
\definecolor{lightgray}{RGB}{211,211,211} % 淡灰色
\definecolor{lightred}{RGB}{253, 207, 243}% 定义淡红色
\usepackage{array}
\usepackage{array}
\usepackage{array}
\usepackage{multirow}

\ifdefined\showcaptionsetup
 \PassOptionsToPackage{caption=false}{subfig}
\fi
\usepackage{subfig}
\makeatother

\begin{document}
\title{Quantum-limited superresolution of two arbitrary incoherent point
sources: beating the resurgence of Rayleigh's curse}
\author{Junyan Li\orcidlink{0009-0008-1357-6590}}
\author{Shengshi Pang\orcidlink{0000-0002-6351-539X}}
\email{pangshsh@mail.sysu.edu.cn}

\affiliation{School of Physics, Sun Yat-sen University, Guangzhou, Guangdong 510275,
China}
\begin{abstract}
Superresolution has been demonstrated to overcome the limitation of
the Rayleigh's criterion and achieve significant improvement of the
precision in resolving the separation of two incoherent optical point
sources. However, in recent years, it was found that if the photon
numbers of the two incoherent optical sources are unknown, the precision
of superresolution vanishes when the two photon numbers are actually
different. In this work, we first analyze the estimation precision
of the separation between two incoherent optical sources with the
same point-spread functions in detail, and show that when the photon
numbers of the two optical sources are different but sufficiently
close, the superresolution can still realized but with different precisions.
We find the condition on how close the photon numbers of two optical
sources need to be to realize the superresolution, and derive the
precision of superresolution in different regimes of the photon number
difference. We further consider the superresolution for two incoherent
optical sources with different point-spread functions, and show that
the competition between the difference of photon numbers, the difference
of the two point-spread functions and the separation of the two optical
sources determines the precision of superresolution. The results exhibit
precision limits distinct from the case of two point sources with
identical point-spread functions and equal photon numbers, and extend
the realizable regimes of the quantum superresolution technique. The
results are finally illustrated by Gaussian point-spread functions.
\end{abstract}
\maketitle
\global\long\def\e{{\rm e}}%
\global\long\def\bra#1{\langle#1|}%
\global\long\def\ket#1{|#1\rangle}%
\global\long\def\hp{\hat{P}}%
\newcommandx\avg[3][usedefault, addprefix=\global, 1=\hp{}^{2}, 2=, 3=]{\langle#1\rangle_{#2}^{#3}}%
\global\long\def\qo{\langle\psi_{1}|\psi_{2}\rangle}%
\global\long\def\norm#1{\left|#1\right|^{2}}%
\global\long\def\ei#1{\e^{i#1\hp}}%
\newcommandx\xk[1][usedefault, addprefix=\global, 1=k]{X_{#1}}%
\global\long\def\nei#1{\e^{-i#1\hp}}%
\global\long\def\hro{\hat{\rho}}%
\newcommandx\nk[1][usedefault, addprefix=\global, 1=k]{N_{#1}}%
\newcommandx\px[1][usedefault, addprefix=\global, 1=x]{\partial_{#1}}%
\newcommandx\nt[1][usedefault, addprefix=\global, 1=]{N_{{\rm tot}}^{#1}}%
\newcommandx\nd[1][usedefault, addprefix=\global, 1=]{N_{{\rm diff}}^{#1}}%
\global\long\def\hl#1{\hat{\mathcal{L}}_{#1}}%
\global\long\def\xb{\bar{X}}%
\newcommandx\gi[1][usedefault, addprefix=\global, 1=i]{g_{#1}}%
\newcommandx\ksi[1][usedefault, addprefix=\global, 1=]{\ket{\psi_{#1}}}%
\newcommandx\bsi[1][usedefault, addprefix=\global, 1=]{\bra{\psi_{#1}}}%
\global\long\def\var{{\rm Var}}%
\global\long\def\hd{\mathcal{H}_{d}}%
\global\long\def\dlm#1{\left.#1\right|_{d\rightarrow0}}%
\global\long\def\avv#1{\avg[#1][12]}%
\global\long\def\sb{\overline{\sigma}}%
\global\long\def\sd{\sigma_{{\rm diff}}}%
\global\long\def\dn{\epsilon}%
\newcommandx\kt[1][usedefault, addprefix=\global, 1=]{\kappa_{{\rm tot}}^{#1}}%
\renewcommandx\sd[1][usedefault, addprefix=\global, 1=]{\sigma_{{\rm diff}}^{#1}}%
\newcommandx\st[1][usedefault, addprefix=\global, 1=]{\sigma_{{\rm tot}}^{#1}}%
\newcommandx\dt[1][usedefault, addprefix=\global, 1=]{\widetilde{d}^{#1}}%
\global\long\def\hdo{\mathcal{H}_{d}^{({\rm opt})}}%
\newcommandx\nb[1][usedefault, addprefix=\global, 1=]{N_{#1}}%

\section{Introduction}

Optical imaging is an important sensing technique that has contributed
to various research areas, such as medical technology, astronomy,
nanotechnology and biochemistry, etc. Typical imaging methods include
classical and quantum imaging \cite{Rushforth1968,Perez-Delgado2012},
fluorescence imaging \cite{Denk1990,Zipfel2003,Helmchen2005}, the
superlens \cite{Durant2006,Jacob2006}, two-photon microscopy \cite{Fei1997,Abouraddy2001},
quantum holography \cite{Sokolov2001,Song2013} and quantum lithography\cite{Boto2000,Liao2010},
etc. However, by the Rayleigh's criterion\cite{Rayleigh1880}, when
the separation of the two sources is smaller than the widths of two
closely located point sources, it becomes extremely difficult to distinguish
between the two sources. This imposes a severe limitation on the resolution
of imaging.

The Rayleigh's criterion indicates the insufficiency of direct measurement
for high-resolution imaging in the positional space of optical sources.
To precisely measure the separation of two point sources, it is critical
to find a suitable measurement basis beyond the positional basis and
a proper estimation method so that the imaging precision can surpass
the classical precision limit of direct imaging. The quantum parameter
estimation theory provides the mathematical tool to solve this problem.
The quantum parameter estimation theory is rooted in its classical
counterpart. The classical parameter estimation theory optimizes the
estimation precision of unknown parameters over all possible estimation
strategies and gives the optimal estimation precision for parameters
in a given parameter-dependent probability distribution by the well-known
Cramér-Rao's bound. The quantum parameter estimation theory considers
the parameters in a quantum system, and further optimizes the estimation
precision over all possible measurement basis in addition to the optimization
of estimation strategies, since different quantum measurement bases
lead to different probability distributions of the measurement results
and thus to different estimation precisions of the unknown parameters.
The precision limit of quantum estimation is characterized by the
quantum Cramér-Rao bound, which is determined by the inverse of the
quantum Fisher information matrix \cite{Helstrom1969a,Giovannetti2011,Liu2016,Yang2019,Liu2020}.

Inspired by the quantum estimation theory, Tsang et al. proposed a
technique of superresolution for two incoherent optical sources \cite{Tsang2016}
to surpass the Rayleigh's criterion. While the measurement precision
of the separation vanishes by the classical direct imaging method
when the separation is close to zero, which is known as Rayleigh’s
curse, the precision of estimating the separation of the two sources
does not vanish by optimizing the measurement basis. This is in sharp
contrast to the Rayleigh's criterion, which shows the power of quantum
estimation theory and opens up a new avenue for precision resolution
of two incoherent optical point sources.

The field of superresolution for incoherent optical sources has undergone
a rapid and extensive development in recent years since Tsang et al.'s
seminal work, in both theories and experimental implementations \cite{Tsang2019r}.
On one hand, the theory of multi-parameter estimation, essential to
the task of resolving incoherent optical sources, has made significant
progress in recent years, which has continuously stimulated the researches
on the superresolution imaging \cite{Tsang2019b,Albarelli2020,Qian2021,Zanforlin2022}.
The superresolution theory has been extended for two-dimensional optical
sources \cite{Ang2017,Napoli2019,Prasad2020a,Prasad2020b}, three-dimensional
optical sources \cite{Yu2018,Prasad2020}, multiple point sources
\cite{Zhou2019,Tsang2019b,Bisketzi2019}, thermal sources \cite{Lupo2016,Nair2016,Wang2021},
far field and near field \cite{Ben-Abdallah2019,Pushkina2021,Matlin2022},
dark field \cite{Brennan2022}, and various types of optical media
\cite{Kolesov2018,Horodynski2021,Xie2022}. The influence of noises
from the environments on the superresolution imaging has also been
investigated \cite{Bai2019,Gessner2020,Oh2021,Mikhalychev2021,Tsang2023,Gorecki2022a}.
Moreover, the superresolution theory has been introduced to localizing
and identifying objects \cite{Sajjad2021,Wang2021a,Prasad2019a,Grace2022},
discrete quantum imaging \cite{Fiderer2021}, spectral superresolution
\cite{Mazelanik2022,Zheltikov2022}, distributed quantum sensing \cite{Zhao2021,Kose2023},
superresolution microscopy \cite{Huszka2019,Cremer2013}, etc.

In addition to the theoretical development, physical implementations
of quantum superresolution have been proposed \cite{Patterson2010,Sheng2016,Zhou2019a,Fabre2020,Lupo2020,Huang2021,Datta2021a,Gorecki2022}
and experimentally realized in various ways, e.g., by optical interferometry
\cite{Nair2016a,Hassett2018}, with the assistance of phase information
\cite{Tham2017} or point-spread function engineering \cite{Paur2018},
using heterodyne detection \cite{Yang2016}, holographic technique
\cite{Paur2016}, parity sorting \cite{Wadood2021}, mode-selective
measurement \cite{Donohue2018}, spectrotemporal shaping \cite{Shah2021},
etc. Meanwhile, quantum superresolution has also been applied to optical
localization and microscopy \cite{Tan2017,Backlund2018,Rehacek2019,Deist2022,Cheng2020},
imaging of remote bodies and stars \cite{Howard2019,Huang2022}, etc.

An implicit assumption in the original superresolution protocol is
that the intensities of the two incoherent optical sources are known
and equal. This assumption looks trivial at first glance, but turns
out to play a critical role in the superresolution: if the photon
numbers are unknown and need to be estimated simultaneously with the
separation of the two sources, the superresolution can work only when
the intensities are equal and the estimation precision of the separation
will vanish otherwise \cite{Rehacek2017}. As the intensities could
vary or fluctuate due to noises and technical imperfections in reality,
this imposes challenges to the application of superresolution in practice.
In recent years, the role of partial coherence in the superresolution
has also been investigated intensively \cite{Larson2018,Tsang2019c,Larson2019a,Tsang2021a,De2021,Wadood2021,Karuseichyk2022,Liang2023},
and entanglement has been found useful to enhance the precision of
superresolution \cite{Sajia2022,Sajia2023}.

In this work, we analyze the effect of unknown photon numbers on the
precision of superresolution from the viewpoint of quantum multiparameter
estimation in detail, and obtain the constraint condition on the difference
of the photon numbers between the two incoherent point sources to
avoid the vanishing of estimation precision. The results show the
precision may deviate from that with identical photon numbers, depending
on the magnitude of the difference of the photon numbers. We further
consider two incoherent point sources with different point-spread
functions, and find that the competition between the difference of
the point-spread functions and the difference of the photon numbers
determines the precision of superresolution for the separation. We
analyze the competition between these parameters in detail, and show
that the superresolution can have various precision limits depending
on the magnitudes of these parameters, beyond the precision limit
of two optical sources with identical photon numbers and point-spread
functions.

\section{Preliminaries}

Suppose two incoherent optical pointlike sources are located at unknown
positions $\xk[1]$ and $\xk[2]$ on the one-dimensional object plane.
Neglecting the probability that more than one photon simultaneously
arrives at the image plane in the Poisson limit, the imaging process
can be regarded as the events of detecting a series of single photons
over multiple coherence time intervals of the sources \cite{Chrostowski2017a}.
Assume the intensities of the two optical sources can be different.
Denote the average numbers of photons emitted by the two sources within
a given observation time as $\nb[1]$ and $\nb[2]$ respectively.
The average single-photon state of the optical field arriving at the
image plane can be described by a density operator $\hro$ \cite{Mandel1959},
\begin{equation}
\hro=\frac{1-\dn}{2}\nei{\xk[1]}\ksi\bsi\ei{\xk[1]}+\frac{1+\dn}{2}\nei{\xk[2]}\ksi\bsi\ei{\xk[2]}.\label{eq:rho}
\end{equation}
The state $|\psi\rangle$ is defined as
\begin{equation}
\ksi=\int_{-\infty}^{\infty}dx\psi(x)\ket x,
\end{equation}
where $\psi(x)$ is the point-spread functions of the imaging system
assumed to be real and symmetric about the origin point \cite{Goodman1996}
throughout this paper, $\epsilon$ is the ratio between the difference
and the total of the photon numbers of the two optical sources,
\begin{equation}
\dn=\frac{\nb[2]-\nb[1]}{\nb[2]+\nb[1]},\label{eq:eps}
\end{equation}
and $\nei{\xk[i]}$, $i=1,2$, are the displacement operators with
$\hp$ defined as
\begin{equation}
\hp=-i\px.
\end{equation}
The probability distribution of a single photon arriving on the image
plane is \cite{Tsang2016}
\begin{equation}
\Lambda(x)=\frac{1-\dn}{2}\norm{\langle x|\nei{\xk[1]}\ksi}+\frac{1+\dn}{2}\norm{\langle x|\nei{\xk[2]}\ksi},\label{eq:pro}
\end{equation}
and the photon count at a width $dx$ around the position $x$ of
the detector is approximately $(N_{1}+N_{2})\Lambda(x)dx$.

Now, the question is: how precise can one measure and estimate the
separation,
\begin{equation}
d=X_{2}-X_{1},
\end{equation}
between the two incoherent point sources?

The well-known Rayleigh's criterion in optical imaging and the quantum
superresolution scheme proposed in the seminal paper \cite{Tsang2016}
concern the distinguishability of two pointlike optical sources with
identical point-spread functions up to a small spatial translation,
for which the quantum model is given by Eq. \eqref{eq:rho}. In this
paper, we will revisit the quantum superresolution scheme for two
point sources with identical point-spread functions to investigate
the validity condition of quantum superresolution, and extend it to
the point sources with different point-spread functions in Sec. \ref{sec:Superresolution-for-two-1}.

\subsection{Direct imaging and Rayleigh's curse\protect\label{sec:direct=000020imaging}}

The most straightforward method to measure the separation is directly
detecting and counting photons arrived on the image plane from the
two point sources \cite{VanTrees2001}.

In order to quantify the precision limit of estimating the separation
of two incoherent optical sources, the well-known Cramér-Rao bound
\cite{Helstrom1969a,Yang2019,Liu2020,Albarelli2020} from the classical
parameter estimation theory needs to be invoked, which is reviewed
in Appendix \ref{sec:Precision-limit-of}. For the task of resolving
two point sources, the positions of the two sources, $X_{1}$ and
$X_{2}$, or equivalently the centroid $\bar{X}=\frac{X_{1}+X_{2}}{2}$
and the separation $d=X_{2}-X_{1}$ of the two sources, are unknown,
so $\bar{X}$ and $d$ are the parameters to measure and estimate.

The Cramér-Rao bound demonstrates that over all possible estimation
strategies, the covariance matrix $C$ for the estimates of the parameters
$\bar{X}$ and $d$ is lower bounded by 
\begin{equation}
C[\xb,d]\geq\nt[-1]\mathcal{J}^{-1}[\xb,d],\label{eq:cramer-rao}
\end{equation}
where $\nt$ is total number of photons from the two optical sources
within a given detection time,
\begin{equation}
\nt=N_{1}+N_{2},\label{eq:ntot}
\end{equation}
the ``$\geq$'' sign represents semi-definite positivity of matrix,
and $\mathcal{J}[\xb,d]$ denotes the Fisher information matrix,
\begin{equation}
\mathcal{J}_{ij}=\int_{-\infty}^{\infty}\frac{1}{\Lambda(x)}\frac{\partial\Lambda(x)}{\partial g_{i}}\frac{\partial\Lambda(x)}{\partial g_{j}}dx,\;g_{i},g_{j}=\xb,d.\label{eq:classical=000020fisher}
\end{equation}
Note that the Cramér-Rao bound can always be asymptotically saturated
by the maximum likelihood estimation strategy when the total number
of photons $\nt$ is sufficiently large \cite{Helstrom1969a}.

For the direct imaging with an ideal continuum photodetector on the
image plane, if the intensities and the point-spread functions of
the two optical sources are the same, i.e., $\nb[1]=\nb[2]=\nb$,
the probability distribution of a single photon arrived on the image
plane can be written as
\begin{equation}
\Lambda(x)=\frac{1}{2}I(x-\bar{X}+d/2)+\frac{1}{2}I(x-\bar{X}-d/2),\label{eq:probability-1}
\end{equation}
where $I(x)$ is defined as
\begin{equation}
I(x)=|\psi(x)|^{2}.
\end{equation}

When the separation $d$ is small, it can be shown \cite{Paur2018}
by the Cramér-Rao bound \eqref{eq:cramer-rao} that the estimation
precision of the separation $d$ by direct imaging is
\begin{equation}
\mathcal{H}_{d}^{({\rm direct})}=\nt/(\mathcal{J}^{-1})_{22}=\frac{1}{8}d^{2}\alpha_{2}\nb,
\end{equation}
where
\begin{equation}
\alpha_{2}=\int_{-\infty}^{\infty}dx\frac{1}{I(x-\bar{X})}\left[\partial_{x}^{2}I(x-\bar{X})\right]^{2}.
\end{equation}
The detail of derivation is provided in Appendix \ref{sec:Precision-limit-of}.
Obviously, when $d\rightarrow0$, the estimation separation precision
vanishes, so one can hardly obtain any information about the separation
by direct imaging when the two point sources are sufficiently close
to each other. This is the Rayleigh's curse for direct imaging of
two identical optical point sources \cite{Tsang2016}.

The above result can be generalized to the case that the two point-spread
functions are the same but the intensities are unequal. In this case,
the probability distribution of a single\textcolor{blue}{{} }photon
on the image plane can be written as 
\begin{equation}
\Lambda(x)=\frac{1-\dn}{2}I(x-\bar{X}+d/2)+\frac{1+\dn}{2}I(x-\bar{X}-d/2).
\end{equation}
The estimation precision of the separation\textcolor{blue}{{} }by direct
imaging can be found \cite{Paur2018} to be
\begin{equation}
\mathcal{H}_{d}^{({\rm direct})}=\nt/(\mathcal{J}^{-1})_{22}=\frac{d^{2}\alpha_{2}N_{\text{tot}}}{16}\left(1-\epsilon^{2}\right)^{2}.
\end{equation}
(See Appendix \ref{sec:Precision-limit-of} for detail.) This indicates
that one cannot get any information about the separation $d$ by direct
imaging in the limit $d\rightarrow0$ in the generalized case either.

\subsection{Quantum-limited superresolution}

To surpass the Rayleigh's criterion, one needs to further resort to
the quantum estimation theory, which optimizes the Cramér-Rao bound
over all possible quantum measurement schemes, including generalized
measurements that are described by Kraus operators \cite{Nielsen2012}.

For the current quantum superresolution problem, the positions of
the two point sources, $\xk[1]$ and $\xk[2]$, or equivalently the
centroid $\xb=\frac{\xk[1]+\xk[2]}{2}$ and the separation $d=\xk[2]-\xk[1]$
are unknown, so $\xb$ and $d$ are the parameters to estimate.

The quantum estimation theory for the superresolution problem is briefly
reviewed in Appendix \ref{sec:Derivation=000020f}, which tells that
the estimation precision of $\xb$ and $d$ is lower bounded by the
following quantum Cramér-Rao's inequality,
\begin{equation}
C[\bar{X},\,d]\geq\nt[-1]\mathcal{Q}^{-1}[\bar{X},\,d],
\end{equation}
where $C[\bar{X},\,d]$ is the covariance matrix for the estimates
of $\xb,\,d$, $\mathcal{Q}[\xb,\,d]$ is the quantum Fisher information
matrix about $\xb,\,d$ with entries
\begin{equation}
\mathcal{Q}_{ij}(\hat{\rho})=\frac{1}{2}{\rm Tr}[\hro\{\hl i,\hl j\}],i,j=1,2,\label{eq:fisher0}
\end{equation}
and $\hl i$, $\hl j$ are the symmetric logarithmic derivatives (SLD)
of the density operator $\hro$,
\begin{equation}
\frac{1}{2}(\hl i\hro+\hro\hl i)=\partial_{\gi}\hro,\;\gi=\xb,d.\label{eq:sld}
\end{equation}

Note that the quantum Cramér-Rao bound for multi-parameter estimation
is not always achievable due to the potential noncommutativity between
the optimal measurements for different parameters. But if one is interested
in the overall estimation precision of different parameters, a weight
matrix $W$ can be imposed on the unknown parameters and the matrix
quantum Cramér-Rao's bound can be turned to a scalar form, 
\begin{equation}
{\rm Tr}(WC)\geq\nt[-1]{\rm Tr}(W\mathcal{Q}^{-1}).\label{eq:scalar}
\end{equation}
The saturation condition for the above scalar quantum Cramér-Rao bound
is 
\begin{equation}
{\rm Tr}(\hro[\hl i,\hl j])={\rm ImTr}(\hro\hl i\hl j)=0,\label{eq:compatibility}
\end{equation}
by recognizing that the quantum Cramér-Rao bound is equivalent to
the Holevo bound when this condition is satisfied and the latter is
always achievable asymptotically with a sufficiently large number
of states \cite{Ragy2016,Suzuki2020}.

For the current problem of quantum superresolution, the interested
parameter is the separation $d$, so one can choose 
\begin{equation}
W=\begin{bmatrix}0 & 0\\
0 & 1
\end{bmatrix},
\end{equation}
assuming $\gi[1]=\xb$ and $\gi[2]=d$, and ${\rm Tr}(WC)$ gives
the estimation variance of the separation. And the above saturation
condition can indeed be satisfied by the current problem, due to the
reality of $\psi(x)$ which makes the symmetric logarithmic derivatives
$\hl i$ and $\hl j$ also real according to Eq. \eqref{eq:sld} and
thus the compatibility condition ${\rm ImTr}(\hro\hl i\hl j)=0$ can
be satisfied. So the quantum Cramér-Rao bound provides an achievable
precision limit for the estimation of the separation allowed by quantum
mechanics.

A striking result found by Tsang et al. \cite{Tsang2016} is that
the best estimation precision of the separation $d$ over all quantum
measurements for two incoherent point sources with the same photon
numbers, determined by the quantum Cramér-Rao's bound, keeps nonzero
even when the separation goes to zero, implying the Rayleigh's criterion
is violated when the quantum measurement is optimized. The optimal
precision of estimating the separation $d$ in the limit $d\rightarrow0$
is found to be
\begin{equation}
\dlm{\hd}=2\nb\avg,
\end{equation}
where the expectation value $\avg[\cdot]$ is taken over the state
$\ksi$, $\avg[\cdot]=\langle\psi|\cdot|\psi\rangle$, and $\nb$
is the photon number of each point source. If the photon numbers of
the two point sources are allowed to be different, the above precision
limit can be generalized to
\begin{equation}
\dlm{\hd}=\nt\left(1-\epsilon^{2}\right)\avg.
\end{equation}

For two Gaussian point-spread functions with the same width $\sigma$,
the above precision limit can be simplified to 
\begin{equation}
\dlm{\mathcal{H}_{d}}=\frac{\nt\left(1-\epsilon^{2}\right)}{4\sigma^{2}}.
\end{equation}
The derivation of the precision limits for a general $d$ and for
the above $d\rightarrow0$ case is presented in Appendix \ref{subsec:With-known-arbitrary}.

An interesting finding in recent years is the vanishing of estimation
precision when the photon numbers $\nb[1]$, $\nb[2]$, of two point
sources are unknown and need to be estimated as well \cite{Rehacek2017}.
In this case, one needs to estimate three parameters, $\xb,\,d,\,\dn$,
simultaneously. If the photon numbers of the two point sources are
different, i.e., $\dn\neq0$, it was found that up to the second order
of $d$, the precision of estimating $d$ was 
\begin{equation}
\hd=\frac{\nt\left(1-\epsilon^{2}\right)}{4\epsilon^{2}}\var(\hp^{2})d^{2},
\end{equation}
where $\var(\hp^{2})=\avg[\hp^{4}]-\avg[\hp^{2}][][2]$.

It can be immediately seen that in this case, $\hd$ is of order $O(d^{2})$
and will vanish as $d$ goes to zero, implying the superresolution
no longer works and the Rayleigh's curse applies again. This is sometimes
termed as the resurgence of Rayleigh's curse for two incoherent point
sources with unknown and unbalanced intensities \cite{Rehacek2017,Rehacek2018}.

\section{Superresolution for two identical point sources\protect\label{sec:Superresolution-for-two}}

In this work, we investigate the problem of precision vanishing in
resolving two incoherent point sources with unknown and unequal intensities
and study the condition for the quantum superresolution to address
this problem in detail.

As a finite difference between the photon numbers of the two sources
can cause the resurgence of Rayleigh's curse while the equality between
the two photon numbers leads to superresolution, we are interested
in the condition on the photon numbers to realize the superresolution:
is it possible to realize the superresolution with a sufficiently
small but nonzero difference between the photon numbers of the two
sources and how close the two photon numbers need to be? In addition
to the photon numbers, we will also investigate the effect of slight
difference between the point-spread functions of the two optical sources
on suppressing the Rayleigh's criterion, and show its competition
with the difference of the photon numbers in determining the precision
of resolving the two point sources.

We consider two identical incoherent point sources located at $X_{1}$,
$X_{2}$ with unknown photon numbers $\nb[1]$, $\nb[2]$ in this
section. The unknown parameters to estimate in this case is $\xb$,
$d$, $\epsilon$, and $\dn$ is the ratio between the difference
and the total of the photon numbers defined in Eq. \eqref{eq:eps}.
The density matrix $\hro$ of a single photon is given by Eq. \eqref{eq:rho}.
By working out the eigenvalues and eigenstates of $\hro$ and applying
the formula of the quantum Fisher information matrix, the quantum
Fisher information matrix with respect to these three unknown parameters
can be obtained \cite{Rehacek2017} as
\begin{equation}
\mathcal{Q}=\left[\begin{array}{ccc}
4\left[\kappa-\gamma^{2}\left(1-\epsilon^{2}\right)\right] & 2\kappa\epsilon & 2\gamma\delta\\
2\kappa\epsilon & \kappa & 0\\
2\gamma\delta & 0 & \frac{1-\delta^{2}}{1-\epsilon^{2}}
\end{array}\right],
\end{equation}
where
\begin{equation}
\kappa=\avg,\;\delta=\avg[\cos(d\hp)],\;\gamma=\avg[\sin(d\hp)\hp].
\end{equation}
According to the quantum Cramér-Rao bound, the estimation precision
of the separation $d$ can be obtained as
\begin{equation}
\hd=\nt/(\mathcal{Q}^{-1})_{22}=\frac{\left(1-\epsilon^{2}\right)\left[\gamma^{2}-\left(1-\delta^{2}\right)\kappa\right]}{\gamma^{2}\left(1-\epsilon^{2}\right)-(1-\delta^{2})\kappa}\hdo,\label{eq:sy}
\end{equation}
where $\hdo$ is the optimal estimation precision when the photon
numbers of the two sources are equal, i.e., $\dn=0$, $\nb[1]=\nb[2]=\nb$,
\begin{equation}
\hdo=2\kappa\nb,\label{eq:hopt}
\end{equation}
The detail of derivation is provided in Appendix \ref{subsec:Unknown-arbitrary-photon}.

To investigate the mechanism underlying the vanishing of estimation
precision when the photon numbers $\nb[1]$ and $\nb[2]$ of the two
sources are different, we expand the numerator and denominator of
the precision $\hd$ \eqref{eq:sy} to the fourth order of $d$ respectively,
and the result turns out to be
\begin{equation}
\hd=\frac{Cd^{2}}{D_{1}+D_{2}d^{2}},\label{eq:w=00003D00003D0}
\end{equation}
where
\begin{equation}
\begin{aligned}C= & 3\nt(1-\dn^{2})\avg\var(\hp^{2}),\\
D_{1}= & 12\dn^{2}\avg,\\
D_{2}= & 3\var(\hp^{2})-4\dn^{2}\avg[\hp^{4}],
\end{aligned}
\end{equation}
and a common $d^{2}$ has been reduced from both the numerator and
denominator.

Obviously, when the intensities of the two optical sources are unequal,
i.e., $\nb[1]\neq\nb[2]$, the coefficient $D_{1}$ is finite, so
$\mathcal{H}_{d}=O(d^{2})$, thus $\mathcal{H}_{d}$ vanishes when
$d\rightarrow0$, and the Rayleigh's curse takes effect again. In
contrast, when the intensities of the two optical sources are equal,
i.e., $\epsilon\rightarrow0$ and $\nt=\nb[1]+\nb[2]=2\nb$, $D_{1}=0$,
so the estimation precision of $d$ \eqref{eq:w=00003D00003D0} becomes
\begin{equation}
\dlm{\mathcal{H}_{d}}=\frac{C}{D_{2}}=2\nb\avg,
\end{equation}
and the superresolution scheme can work in this case, which is the
result by Ref. \cite{Tsang2016}.

Nevertheless, a further observation of $\mathcal{H}_{d}$ in Eq. \eqref{eq:w=00003D00003D0}
suggests that even when $\nb[1]\neq\nb[2]$ there is still opportunity
to have nonvanishing estimation precision for the separation $d$
if the coefficient $D_{1}$ is sufficiently small. It can be seen
that if $D_{1}$ has the order $O\left(d^{2}\var(\hp^{2})\right)$,
i.e., the order of $\dn$ is $O\left(d\sqrt{\var(\hp^{2})/\avg}\right)$,
the precision $\mathcal{H}_{d}$ keeps finite when $d$ is small,
implying the superresolution scheme can still work in this case. Furthermore,
if the order of the coefficient $D_{1}$ is higher than $O(d^{2}\var(\hp^{2}))$,
or equivalently the order of $\dn$ is higher than $O\left(d\sqrt{\var(\hp^{2})/\avg}\right)$,
$D_{1}$ can be dropped when $d\rightarrow0$, and $\mathcal{H}_{d}$
is equal to $C/D_{2}=\nt\avg$ which restores the original result
for the precision of superresolution in \cite{Tsang2016}.

The above observation immediately leads to the condition for the superresolution
of two identical incoherent optical sources: the order of the ratio
$\epsilon$ between the difference and the total of the photon numbers
of the two optical sources must be higher than or equal to $O\left(d\sqrt{\var(\hp^{2})/\avg}\right)$.
The original result in \cite{Tsang2016} was obtained for the case
$\nb[1]=\nb[2]$, but is effective for the range of $\dn$ that is
higher than $O\left(d\sqrt{\var(\hp^{2})/\avg}\right)$, while the
superresolution scheme can actually work in a broader range including
the case that the order of $D_{1}$ is equal to $O\left(d^{2}\var(\hp^{2})\right)$.
But the latter has a lower precision than the former, as the ratio
between the two precisions is
\begin{equation}
\frac{D_{2}d^{2}}{D_{1}+D_{2}d^{2}}<1.\label{eq:ratio}
\end{equation}

It must be noted that this condition for the validity of superresolution
with unknown photon numbers does not mean that the photon number changes
with $d$, but rather impose a limitation on the magnitude of photon
number difference between the two optical sources to realize the superresolution.

\section{Superresolution for two arbitrary point sources\protect\label{sec:Superresolution-for-two-1}}

While the point-spread functions can be identical for the point sources
at different positions of the object plane if the imaging system is
spatially invariant with specific conditions such as the paraxial
approximation \cite{Tsang2016,Napoli2019,Yang2023}, the point-spread
functions can vary with the positions of the point sources in general
\cite{Xiao2024}. For example, the longitudinal shifts from the focused
object plane will make the locations of two point sources deviate
in the depth from each other, and thus lead to different point-spread
functions due to the diffraction and geometrical blurring effects
\cite{Pawley2008,Shechtman2014,Li2018}. Such a difference in the
point-spread functions may reduce the overlap between the two optical
sources, so an intuitive idea is that it may increase the distinguishability
of the two sources which could be helpful to the resolution of the
two sources. In the following, we generalize the result of the previous
section and study the condition of quantum superresolution for two
optical point sources with arbitrary point-spread functions.

Suppose the point-spread functions of two point sources are $\psi_{1}(x)$
and $\psi_{2}(x)$ with displacements $\xk[1]$ and $\xk[2]$ from
the origin point respectively. Denote $\ksi[k]=\int_{-\infty}^{\infty}dx\psi_{k}(x)\ket x$,
$k=1,2$. The density matrix of a single photon arrived on the image
plane can then be written as
\begin{equation}
\begin{aligned}\hro= & \frac{1-\dn}{2}\nei{\xk[1]}\ksi[1]\bsi[1]\ei{\xk[1]}\\
 & +\frac{1+\dn}{2}\nei{\xk[2]}\ksi[2]\bsi[2]\ei{\xk[2]},
\end{aligned}
\label{eq:1-1}
\end{equation}
where $\nt$ is the total number of the photons, $\nt=\nb[1]+\nb[2]$,
and $\epsilon$ is the ratio between the difference and the total
of the photon numbers of the two point sources defined in Eq. \eqref{eq:eps}.

The estimation precision for the separation can still be obtained
by the scalar quantum Cramér-Rao bound \eqref{eq:scalar}. The quantum
Fisher information matrix with respect to the unknown parameters $\xb,d,\epsilon$
turns out to be 
\begin{equation}
\begin{aligned} & \mathcal{Q}=\\
 & \left[\begin{array}{ccc}
2\kappa_{\text{tot}}(1+\chi\epsilon)-4\gamma^{2}\left(1-\epsilon^{2}\right) & \kappa_{\text{tot}}(\chi+\epsilon) & 2\gamma\delta\\
\kappa_{\text{tot}}(\chi+\epsilon) & \frac{1}{2}\kappa_{\text{tot}}(1+\chi\epsilon) & 0\\
2\gamma\delta & 0 & \frac{1-\delta^{2}}{1-\epsilon^{2}}
\end{array}\right],
\end{aligned}
\end{equation}
leading to the exact result for the estimation precision of $d$ as
\begin{equation}
\hd=\frac{\left(1-\epsilon^{2}\right)\left[\left(1-\delta^{2}\right)\left(1-\chi^{2}\right)\kt-2\gamma^{2}(1+\chi\epsilon)\right]}{\left(1-\delta^{2}\right)(1+\chi\epsilon)\kt-2\gamma^{2}\left(1-\epsilon^{2}\right)}\hdo,\label{eq:pre}
\end{equation}
where 
\begin{equation}
\begin{aligned}\kappa_{i}= & \avg[][i],\,i=1,2,\\
\delta= & \avv{\cos(d\hp)},\;\gamma=\avv{\sin(d\hp)\hp},\\
\kt= & \avg[][1]+\avg[][2],\;\chi=\frac{\avg[][2]-\avg[][1]}{\avg[][1]+\avg[][2]},
\end{aligned}
\label{eq:kdx}
\end{equation}
and $\avg[\cdot][i]=\bsi[i]\cdot\ksi[i]$, $\avv{\cdot}=\bsi[1]\cdot\ksi[2]$.
$\hdo$ is the optimal estimation precision when the photon numbers
of the two sources are equal, i.e., $\dn=0$, in the limit $d\rightarrow0$,
\begin{equation}
\hdo=\frac{1}{2}\nt\kt,\label{eq:hdo-1}
\end{equation}
which can be simplified to Eq. \eqref{eq:hopt} when the point-spread
functions of the two sources are the same. The derivation is provided
in Appendix \ref{sec:Quantum-fisher-information=000020of=000020unequal=000020source}.

Considering the separation $d$ is small, we expand the numerator
and the denominator of $\mathcal{H}_{d}$ to the fourth order of $d$
respectively, 
\begin{equation}
\mathcal{H}_{d}=\frac{A_{0}+A_{2}d^{2}+A_{4}d^{4}}{B_{0}+B_{2}d^{2}+B_{4}d^{4}}.\label{eq:hdexpand}
\end{equation}
The coefficients $A_{i}$, $B_{i}$, $i=0,2,4$, are quite complex,
so they are provided in Eq. \eqref{eq:a0b0a2b2a4b4} of Appendix \ref{sec:Quantum-fisher-information=000020of=000020unequal=000020source}.

It is obtained in Appendix \ref{sec:Quantum-fisher-information=000020of=000020unequal=000020source}
that
\begin{equation}
A_{0},B_{0}\propto1-|\langle\psi_{1}|\psi_{2}\rangle|^{2}.
\end{equation}
So, if the difference between the two point-spread functions is not
small, i.e., the magnitudes of $A_{0}$ and $B_{0}$ are of order
$O(1)$, only the constant terms $A_{0}$ and $B_{0}$ remain in $\mathcal{H}_{d}$
when $d\rightarrow0$. The estimation precision of the separation
$d$ in the limit $d\rightarrow0$ can be simplified to 
\begin{equation}
\dlm{\mathcal{H}_{d}}=\frac{\nt\kt\left(1-\chi^{2}\right)\left(1-\epsilon^{2}\right)}{2(1+\chi\epsilon)},\label{eq:hdd0}
\end{equation}
implying that the superresolution can work even if the photon numbers
of the two point sources are different in this case.

A more subtle case is when the point-spread functions of the two optical
sources are different but close to each other. In this case, the constant
terms $A_{0}$ and $B_{0}$ in $\mathcal{H}_{d}$ \eqref{eq:hdexpand}
will become small though still nonzero. If the magnitudes of these
leading terms are comparable to $O(d^{2})$ or even $O(d^{4})$ (although
they are independent of $d$ in physics), they will compete with the
second-order and fourth-order terms in Eq. \eqref{eq:hdexpand} to
determine $\mathcal{H}_{d}$.

Similarly, it is shown in Appendix \ref{sec:Quantum-fisher-information=000020of=000020unequal=000020source}
that the coefficients $A_{2}$ and $B_{2}$ of the quadratic terms
in $\mathcal{H}_{d}$ may also become small but nonzero when the two
point-distribution functions differ slightly from each other, so if
they are of order $O(d^{2})$, they will also compete with the fourth-order
terms in determining $\mathcal{H}_{d}$. Moreover, the difference
between the photon numbers of the two optical sources are also involved
in $\mathcal{H}_{d}$ which may compete with the higher-order terms
in $\hd$ if the difference in the photon numbers of the two point
sources is small. Combining all these together, the estimation precision
of the separation $d$ and thus the validity of superresolution are
highly dependent on the competition between the difference in the
two point-spread functions, the difference in the photon numbers of
the two point sources and the magnitude of the separation.

\begin{table}[!th]
\caption{\protect\label{tab:con}Summary of the ratios $\protect\hd/\protect\hdo$
between the estimation precision $\protect\hd$ of the separation
$d$ and the optimal estimation precision $\protect\hdo$ \eqref{eq:hdo-1}
in different regimes of the parameters. The difference between the
point-spread functions and that between the photon numbers are characterized
by Eq. \eqref{eq:ulnk} and we denote $1-\protect\qo=b\text{\ensuremath{\protect\dt[h]}}$,
$\frac{1}{2}-\xi=a\protect\dt[f]$, $\chi=c\protect\dt[e]$, $\epsilon=y\protect\dt[s]$,
where $y,a,b,c$ are real coefficients, and $\protect\dt$ is the
dimensionless separation \eqref{eq:dt}. The exponents $s,f,h,e$
are nonnegative integers, and $h,\,f,\,e$ satisfy the constraint
relation \eqref{eq:re}. The blue region represents the cases that
attain the optimal estimation precision $\protect\hdo$, the red region
represents the cases that the estimation precision vanishes, and the
yellow region represents the cases that the superresolution can still
work but with a lower precision. The estimation precision of the separation\textcolor{blue}{{}
$d$} given by the original superresolution scheme in \cite{Tsang2016}
corresponds to $h\protect\geq3$, $s\protect\geq2$, and the resurgence
of Rayleigh's curse \cite{Rehacek2017} corresponds to $h\protect\geq3,\,s=0$.
The total photon number $\protect\nt$ is sufficiently large, i.e.,
$\protect\nt\gg1$. The results are evaluated in the limit $d\rightarrow0$.
As the two point-spread functions approach to be the same in this
limit, the subscripts of $\left\langle P^{4}\right\rangle {}_{12}$
and $\left\langle P^{2}\right\rangle {}_{12}$ are dropped in the
table.}

\centering{}%
\begin{tabular}{|>{\centering}p{1cm}|>{\centering}p{1.1cm}cc|}
\hline 
 & \multicolumn{1}{c|}{$\begin{gathered}\\s\geq2\\
\\\end{gathered}
$} & \multicolumn{1}{c|}{$\begin{gathered}\\s=1\\
\\\end{gathered}
$} & $\begin{gathered}\\s=0\\
\\\end{gathered}
$\tabularnewline
\hline 
$\begin{gathered}\\h\geq5\\
\\\end{gathered}
$ & \cellcolor{lightyellow} & \cellcolor{lightblue}$\begin{gathered}\\\frac{\var(\hp^{2})}{4y^{2}\avg[][][2]+\var(\hp^{2})}\\
\\\end{gathered}
$ & \cellcolor{lightred}\tabularnewline
\cline{1-1}
$\begin{gathered}\\h=4\\
\\\end{gathered}
$ & \cellcolor{lightyellow} & \cellcolor{lightblue}$\begin{gathered}\\\frac{8b\avg[][][2]+\var(\hp^{2})}{4(2b+y^{2})\avg[][][2]+\var(\hp^{2})}\\
\\\end{gathered}
$ & \cellcolor{lightred}$\begin{gathered}0\\
\\\\\end{gathered}
$\tabularnewline
\cline{1-1}
$\begin{gathered}\\h=3\\
\\\end{gathered}
$ & \multicolumn{2}{c}{\cellcolor{lightyellow}} & \cellcolor{lightred}\tabularnewline
\cline{1-1}
$\begin{gathered}\\h=2\\
\\\end{gathered}
$ & \multicolumn{2}{c}{$\begin{gathered}\\1\\
\\\end{gathered}
$\cellcolor{lightyellow}} & \cellcolor{lightblue}$\begin{gathered}\frac{2b\left(1-y^{2}\right)}{2b+y^{2}}\end{gathered}
$\tabularnewline
\cline{1-1}
$\begin{gathered}\\h=1\\
\\\end{gathered}
$ & \cellcolor{lightyellow} & \cellcolor{lightyellow} & \cellcolor{lightblue}$\begin{gathered}1-y^{2}\end{gathered}
$\tabularnewline
\cline{1-1}
$\begin{gathered}\\h=0\\
\\\end{gathered}
$ & \multicolumn{2}{c}{\cellcolor{lightblue}$\begin{gathered}1-c^{2}\end{gathered}
$} & \cellcolor{lightblue}$\begin{gathered}\frac{\left(1-c^{2}\right)\left(1-y^{2}\right)}{1+cy}\end{gathered}
$\tabularnewline
\hline 
\end{tabular}
\end{table}

As the point-spread functions are continuous, it generally requires
an infinite number of parameters to characterize the difference between
the two point-spread functions in principle, which seems to impose
significant difficulty in analyzing the effect of the difference of
the point-spread functions on the precision of quantum superresolution.
But in fact, one can figure out just a few quantities from $\mathcal{H}_{d}$
\eqref{eq:hdexpand} that are dependent on these differences and will
determine $\hd$. In Appendix \ref{subsec:Critical-quantities}, it
is shown that only the following three quantities, $1-\langle\psi_{1}|\psi_{2}\rangle$,
$\chi$ as defined in Eq. \eqref{eq:kdx}, and $\frac{1}{2}-\xi$
with $\xi$ defined as
\begin{equation}
\xi=\frac{\avg[][12]}{\avg[][2]+\avg[][1]},
\end{equation}
are relevant to the difference between the two point-spread functions
and will determine the estimation precision $\hd$. And they tend
to be zero as the point-spread functions of the two sources get close
to each other, so these quantities will compete with each other and
with the difference of photon numbers and play a key role in determining
the validity of superresolution.

But note that the photon numbers and the closeness between the point-spread
functions are dimensionless while the separation $d$ is not, so for
the sake of convenience to make these quantities comparable and characterize
the magnitudes of these quantities, we rescale the separation $d$
to make it dimensionless. We define the following dimensionless separation
$\widetilde{d}$,
\begin{equation}
\dt=\frac{d}{V},\label{eq:dt}
\end{equation}
where $V=1/\sqrt{\avg}$. The subscript is dropped from $\avg$ in
the definition of $V$, as the following results of the estimation
precision $\hd$ are evaluated in the limit $d\rightarrow0$ which
leads to $\psi_{1}(x)=\psi_{2}(x)$ (as the difference between the
two point-spread functions is characterized in terms of the power
of $\dt$ later) and thus $\avg[][1]=\avg[][2]=\avg[][12]$. In order
to represent the magnitudes of these quantities which are all considered
to be small, we denote them in the orders of $\dt$, 
\begin{equation}
\begin{aligned}1-\langle\psi_{1}|\psi_{2}\rangle=O(\dt[h]),\; & \frac{1}{2}-\xi=O(\dt[f]),\\
\epsilon=O(\dt[s]),\; & \chi=O(\dt[e]),
\end{aligned}
\label{eq:ulnk}
\end{equation}
where $s,f,h,e$ are nonnegative integers. Note that the exponents
$h,f,e$ cannot be arbitrary but are constrained by the relation 
\begin{equation}
h/2\leq e\leq h\leq f\leq2h.\label{eq:re}
\end{equation}
A detailed proof of this relation can be found in Appendix \ref{sec:Derivation=000020hfe}.
This constraint will be taken into account in the following computation
of $\mathcal{H}_{d}$. When the exponents $h,f,e,s$ vary, the estimation
precision $\mathcal{H}_{d}$ of the separation $d$ will be different.
The estimation precision $\mathcal{H}_{d}$ can be computed by Eq.
\eqref{eq:hdexpand} for different ranges of these exponents.

The results of the estimation precision $\hd$ are summarized in Table
\ref{tab:con} in terms of the ratio between $\hd$ and the optimal
estimation precision $\hdo$ \eqref{eq:hdo-1}, which covers all the
possible ranges of the difference between the two point-spread functions
and the difference between the two photon numbers and rules out the
cases that the exponents do not satisfy the constraint relation \eqref{eq:re}.
It can be seen that the superresolution protocol can actually work
for a wide range of point-spread functions with various magnitudes
of the photon numbers, and the estimation precision $\hd$ vanishes
only when the order of $1-\langle\psi_{1}|\psi_{2}\rangle$ is higher
than or equal to $O(\dt[3])$ and $\epsilon=O(1)$, which is the cases
considered in Ref. \cite{Rehacek2017}. The case $h\geq3$ with $s\geq2$
is the regime that was considered by the original superresolution
scheme \cite{Tsang2016}. The case $h\geq5$ with $s=1$ is the regime
where the effective condition of quantum superresolution for two identical
point-spread functions was obtained in Sec. \ref{sec:Superresolution-for-two}.
For the other cases, the superresolution scheme can still work, but
the estimation precision of the separation $d$ may deviate from that
of the original superresolution scheme \cite{Tsang2016}.

It needs to be stressed again that the orders of the difference between
the point-spread functions and the difference between the photon numbers
of the two point sources in terms of the (dimensionless) separation
do not mean that they change with $d$ but rather impose a limitation
on the magnitudes of those differences to realize the superresolution.

\section{Example: Point sources of Gaussian distributions}

\begin{figure}[!th]
\centering{}\includegraphics[scale=0.9]{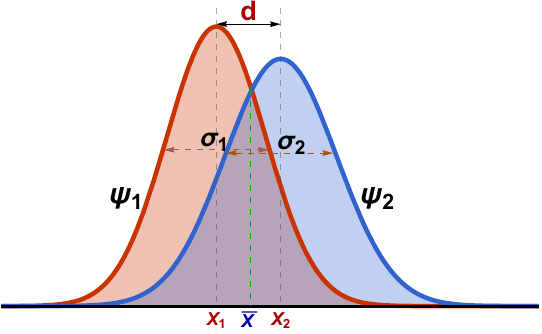}\caption{\protect\label{fig:Illustration-of-two}Illustration of two Gaussian
point-spread functions $\psi_{1}(x)$ and $\psi_{2}(x)$ with positions
$X_{1}$, $X_{2}$ and widths $\sigma_{1}$, $\sigma_{2}$ respectively.
The purpose of the superresolution scheme is to estimate the separation
$d=X_{2}-X_{1}$ when $d$ is small.}
\end{figure}

To illustrate the above results, we consider the Gaussian point-spread
functions as a typical example. Assume $\psi_{1}(x)$ and $\psi_{2}(x)$
are Gaussian point-spread functions,
\begin{equation}
\psi_{i}(x)=\frac{1}{\sqrt[4]{2\pi\sigma_{i}^{2}}}\exp\left[-\frac{x^{2}}{4\sigma_{i}^{2}}\right],\;i=1,2,
\end{equation}
where $\sigma_{i}=\frac{1}{2\sqrt{\kappa_{i}}}=\lambda_{i}/2\pi{\rm NA},i=1,2$,
$\lambda$ is the free-space wavelength, ${\rm NA}$ is the effective
numerical aperture \cite{Pawley2006}, and the spatial-frequency variance
$\kappa_{i}$ for the Gaussian point-spread function is $\frac{1}{4\sigma_{i}^{2}},\,i=1,2$.
The two general Gaussian point-spread functions are illustrated in
Fig. \ref{fig:Illustration-of-two}.

\subsection{Gaussian point-spread functions with the same width}

For two Gaussian point-spread functions with the same width $\sigma$,
the estimation precision of the separation is given by Eq. \eqref{eq:w=00003D00003D0},
where the coefficients can be simplified by the Gaussian point-spread
function and the estimation precision of $d$ becomes
\begin{equation}
\hd=\frac{\left(1-\epsilon^{2}\right)d^{2}/\sigma^{2}}{32\epsilon^{2}+4d^{2}/\sigma^{2}}\hdo,\label{eq:eta0}
\end{equation}
where $\hdo$ is the optimal estimation precision of $d$ when the
 photon numbers of the two point sources are equal and $d\rightarrow0$,
\begin{equation}
\hdo=\frac{\nt}{\sigma^{2}}.\label{eq:hdo-2}
\end{equation}
It can be seen that the ratio between $\hd$ and $\hdo$ is determined
by two parameters only, $\epsilon$, i.e., the ratio between the difference
and the total of the two photon numbers, and $d/\sigma$, i.e., the
ratio between the separation and the width of the two point sources.

Hence, if the difference between the two photon numbers keeps finite
compared to the total photon number, i.e., $\epsilon=O(1)$, the estimation
precision $\mathcal{H}_{d}$ vanishes as $d$ goes to $0$, implying
the Rayleigh's curse applies again. But if the order of the ratio
$\dn$ is higher than or equal to $O(d/\sigma)$, the magnitude of
the denominator of $\hd$ becomes comparable to the numerator, so
the estimation precision $\mathcal{H}_{d}$ will be independent of
$d$, and one can still acquire information about the separation $d$
in this case. In particular, if the order of $\epsilon$ is higher
than $O(d/\sigma)$, $\hd$ will be restored to the original estimation
precision $\hdo$ with equal photon numbers obtained in \cite{Tsang2016}.
If the order of $\epsilon$ is equal to $O(d/\sigma)$, the precision
$\hd$ does not vanish, but will be decreased by a factor \eqref{eq:ratio},
\begin{equation}
\frac{1-\epsilon^{2}}{8\epsilon^{2}\left(\frac{\sigma}{d}\right)^{2}+1}.
\end{equation}
This explicitly shows the influence of the photon number difference
in the estimation precision of superresolution.

To illustrate the above result numerically, the ratio of the precision
$\mathcal{H}_{d}$ to the optimal precision $\hdo$ \eqref{eq:hdo-2}
for two Gaussian point-spread functions with the same width is plotted
with respect to $\epsilon$ for different $d/\sigma$ in Fig. \ref{fig:w=00003D00003D0}.
It can be seen that for a fixed separation, the precision of the separation
reaches the maximum when $\epsilon=0$, i.e., the photon numbers of
the two point sources are equal, which is the regime of the original
superresolution in \cite{Tsang2016}, and decreases to zero when the photon
number difference increases, which is the reoccurrence of Rayleigh's
curse found in \cite{Rehacek2017}. It also shows that when $\epsilon$
is nonzero but sufficiently small, the precision $\mathcal{H}_{d}$
will still be significant and the superresolution can work, showing
the effective range of quantum superresolution which is broader than
$\epsilon=0$. Moreover, the figure also suggests that for the same
$\epsilon$, a smaller separation compared to the width of the Gaussian
point-spread function leads to a lower estimation precision for $d$.
This can be understood as for a smaller $d$, the magnitude of $\epsilon$
becomes larger compared to $d/\sigma$, so it is more difficult to
satisfy the effective condition of the superresolution and the estimation
precision of $d$ decreases as a result.

\subsection{Gaussian point-spread functions with arbitrary widths $\sigma_{1}$
and $\sigma_{2}$}

For two Gaussian point-spread functions with arbitrary widths $\sigma_{1}$
and $\sigma_{2}$, the constant terms in Eq. \eqref{eq:w=00003D00003D0}
can be obtained as
\begin{equation}
\begin{aligned}A_{0}= & 8\nt\left(1-\epsilon^{2}\right)\left(\sigma_{1}-\sigma_{2}\right){}^{2}\left(\sigma_{1}^{2}+\sigma_{2}^{2}\right){}^{2},\\
B_{0}= & 16\left(\sigma_{1}-\sigma_{2}\right){}^{2}\left(\sigma_{1}^{2}+\sigma_{2}^{2}\right){}^{2}\left[\sigma_{1}^{2}(1+\epsilon)+\sigma_{2}^{2}(1-\epsilon)\right].
\end{aligned}
\end{equation}

\begin{figure}[!h]
\begin{centering}
\includegraphics[scale=0.92]{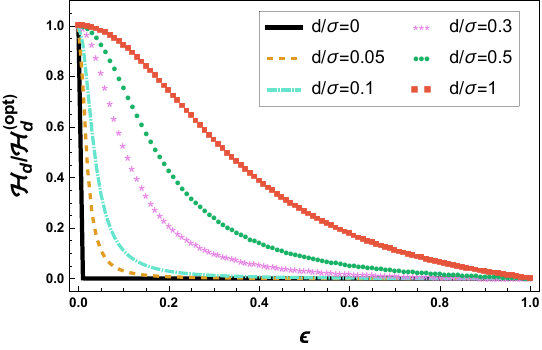}
\par\end{centering}
\caption{\protect\label{fig:w=00003D00003D0}Plot of the ratio of the estimation
precision $\mathcal{H}_{d}$ to the optimal precision $\protect\hdo$
with respect to the ratio $\protect\dn$ for two Gaussian point-spread
functions given different ratios $d/\sigma$. $d$ is the separation
of the two point sources, and $\sigma$ is the width of the two Gaussian
point-spread functions. It can be observed from the figure that when
the photon numbers of the two sources are the same, i.e., $\protect\dn=0$,
the estimation precision can always reach the optimum, and when $\epsilon>0$,
the estimation precision decreases. The ratio $d/\sigma$ influences
how fast the precision $\protect\hd$ decays to zero, i.e., the speed
that $\protect\hd$ transits from the regime of quantum superresolution
to the regime of Rayleigh's curse, when the photon number difference
increases. It verifies Eq. \eqref{eq:r3} that $d/\sigma=0$ is the
minimal point of $\protect\hd/\protect\hdo$, so the large the ratio
$d/\sigma$ is, the higher the precision $\protect\hd$ is and hence
the slower $\protect\hd$ decays to zero, and vice versa.}
\end{figure}

If we take the limit $d\rightarrow0$ immediately in $\hd$ \eqref{eq:hdexpand},
the estimation precision $\hd$ turns out to be 
\begin{equation}
\dlm{\hd}=\frac{1-\epsilon^{2}}{1-2\eta\epsilon+\eta^{2}}\hdo,\label{eq:op}
\end{equation}
where 
\begin{equation}
\eta=\frac{\sigma_{2}-\sigma_{1}}{\sigma_{1}+\sigma_{2}},\;\st=\sigma_{1}+\sigma_{2},\label{eq:eta}
\end{equation}
and $\hdo$ is the optimal estimation precision when the widths of
the two Gaussian point-spread functions are the same, $\sigma_{1}=\sigma_{2}=\sigma$
and the photon numbers of the two point-sources are equal, $\nb[1]=\nb[2]=\nb$,
\begin{equation}
\hdo=\frac{\nt}{\st[2]}=\frac{\nb}{2\sigma^{2}}.\label{eq:hdo}
\end{equation}

Eq. \eqref{eq:op} tells that the ratio $\hd/\hdo$ is determined
by the ration $\eta$ between the difference and the total of the
widths of the two Gaussian point-spread functions and the ratio $\dn$
between the difference and the total of the two photon numbers only.

It is worth mentioning that there is a trade-off relation between
the ratios $\dn$ and $\eta$ in the estimation precision $\hd$ \eqref{eq:op}.
By optimizing $\hd$ \eqref{eq:op} over all $\dn$ and $\eta$, it
can found that the estimation of the separation $d$ reaches the highest
precision when $\dn=\eta$, and the precision $\hd$ is exactly the
optimal precision $\hdo$ \eqref{eq:hdo} in this case. This implies
that the optimal precision of estimating the separation $d$ can even
be achieved when the photon numbers of the two point sources are different,
if the widths of the two Gaussian point-spread functions are also
different and the condition $\dn=\eta$ is satisfied.

It should be noted that such a result of $\hd$ in the limit $d\rightarrow0$
implicitly assumes that the order of the leading terms $A_{0}$ and
$B_{0}$ in $\hd$ \eqref{eq:hdexpand} is $O(1)$, i.e., the width
difference $\sigma_{2}-\sigma_{1}$ is sufficiently large compared
to the separation $d$. When the two Gaussian optical sources are
just slightly different in the widths, i.e., $\sigma_{2}-\sigma_{1}$
is small, higher-order terms of $d$ in $\hd$ can also come into
play in determining $\hd$ when $d\rightarrow0$. Similar to the preceding
section, the precision $\mathcal{H}_{d}$ for the separation $d$
will be determined by the competition between the difference in the
widths of Gaussian point-spread functions and the difference in the
photon numbers in this case.

To characterize the magnitudes of the difference between the two widths
and the difference between the two photon numbers, we denote the ratio
between the difference and the total of the photon numbers and the
ratio between the difference and the total of the two widths in the
powers of $\tilde{d}$ as
\begin{equation}
\dn=\frac{\nb[2]-\nb[1]}{\nt}=O(\dt[s]),\;\eta=\frac{\sigma_{2}-\sigma_{1}}{\st}=O(\dt[t]),\label{eq:eeta}
\end{equation}
so that we can compare the magnitudes of the photon number difference
and the width difference and obtain the estimation precision $\hd$
of the separation $d$ in different ranges of the these parameters\textcolor{blue}{.
}For the simplicity of computation, we redefine the dimensionless
separation $\dt$\eqref{eq:dt} as
\begin{equation}
\dt=\frac{d}{(\sigma_{1}+\sigma_{2})/2}.\label{eq:dt0}
\end{equation}
In the limit $d\rightarrow0$, $\sigma_{1}=\sigma_{2}$ as the difference
between the two point-spread functions is quantified in terms of the
power of $\dt$ \eqref{eq:eeta}, and $\dt$ can now be simplified
to
\begin{equation}
\dt=\frac{d}{\sigma}.\label{eq:dt1}
\end{equation}

\begin{table}[!t]
\begin{centering}
\caption{\protect\label{tab:The-detailed-results-1}Summary of the ratios $\protect\hd/\protect\hdo$
\eqref{eq:hdo} for Gaussian point-spread functions with width difference
$\eta=z\protect\dt[t]$ and photon number difference $\epsilon=y\protect\dt[s]$.
The blue region denotes the cases that reach the optimal estimation
precision $\protect\hdo$, the red region denotes the cases that the
estimation precision vanishes, and the yellow region denotes the cases
that the superresolution work with a lower precision. The case $t\protect\geq2$,
$s\protect\geq2$, where the magnitudes of the coefficients $z$ and
$y$ are $O(1)$, includes the original superresolution scheme for
two Gaussian point spread functions, and the case $t\protect\geq2$,
$s=0$ is the resurgence of Rayleigh's curse where the estimation
precision of the separation vanishes. All the results are evaluated
in the limit $d\rightarrow0$, and the dimensionless separation $\protect\dt$
is given by Eq. \eqref{eq:dt1}.}
\LyXZeroWidthSpace{}
\par\end{centering}
\noindent\centering{}%
\begin{tabular}{|>{\centering}p{1.5cm}|>{\centering}p{2cm}cc|}
\hline 
 & \multicolumn{1}{>{\centering}p{2cm}|}{$\begin{gathered}\\s\geq2\\
\\\end{gathered}
$} & \multicolumn{1}{c|}{$\begin{gathered}\\s=1\\
\\\end{gathered}
$} & $\begin{gathered}\\s=0\\
\\\end{gathered}
$\tabularnewline
\hline 
\begin{spacing}{0}
$\begin{gathered}\\t>2\\
\\\end{gathered}
$
\end{spacing}
 & \cellcolor{lightyellow} & \cellcolor{lightblue}$\begin{gathered}\\\frac{1}{1+8y^{2}}\\
\\\end{gathered}
$ & \cellcolor{lightred}$\begin{gathered}\\\\\\0
\end{gathered}
$\tabularnewline
\cline{1-1}
$\begin{gathered}\\t=2\\
\\\end{gathered}
$ & \cellcolor{lightyellow}$\begin{gathered}\\\\1
\end{gathered}
$ & \cellcolor{lightblue}$\begin{gathered}\\\frac{1+64z^{2}}{1+64z^{2}+8y^{2}}\\
\\\end{gathered}
$ & \cellcolor{lightred}\tabularnewline
\cline{1-1}
$\begin{gathered}\\t=1\\
\\\end{gathered}
$ & \multicolumn{2}{c}{\cellcolor{lightyellow}} & \cellcolor{lightblue}$\begin{gathered}\\\frac{8\left(1-y^{2}\right)z^{2}}{y^{2}+8z^{2}}\\
\\\end{gathered}
$\tabularnewline
\cline{1-1}
$\begin{gathered}\\t=0\\
\\\end{gathered}
$ & \multicolumn{2}{c}{\cellcolor{lightblue}$\begin{gathered}\\\frac{1}{1+z^{2}}\\
\\\end{gathered}
$} & \cellcolor{lightblue}$\begin{gathered}\\\frac{1-y^{2}}{1-2yz+z^{2}}\\
\\\end{gathered}
$\tabularnewline
\hline 
\end{tabular}
\end{table}

Table \ref{tab:The-detailed-results-1} summarizes the results of
$\mathcal{H}_{d}$ in terms of the ratio between $\hd$ and the optimal
estimation precision $\hdo$ \eqref{eq:hdo} for two arbitrary Gaussian
point-spread functions in the limit $d\rightarrow0$ for all feasible
ranges of the exponents $t$ and $s$, which shows that different
magnitudes of the difference in the photon numbers and the difference
in the widths lead to different estimation precisions of the separation.
Similar as the general results in the preceding section, the Rayleigh's
criterion applies only in the case that the widths of the two Gaussian
point-distribution functions are sufficiently close, i.e., $t\geq2$,
and the difference of the photon numbers keeps finite compared to
the total photon number, i.e., $\epsilon=O(1)$. For all the other
cases, the superresolution scheme can work. The case $t\geq2$, i.e.,
the ratio of the difference of the widths to the total of the widths
is of order $O(\dt[2])$ or higher, with $\epsilon\leq O(\dt[2])$,
includes the case considered by the original superresolution scheme
\cite{Tsang2016} for two identical Gaussian optical sources with
equal photon numbers. For the other regimes of the above two ratios,
the quantum resolution protocol is still effective, but the estimation
precision may vary. In particular, when $t\leq1$, i.e., the order
of the ratio between the difference and the total of the two Gaussian
widths is lower than or equal to $O(\dt)$, the superresolution can
always work regardless of the photon number difference, implying the
difference between the two Gaussian point-spread functions is sufficient
to make the superresolution work in this case, which is a manifestation
of the effect of the difference between the point-spread functions
in extending the effective regime of superresolution.

\begin{figure}[H]
\begin{centering}
\subfloat[\centering\label{fig:a}]{\centering{}\includegraphics[scale=0.92]{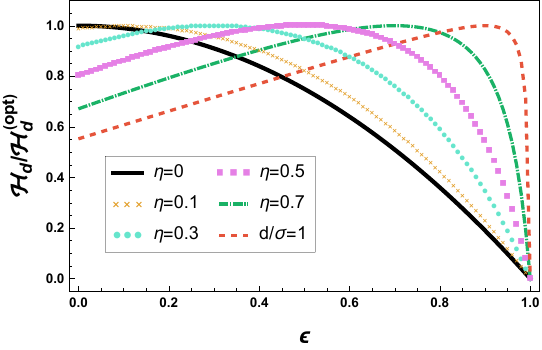}

}
\par\end{centering}
\begin{centering}
\subfloat[\centering\label{fig:b}]{\centering{}\includegraphics[scale=0.92]{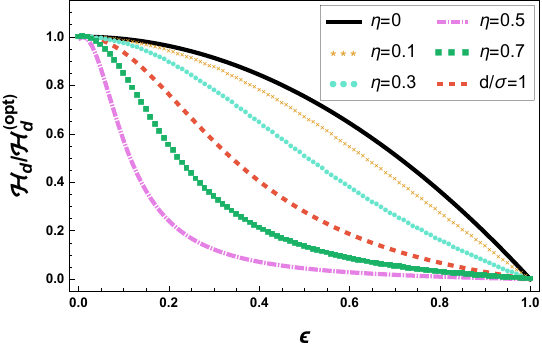}

}
\par\end{centering}
\caption{\protect\label{fig:different=000020width}Plot of the ratios between
the estimation precision $\mathcal{H}_{d}$ and the optimal estimation
precision $\protect\hdo$ with respect to the ratio $\protect\dn$
for two different Gaussian point-spread functions. (a) The estimation
precision $\protect\hd$ for different ratios $\eta$ of the width
difference to the total width of the two sources in the limit $d\rightarrow0$.
It can be observed that the precision $\protect\hd$ can still reach
the maximum though the photon numbers of the two point sources are
different if the point-spread functions of the two point sources are
not identical, and the highest precision is achieved when $\protect\dn=\eta$.
The transition of $\protect\hd$ to the regime of Rayleigh's curse
starts when the difference of the photon numbers increases beyond
the point $\protect\dn=\eta$, and the larger $\eta$ is, the faster
the estimation precision $\protect\hd$ decays with $\protect\dn$.
The exact effect of $\eta$ on the decay of the estimation precision
is determined by Eq. \eqref{eq:op}. (b) The estimation precision
$\protect\hd$ for different dimensionless separations $\protect\dt$
\eqref{eq:dt0}. The ratio $\eta$ \eqref{eq:eeta} is $0.005$. The
estimation precision can always reach the maximum when the photon
numbers of the two sources are equal, i.e., $\protect\dn=0$, and
decays to zero as $\protect\dn$ increases. The dimensionless separation
$\protect\dt$ changes the decay speed of the estimation precision.
$\protect\hd/\protect\hdo$ has two extreme points in this case: the
maximal point is $\protect\dt=0$, and the minimal point is $\protect\dt\approx0.2$,
which can be determined by Eq. \eqref{eq:extreme}. So, when $0\protect\leq\protect\dt\apprle0.2$,
$\protect\hd$ becomes smaller and decays faster with $\epsilon$
when $\protect\dt$ increases, and when $\protect\dt\apprge0.2$,
$\protect\hd$ behaves in the opposite way.}
\end{figure}

The superresolution precision $\mathcal{H}_{d}$ for two different
Gaussian point-spread functions is plotted in terms of the ratio between
$\hd$ and the optimal estimation precision $\hdo$ with respective
to $\epsilon$ in Fig. \ref{fig:different=000020width}. It can be
seen from Fig. \ref{fig:a} that given the separation $d=0$, for
each $\eta$, the optimal estimation precision can always be achieved,
but by different $\dn$. In fact, the estimation precision reaches
the maximum when $\dn=\eta$, which can be obtained by optimizing
Eq. \eqref{eq:op} over $\dn$ and $\eta$ as discussed above. And
the figure verifies that optimization result. Fig. \ref{fig:b} considers
the case $\eta=0.005$ for different dimensionless separations $\dt$
of the two Gaussian point-spread functions. It shows that the estimation
can always reach the highest precision and the quantum superresolution
works when the photon numbers of the two sources are equal, i.e.,
$\dn=0$. And as $\dn$ increases, the estimation precision $\hd$
will decrease to zero and the Rayleigh's curse takes effect again.

Moreover, one can see that the precision $\hd$ does not change monotonically
with the dimensionless separation $\dt$ and the ratio $\eta$. In
fact, there exist extreme points of $\hd$ with respect to $\dt$
and $\eta$, when $\epsilon$ is fixed.

When $\eta\neq0$, i.e., the widths of the two Gaussian point-spread
functions are different, the extreme points of $\hd$ with respect
to $\dt$ can be worked out as
\begin{equation}
\begin{aligned}\dt_{1}= & 0,\\
\dt_{2}= & 2\sqrt{\eta^{2}+1}\sqrt{w_{0}+1},
\end{aligned}
\label{eq:extreme}
\end{equation}
where $w_{0}$ is the value of the principal branch $W_{0}$ of the
Lambert W function at $\frac{\eta^{2}-1}{e\left(\eta^{2}+1\right)}$,
i.e., $w_{0}=W_{0}\left[\frac{\eta^{2}-1}{e\left(\eta^{2}+1\right)}\right]$.
When the ratio $\eta$ is small, we can expand the above second solution
to $\dt$ to the lowest order of $\eta$,
\begin{equation}
\dt_{2}\approx2\sqrt{2\eta}.
\end{equation}

In order to determine the maximal point and the minimal point from
the above two extremal solutions, we compute the second-order derivative
of $\hd$ with respect to $\dt$, and it turns out that
\begin{equation}
\left.\partial_{\dt}^{2}\hd\right|_{\dt=r_{1}}=-\frac{\left(1-\eta^{2}\right)\left(1-\epsilon^{2}\right)\left(-2\eta+\eta^{2}\epsilon+\epsilon\right)^{2}}{4\eta^{2}\left(\eta^{2}+1\right)^{2}\left(\eta^{2}-2\eta\epsilon+1\right)^{2}}<0,
\end{equation}
as $|\eta|<1$ and $|\epsilon|<1$. Therefore, $\dt_{1}$ is the maximal
point and $\dt_{2}$ is the minimum point.

This is why $\dt=0$ gives the maximal value in Fig. \ref{fig:b}.
And in that figure, $\eta=0.005$, which leads to the minimal point
$\dt\approx0.2$, so $\hd/\hdo$ decreases before $\dt=0.2$ and increases
after that point. The maximal and the minimal values of $\hd$ can
be obtained as
\begin{equation}
\begin{aligned}\left.\hd\right|_{\dt=\dt_{1}}= & \frac{1-\epsilon^{2}}{\eta^{2}-2\eta\epsilon+1}\hdo,\\
\left.\hd\right|_{\dt=\dt_{2}}= & \frac{\left(1-\epsilon^{2}\right)\left[\eta^{2}+w_{0}\left(\eta^{2}-2\eta\epsilon+1\right)+1\right]\hdo}{\left(1-\eta^{2}\right)^{2}\left(1-\epsilon^{2}\right)w_{0}+\left(1+\eta^{2}\right)\left(\eta^{2}-2\eta\epsilon+1\right)}.
\end{aligned}
\end{equation}
Interestingly, it can be verified that when $\dt\rightarrow\pm\infty$,
i.e., the separation between the two point sources is sufficiently
large, the limit of $\hd$ is
\begin{equation}
\lim_{\dt\rightarrow\pm\infty}=\frac{1-\epsilon^{2}}{\eta^{2}-2\eta\epsilon+1}\hdo,\label{eq:rinf}
\end{equation}
which coincides with the maximal value of $\hd$ at $\dt=0$. This
is why the value of $\hd/\hdo$ cannot be larger than that at $\dt=0$
when $\dt$ goes beyond the minimal point in Fig. \ref{fig:b} though
it increases with $\dt$ in that regime.

When $\eta=0$, i.e., the widths of the two Gaussian point-spread
functions are the same, it turns out that $\hd$ has only one extreme
point,
\begin{equation}
\dt_{3}=0.\label{eq:r3}
\end{equation}
If we compute the second-order derivative of $\hd$ with respect to
$\dt$, it shows that
\begin{equation}
\left.\partial_{\dt}^{2}\hd\right|_{\dt=\dt_{3}}=\left(\frac{1}{\epsilon^{2}}-1\right)\hdo>0,
\end{equation}
as $|\epsilon|<1$. So, in sharp contrast to the case $\eta\neq0$,
the point $\dt=0$ becomes the minimal point in the case $\eta=0$.
This is why $\hd/\hdo$ is the smallest at $\dt=0$ in Fig. \ref{fig:w=00003D00003D0}.

\begin{figure}[!h]
\begin{centering}
\subfloat[\centering$\epsilon=0$]{\centering{}\includegraphics[scale=0.28]{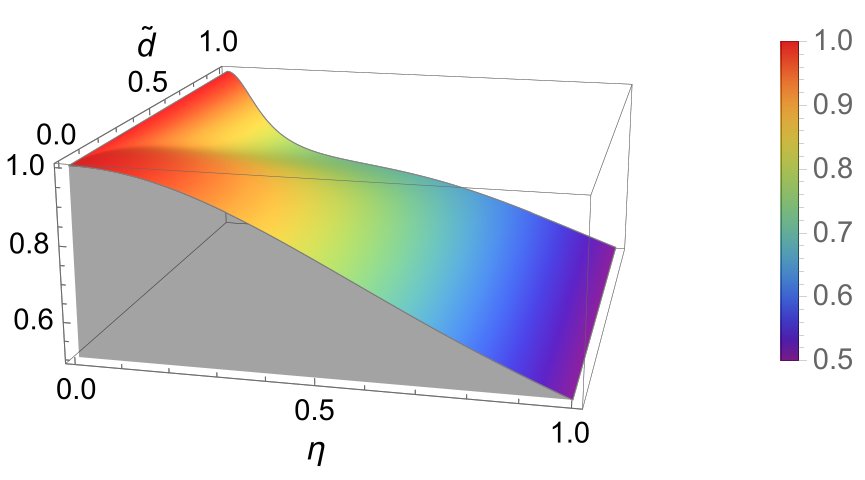}}\ \subfloat[\centering$\epsilon=0.25$]{\centering{}\includegraphics[scale=0.28]{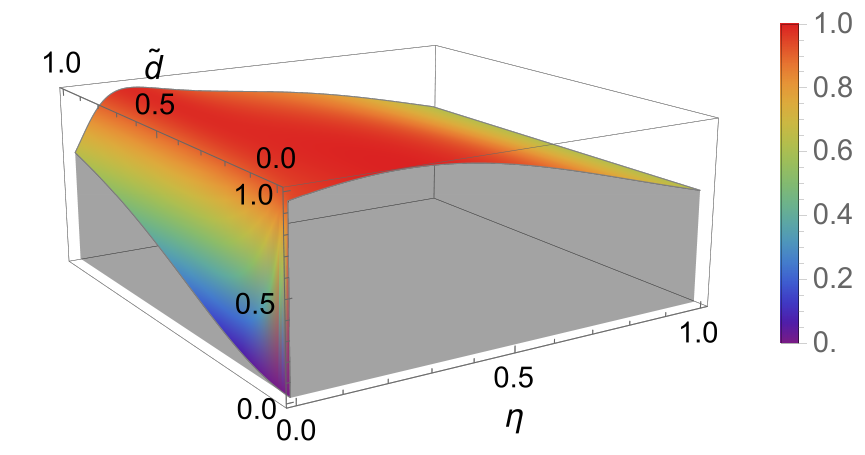}}
\par\end{centering}
\begin{centering}
\smallskip{}
\subfloat[\centering$\epsilon=0.49$]{\includegraphics[scale=0.28]{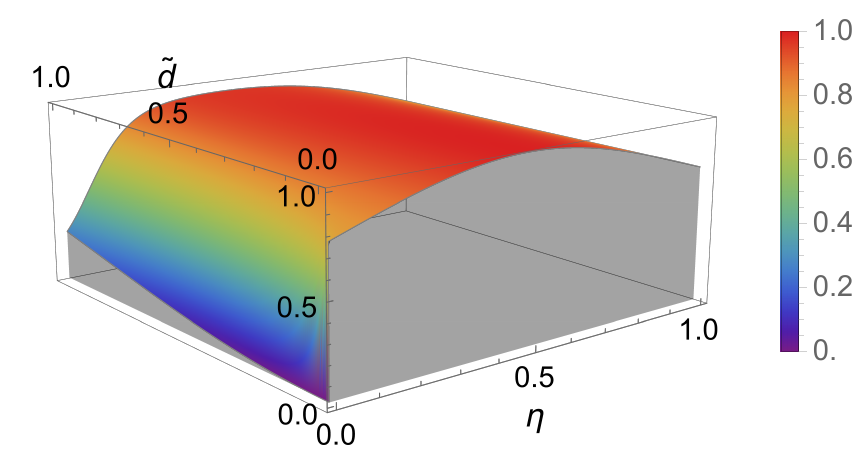}

}\ \subfloat[\centering$\epsilon=0.64$]{\includegraphics[scale=0.28]{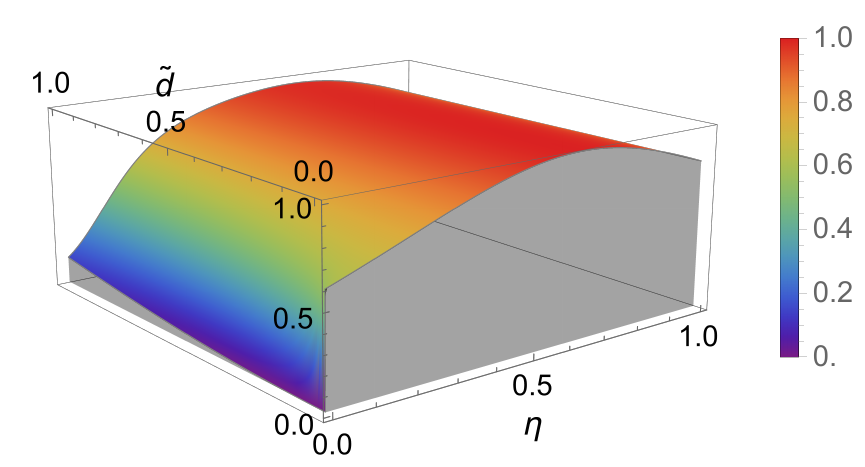}}
\par\end{centering}
\caption{\protect\label{fig:etr}The ratio $\protect\hd/\protect\hdo$ with
respect to the $\eta$ and $\protect\dt$ for different $\epsilon$.
The plots show the discontinuity of $\protect\hd/\protect\hdo$ at
the point $\eta=0$, $\protect\dt=0$. Whenever $\epsilon\protect\neq0$,
the limit of $\protect\hd/\protect\hdo$ is zero when $\protect\dt\rightarrow0$
given $\eta=0$ and nonzero when $\eta\rightarrow0$ given $\protect\dt=0$,
so the limit $\protect\dt\rightarrow0$ and the limit $\eta\rightarrow0$
does not commute for the precision $\protect\hd$ in this case. When
$\epsilon=0$, however, the values of $\protect\hd/\protect\hdo$
in these two limits coincide, so the discontinuity of $\protect\hd/\protect\hdo$
does not occur in this case.}
\end{figure}

Finally, it is worth noting that the precision $\hd$ for the case
$\sigma_{1}=\sigma_{2}$ and $\dt\rightarrow0$ in Fig. \ref{fig:w=00003D00003D0}
behaves dramatically different from that for $\dt=0$ and $\eta\rightarrow0$
in Fig. \ref{fig:a}. The mechanism of this difference lies in the
discontinuity of $\hd$ at $\dt=0$, $\eta=0$. It can be verified
that
\begin{equation}
\begin{aligned}\lim_{\dt\rightarrow0}\lim_{\eta\rightarrow0}\hd= & 0,\\
\lim_{\eta\rightarrow0}\lim_{\dt\rightarrow0}\hd= & (1-\epsilon^{2})\hdo,
\end{aligned}
\end{equation}
according to Eq. \eqref{eq:eta0} and \eqref{eq:op} whenever $\epsilon\neq0$,
so
\begin{equation}
\lim_{\dt\rightarrow0}\lim_{\eta\rightarrow0}\hd\neq\lim_{\eta\rightarrow0}\lim_{\dt\rightarrow0}\hd.
\end{equation}
In Fig. \ref{fig:etr}, the precision $\hd$ is plotted with respect
to $\dt$ and $\eta$ for different $\epsilon$ to illustrate this
discontinuity. Hence, $\hd$ has different behaviors with respect
to $\epsilon$ in the above two cases.

\section{Conclusions}

The superresolution is an important quantum technique to surpass the
Rayleigh's criterion in resolving two closely located point sources,
manifesting the power of quantum estimation theory in the field of
quantum imaging. Nevertheless, the vanishing of the estimation precision
when the photon numbers of the two point sources are unknown imposes
limitations on its applicability. This work studies the condition
for the superresolution technique to work by analyzing the competing
roles of the difference in the photon numbers and the difference in
the point-spread functions between the two optical sources as well
as the separation itself in determining the estimation precision of
the separation. The results not only provide conditions for the realizability
of quantum superresolution, but also show various precision limits
of superresolution in different regimes of parameters, which indicates
that the competition between the difference of the photon numbers,
the difference of the point-spread functions and the magnitude of
the separation plays a key role in the validity of superresolution.
We hope this work can provide new insights into the quantum theory
of superresolution and extend the applications of quantum superresolution
in practice.
\begin{acknowledgments}
The authors acknowledge the helpful discussions with Jingyi Fan and
Yutong Huang. This work is supported by the National Natural Science
Foundation of China (Grant No. 12075323).
\end{acknowledgments}

\bibliographystyle{apsrev4-2}
\bibliography{superresolution}

\begin{thebibliography}{109}%
\makeatletter
\providecommand \@ifxundefined [1]{%
 \@ifx{#1\undefined}
}%
\providecommand \@ifnum [1]{%
 \ifnum #1\expandafter \@firstoftwo
 \else \expandafter \@secondoftwo
 \fi
}%
\providecommand \@ifx [1]{%
 \ifx #1\expandafter \@firstoftwo
 \else \expandafter \@secondoftwo
 \fi
}%
\providecommand \natexlab [1]{#1}%
\providecommand \enquote  [1]{``#1''}%
\providecommand \bibnamefont  [1]{#1}%
\providecommand \bibfnamefont [1]{#1}%
\providecommand \citenamefont [1]{#1}%
\providecommand \href@noop [0]{\@secondoftwo}%
\providecommand \href [0]{\begingroup \@sanitize@url \@href}%
\providecommand \@href[1]{\@@startlink{#1}\@@href}%
\providecommand \@@href[1]{\endgroup#1\@@endlink}%
\providecommand \@sanitize@url [0]{\catcode `\\12\catcode `\$12\catcode `\&12\catcode `\#12\catcode `\^12\catcode `\_12\catcode `\%12\relax}%
\providecommand \@@startlink[1]{}%
\providecommand \@@endlink[0]{}%
\providecommand \url  [0]{\begingroup\@sanitize@url \@url }%
\providecommand \@url [1]{\endgroup\@href {#1}{\urlprefix }}%
\providecommand \urlprefix  [0]{URL }%
\providecommand \Eprint [0]{\href }%
\providecommand \doibase [0]{https://doi.org/}%
\providecommand \selectlanguage [0]{\@gobble}%
\providecommand \bibinfo  [0]{\@secondoftwo}%
\providecommand \bibfield  [0]{\@secondoftwo}%
\providecommand \translation [1]{[#1]}%
\providecommand \BibitemOpen [0]{}%
\providecommand \bibitemStop [0]{}%
\providecommand \bibitemNoStop [0]{.\EOS\space}%
\providecommand \EOS [0]{\spacefactor3000\relax}%
\providecommand \BibitemShut  [1]{\csname bibitem#1\endcsname}%
\let\auto@bib@innerbib\@empty
%</preamble>
\bibitem [{\citenamefont {Rushforth}\ and\ \citenamefont {Harris}(1968)}]{Rushforth1968}%
  \BibitemOpen
  \bibfield  {author} {\bibinfo {author} {\bibfnamefont {C.~K.}\ \bibnamefont {Rushforth}}\ and\ \bibinfo {author} {\bibfnamefont {R.~W.}\ \bibnamefont {Harris}},\ }\href {https://doi.org/10.1364/JOSA.58.000539} {\bibfield  {journal} {\bibinfo  {journal} {Journal of the Optical Society of America}\ }\textbf {\bibinfo {volume} {58}},\ \bibinfo {pages} {539} (\bibinfo {year} {1968})}\BibitemShut {NoStop}%
\bibitem [{\citenamefont {{P{\'e}rez-Delgado}}\ \emph {et~al.}(2012)\citenamefont {{P{\'e}rez-Delgado}}, \citenamefont {Pearce},\ and\ \citenamefont {Kok}}]{Perez-Delgado2012}%
  \BibitemOpen
  \bibfield  {author} {\bibinfo {author} {\bibfnamefont {C.~A.}\ \bibnamefont {{P{\'e}rez-Delgado}}}, \bibinfo {author} {\bibfnamefont {M.~E.}\ \bibnamefont {Pearce}},\ and\ \bibinfo {author} {\bibfnamefont {P.}~\bibnamefont {Kok}},\ }\href {https://doi.org/10.1103/PhysRevLett.109.123601} {\bibfield  {journal} {\bibinfo  {journal} {Physical Review Letters}\ }\textbf {\bibinfo {volume} {109}},\ \bibinfo {pages} {123601} (\bibinfo {year} {2012})}\BibitemShut {NoStop}%
\bibitem [{\citenamefont {Denk}\ \emph {et~al.}(1990)\citenamefont {Denk}, \citenamefont {Strickler},\ and\ \citenamefont {Webb}}]{Denk1990}%
  \BibitemOpen
  \bibfield  {author} {\bibinfo {author} {\bibfnamefont {W.}~\bibnamefont {Denk}}, \bibinfo {author} {\bibfnamefont {J.~H.}\ \bibnamefont {Strickler}},\ and\ \bibinfo {author} {\bibfnamefont {W.~W.}\ \bibnamefont {Webb}},\ }\href {https://doi.org/10.1126/science.2321027} {\bibfield  {journal} {\bibinfo  {journal} {Science}\ }\textbf {\bibinfo {volume} {248}},\ \bibinfo {pages} {73} (\bibinfo {year} {1990})}\BibitemShut {NoStop}%
\bibitem [{\citenamefont {Zipfel}\ \emph {et~al.}(2003)\citenamefont {Zipfel}, \citenamefont {Williams},\ and\ \citenamefont {Webb}}]{Zipfel2003}%
  \BibitemOpen
  \bibfield  {author} {\bibinfo {author} {\bibfnamefont {W.~R.}\ \bibnamefont {Zipfel}}, \bibinfo {author} {\bibfnamefont {R.~M.}\ \bibnamefont {Williams}},\ and\ \bibinfo {author} {\bibfnamefont {W.~W.}\ \bibnamefont {Webb}},\ }\href {https://doi.org/10.1038/nbt899} {\bibfield  {journal} {\bibinfo  {journal} {Nature Biotechnology}\ }\textbf {\bibinfo {volume} {21}},\ \bibinfo {pages} {1369} (\bibinfo {year} {2003})}\BibitemShut {NoStop}%
\bibitem [{\citenamefont {Helmchen}\ and\ \citenamefont {Denk}(2005)}]{Helmchen2005}%
  \BibitemOpen
  \bibfield  {author} {\bibinfo {author} {\bibfnamefont {F.}~\bibnamefont {Helmchen}}\ and\ \bibinfo {author} {\bibfnamefont {W.}~\bibnamefont {Denk}},\ }\href {https://doi.org/10.1038/nmeth818} {\bibfield  {journal} {\bibinfo  {journal} {Nature Methods}\ }\textbf {\bibinfo {volume} {2}},\ \bibinfo {pages} {932} (\bibinfo {year} {2005})}\BibitemShut {NoStop}%
\bibitem [{\citenamefont {Durant}\ \emph {et~al.}(2006)\citenamefont {Durant}, \citenamefont {Liu}, \citenamefont {Steele},\ and\ \citenamefont {Zhang}}]{Durant2006}%
  \BibitemOpen
  \bibfield  {author} {\bibinfo {author} {\bibfnamefont {S.}~\bibnamefont {Durant}}, \bibinfo {author} {\bibfnamefont {Z.}~\bibnamefont {Liu}}, \bibinfo {author} {\bibfnamefont {J.~M.}\ \bibnamefont {Steele}},\ and\ \bibinfo {author} {\bibfnamefont {X.}~\bibnamefont {Zhang}},\ }\href {https://doi.org/10.1364/JOSAB.23.002383} {\bibfield  {journal} {\bibinfo  {journal} {Journal of the Optical Society of America B}\ }\textbf {\bibinfo {volume} {23}},\ \bibinfo {pages} {2383} (\bibinfo {year} {2006})}\BibitemShut {NoStop}%
\bibitem [{\citenamefont {Jacob}\ \emph {et~al.}(2006)\citenamefont {Jacob}, \citenamefont {Alekseyev},\ and\ \citenamefont {Narimanov}}]{Jacob2006}%
  \BibitemOpen
  \bibfield  {author} {\bibinfo {author} {\bibfnamefont {Z.}~\bibnamefont {Jacob}}, \bibinfo {author} {\bibfnamefont {L.~V.}\ \bibnamefont {Alekseyev}},\ and\ \bibinfo {author} {\bibfnamefont {E.}~\bibnamefont {Narimanov}},\ }\href {https://doi.org/10.1364/OE.14.008247} {\bibfield  {journal} {\bibinfo  {journal} {Optics Express}\ }\textbf {\bibinfo {volume} {14}},\ \bibinfo {pages} {8247} (\bibinfo {year} {2006})}\BibitemShut {NoStop}%
\bibitem [{\citenamefont {Fei}\ \emph {et~al.}(1997)\citenamefont {Fei}, \citenamefont {Jost}, \citenamefont {Popescu}, \citenamefont {Saleh},\ and\ \citenamefont {Teich}}]{Fei1997}%
  \BibitemOpen
  \bibfield  {author} {\bibinfo {author} {\bibfnamefont {H.-B.}\ \bibnamefont {Fei}}, \bibinfo {author} {\bibfnamefont {B.~M.}\ \bibnamefont {Jost}}, \bibinfo {author} {\bibfnamefont {S.}~\bibnamefont {Popescu}}, \bibinfo {author} {\bibfnamefont {B.~E.~A.}\ \bibnamefont {Saleh}},\ and\ \bibinfo {author} {\bibfnamefont {M.~C.}\ \bibnamefont {Teich}},\ }\href {https://doi.org/10.1103/PhysRevLett.78.1679} {\bibfield  {journal} {\bibinfo  {journal} {Physical Review Letters}\ }\textbf {\bibinfo {volume} {78}},\ \bibinfo {pages} {1679} (\bibinfo {year} {1997})}\BibitemShut {NoStop}%
\bibitem [{\citenamefont {Abouraddy}\ \emph {et~al.}(2001)\citenamefont {Abouraddy}, \citenamefont {Saleh}, \citenamefont {Sergienko},\ and\ \citenamefont {Teich}}]{Abouraddy2001}%
  \BibitemOpen
  \bibfield  {author} {\bibinfo {author} {\bibfnamefont {A.~F.}\ \bibnamefont {Abouraddy}}, \bibinfo {author} {\bibfnamefont {B.~E.~A.}\ \bibnamefont {Saleh}}, \bibinfo {author} {\bibfnamefont {A.~V.}\ \bibnamefont {Sergienko}},\ and\ \bibinfo {author} {\bibfnamefont {M.~C.}\ \bibnamefont {Teich}},\ }\href {https://doi.org/10.1103/PhysRevA.64.050101} {\bibfield  {journal} {\bibinfo  {journal} {Physical Review A}\ }\textbf {\bibinfo {volume} {64}},\ \bibinfo {pages} {050101} (\bibinfo {year} {2001})}\BibitemShut {NoStop}%
\bibitem [{\citenamefont {Sokolov}\ \emph {et~al.}(2001)\citenamefont {Sokolov}, \citenamefont {Gatti}, \citenamefont {Kolobov},\ and\ \citenamefont {Lugiato}}]{Sokolov2001}%
  \BibitemOpen
  \bibfield  {author} {\bibinfo {author} {\bibfnamefont {I.~V.}\ \bibnamefont {Sokolov}}, \bibinfo {author} {\bibfnamefont {A.}~\bibnamefont {Gatti}}, \bibinfo {author} {\bibfnamefont {M.~I.}\ \bibnamefont {Kolobov}},\ and\ \bibinfo {author} {\bibfnamefont {L.~A.}\ \bibnamefont {Lugiato}},\ }\href {https://doi.org/10.1070/PU2001v044n11ABEH000995} {\bibfield  {journal} {\bibinfo  {journal} {Physics-Uspekhi}\ }\textbf {\bibinfo {volume} {44}},\ \bibinfo {pages} {1199} (\bibinfo {year} {2001})}\BibitemShut {NoStop}%
\bibitem [{\citenamefont {Song}\ \emph {et~al.}(2013)\citenamefont {Song}, \citenamefont {Xu}, \citenamefont {Wang}, \citenamefont {Xiong}, \citenamefont {Zhang}, \citenamefont {Cao},\ and\ \citenamefont {Wang}}]{Song2013}%
  \BibitemOpen
  \bibfield  {author} {\bibinfo {author} {\bibfnamefont {X.-B.}\ \bibnamefont {Song}}, \bibinfo {author} {\bibfnamefont {D.-Q.}\ \bibnamefont {Xu}}, \bibinfo {author} {\bibfnamefont {H.-B.}\ \bibnamefont {Wang}}, \bibinfo {author} {\bibfnamefont {J.}~\bibnamefont {Xiong}}, \bibinfo {author} {\bibfnamefont {X.}~\bibnamefont {Zhang}}, \bibinfo {author} {\bibfnamefont {D.-Z.}\ \bibnamefont {Cao}},\ and\ \bibinfo {author} {\bibfnamefont {K.}~\bibnamefont {Wang}},\ }\href {https://doi.org/10.1063/1.4822423} {\bibfield  {journal} {\bibinfo  {journal} {Applied Physics Letters}\ }\textbf {\bibinfo {volume} {103}},\ \bibinfo {pages} {131111} (\bibinfo {year} {2013})}\BibitemShut {NoStop}%
\bibitem [{\citenamefont {Boto}\ \emph {et~al.}(2000)\citenamefont {Boto}, \citenamefont {Kok}, \citenamefont {Abrams}, \citenamefont {Braunstein}, \citenamefont {Williams},\ and\ \citenamefont {Dowling}}]{Boto2000}%
  \BibitemOpen
  \bibfield  {author} {\bibinfo {author} {\bibfnamefont {A.~N.}\ \bibnamefont {Boto}}, \bibinfo {author} {\bibfnamefont {P.}~\bibnamefont {Kok}}, \bibinfo {author} {\bibfnamefont {D.~S.}\ \bibnamefont {Abrams}}, \bibinfo {author} {\bibfnamefont {S.~L.}\ \bibnamefont {Braunstein}}, \bibinfo {author} {\bibfnamefont {C.~P.}\ \bibnamefont {Williams}},\ and\ \bibinfo {author} {\bibfnamefont {J.~P.}\ \bibnamefont {Dowling}},\ }\href {https://doi.org/10.1103/PhysRevLett.85.2733} {\bibfield  {journal} {\bibinfo  {journal} {Physical Review Letters}\ }\textbf {\bibinfo {volume} {85}},\ \bibinfo {pages} {2733} (\bibinfo {year} {2000})}\BibitemShut {NoStop}%
\bibitem [{\citenamefont {Liao}\ \emph {et~al.}(2010)\citenamefont {Liao}, \citenamefont {{Al-Amri}},\ and\ \citenamefont {Suhail~Zubairy}}]{Liao2010}%
  \BibitemOpen
  \bibfield  {author} {\bibinfo {author} {\bibfnamefont {Z.}~\bibnamefont {Liao}}, \bibinfo {author} {\bibfnamefont {M.}~\bibnamefont {{Al-Amri}}},\ and\ \bibinfo {author} {\bibfnamefont {M.}~\bibnamefont {Suhail~Zubairy}},\ }\href {https://doi.org/10.1103/PhysRevLett.105.183601} {\bibfield  {journal} {\bibinfo  {journal} {Physical Review Letters}\ }\textbf {\bibinfo {volume} {105}},\ \bibinfo {pages} {183601} (\bibinfo {year} {2010})}\BibitemShut {NoStop}%
\bibitem [{\citenamefont {Rayleigh}(1880)}]{Rayleigh1880}%
  \BibitemOpen
  \bibfield  {author} {\bibinfo {author} {\bibfnamefont {L.}~\bibnamefont {Rayleigh}},\ }\href {https://doi.org/10.1093/mnras/40.4.254} {\bibfield  {journal} {\bibinfo  {journal} {Monthly Notices of the Royal Astronomical Society}\ }\textbf {\bibinfo {volume} {40}},\ \bibinfo {pages} {254} (\bibinfo {year} {1880})}\BibitemShut {NoStop}%
\bibitem [{\citenamefont {Helstrom}(1969)}]{Helstrom1969a}%
  \BibitemOpen
  \bibfield  {author} {\bibinfo {author} {\bibfnamefont {C.~W.}\ \bibnamefont {Helstrom}},\ }\href {https://doi.org/10.1007/BF01007479} {\bibfield  {journal} {\bibinfo  {journal} {Journal of Statistical Physics}\ }\textbf {\bibinfo {volume} {1}},\ \bibinfo {pages} {231} (\bibinfo {year} {1969})}\BibitemShut {NoStop}%
\bibitem [{\citenamefont {Giovannetti}\ \emph {et~al.}(2011)\citenamefont {Giovannetti}, \citenamefont {Lloyd},\ and\ \citenamefont {Maccone}}]{Giovannetti2011}%
  \BibitemOpen
  \bibfield  {author} {\bibinfo {author} {\bibfnamefont {V.}~\bibnamefont {Giovannetti}}, \bibinfo {author} {\bibfnamefont {S.}~\bibnamefont {Lloyd}},\ and\ \bibinfo {author} {\bibfnamefont {L.}~\bibnamefont {Maccone}},\ }\href {https://doi.org/10.1038/nphoton.2011.35} {\bibfield  {journal} {\bibinfo  {journal} {Nature Photonics}\ }\textbf {\bibinfo {volume} {5}},\ \bibinfo {pages} {222} (\bibinfo {year} {2011})}\BibitemShut {NoStop}%
\bibitem [{\citenamefont {Liu}\ \emph {et~al.}(2016)\citenamefont {Liu}, \citenamefont {Lu}, \citenamefont {Sun},\ and\ \citenamefont {Wang}}]{Liu2016}%
  \BibitemOpen
  \bibfield  {author} {\bibinfo {author} {\bibfnamefont {J.}~\bibnamefont {Liu}}, \bibinfo {author} {\bibfnamefont {X.-M.}\ \bibnamefont {Lu}}, \bibinfo {author} {\bibfnamefont {Z.}~\bibnamefont {Sun}},\ and\ \bibinfo {author} {\bibfnamefont {X.}~\bibnamefont {Wang}},\ }\href {https://doi.org/10.1088/1751-8113/49/11/115302} {\bibfield  {journal} {\bibinfo  {journal} {Journal of Physics A: Mathematical and Theoretical}\ }\textbf {\bibinfo {volume} {49}},\ \bibinfo {pages} {115302} (\bibinfo {year} {2016})}\BibitemShut {NoStop}%
\bibitem [{\citenamefont {Yang}\ \emph {et~al.}(2019)\citenamefont {Yang}, \citenamefont {Pang}, \citenamefont {Zhou},\ and\ \citenamefont {Jordan}}]{Yang2019}%
  \BibitemOpen
  \bibfield  {author} {\bibinfo {author} {\bibfnamefont {J.}~\bibnamefont {Yang}}, \bibinfo {author} {\bibfnamefont {S.}~\bibnamefont {Pang}}, \bibinfo {author} {\bibfnamefont {Y.}~\bibnamefont {Zhou}},\ and\ \bibinfo {author} {\bibfnamefont {A.~N.}\ \bibnamefont {Jordan}},\ }\href {https://doi.org/10.1103/PhysRevA.100.032104} {\bibfield  {journal} {\bibinfo  {journal} {Physical Review A}\ }\textbf {\bibinfo {volume} {100}},\ \bibinfo {pages} {032104} (\bibinfo {year} {2019})}\BibitemShut {NoStop}%
\bibitem [{\citenamefont {Liu}\ \emph {et~al.}(2020)\citenamefont {Liu}, \citenamefont {Yuan}, \citenamefont {Lu},\ and\ \citenamefont {Wang}}]{Liu2020}%
  \BibitemOpen
  \bibfield  {author} {\bibinfo {author} {\bibfnamefont {J.}~\bibnamefont {Liu}}, \bibinfo {author} {\bibfnamefont {H.}~\bibnamefont {Yuan}}, \bibinfo {author} {\bibfnamefont {X.-M.}\ \bibnamefont {Lu}},\ and\ \bibinfo {author} {\bibfnamefont {X.}~\bibnamefont {Wang}},\ }\href {https://doi.org/10.1088/1751-8121/ab5d4d} {\bibfield  {journal} {\bibinfo  {journal} {Journal of Physics A: Mathematical and Theoretical}\ }\textbf {\bibinfo {volume} {53}},\ \bibinfo {pages} {023001} (\bibinfo {year} {2020})}\BibitemShut {NoStop}%
\bibitem [{\citenamefont {Tsang}\ \emph {et~al.}(2016)\citenamefont {Tsang}, \citenamefont {Nair},\ and\ \citenamefont {Lu}}]{Tsang2016}%
  \BibitemOpen
  \bibfield  {author} {\bibinfo {author} {\bibfnamefont {M.}~\bibnamefont {Tsang}}, \bibinfo {author} {\bibfnamefont {R.}~\bibnamefont {Nair}},\ and\ \bibinfo {author} {\bibfnamefont {X.-M.}\ \bibnamefont {Lu}},\ }\href {https://doi.org/10.1103/PhysRevX.6.031033} {\bibfield  {journal} {\bibinfo  {journal} {Physical Review X}\ }\textbf {\bibinfo {volume} {6}},\ \bibinfo {pages} {031033} (\bibinfo {year} {2016})}\BibitemShut {NoStop}%
\bibitem [{\citenamefont {Tsang}(2019{\natexlab{a}})}]{Tsang2019r}%
  \BibitemOpen
  \bibfield  {author} {\bibinfo {author} {\bibfnamefont {M.}~\bibnamefont {Tsang}},\ }\href {https://doi.org/10.1080/00107514.2020.1736375} {\bibfield  {journal} {\bibinfo  {journal} {Contemporary Physics}\ }\textbf {\bibinfo {volume} {60}},\ \bibinfo {pages} {279} (\bibinfo {year} {2019}{\natexlab{a}})},\ \Eprint {https://arxiv.org/abs/1906.02064} {1906.02064} \BibitemShut {NoStop}%
\bibitem [{\citenamefont {Tsang}(2019{\natexlab{b}})}]{Tsang2019b}%
  \BibitemOpen
  \bibfield  {author} {\bibinfo {author} {\bibfnamefont {M.}~\bibnamefont {Tsang}},\ }\href {https://doi.org/10.1103/PhysRevResearch.1.033006} {\bibfield  {journal} {\bibinfo  {journal} {Physical Review Research}\ }\textbf {\bibinfo {volume} {1}},\ \bibinfo {pages} {033006} (\bibinfo {year} {2019}{\natexlab{b}})},\ \Eprint {https://arxiv.org/abs/1906.04578} {1906.04578} \BibitemShut {NoStop}%
\bibitem [{\citenamefont {Albarelli}\ \emph {et~al.}(2020)\citenamefont {Albarelli}, \citenamefont {Barbieri}, \citenamefont {Genoni},\ and\ \citenamefont {Gianani}}]{Albarelli2020}%
  \BibitemOpen
  \bibfield  {author} {\bibinfo {author} {\bibfnamefont {F.}~\bibnamefont {Albarelli}}, \bibinfo {author} {\bibfnamefont {M.}~\bibnamefont {Barbieri}}, \bibinfo {author} {\bibfnamefont {M.~G.}\ \bibnamefont {Genoni}},\ and\ \bibinfo {author} {\bibfnamefont {I.}~\bibnamefont {Gianani}},\ }\href {https://doi.org/10.1016/j.physleta.2020.126311} {\bibfield  {journal} {\bibinfo  {journal} {Physics Letters A}\ }\textbf {\bibinfo {volume} {384}},\ \bibinfo {pages} {126311} (\bibinfo {year} {2020})}\BibitemShut {NoStop}%
\bibitem [{\citenamefont {Qian}\ \emph {et~al.}(2021)\citenamefont {Qian}, \citenamefont {Cui}, \citenamefont {Luo}, \citenamefont {Zheng}, \citenamefont {Huang}, \citenamefont {Ai}, \citenamefont {He}, \citenamefont {Li},\ and\ \citenamefont {Guo}}]{Qian2021}%
  \BibitemOpen
  \bibfield  {author} {\bibinfo {author} {\bibfnamefont {Z.-H.}\ \bibnamefont {Qian}}, \bibinfo {author} {\bibfnamefont {J.-M.}\ \bibnamefont {Cui}}, \bibinfo {author} {\bibfnamefont {X.-W.}\ \bibnamefont {Luo}}, \bibinfo {author} {\bibfnamefont {Y.-X.}\ \bibnamefont {Zheng}}, \bibinfo {author} {\bibfnamefont {Y.-F.}\ \bibnamefont {Huang}}, \bibinfo {author} {\bibfnamefont {M.-Z.}\ \bibnamefont {Ai}}, \bibinfo {author} {\bibfnamefont {R.}~\bibnamefont {He}}, \bibinfo {author} {\bibfnamefont {C.-F.}\ \bibnamefont {Li}},\ and\ \bibinfo {author} {\bibfnamefont {G.-C.}\ \bibnamefont {Guo}},\ }\href {https://doi.org/10.1103/PhysRevLett.127.263603} {\bibfield  {journal} {\bibinfo  {journal} {Physical Review Letters}\ }\textbf {\bibinfo {volume} {127}},\ \bibinfo {pages} {263603} (\bibinfo {year} {2021})}\BibitemShut {NoStop}%
\bibitem [{\citenamefont {Zanforlin}\ \emph {et~al.}(2022)\citenamefont {Zanforlin}, \citenamefont {Lupo}, \citenamefont {Connolly}, \citenamefont {Kok}, \citenamefont {Buller},\ and\ \citenamefont {Huang}}]{Zanforlin2022}%
  \BibitemOpen
  \bibfield  {author} {\bibinfo {author} {\bibfnamefont {U.}~\bibnamefont {Zanforlin}}, \bibinfo {author} {\bibfnamefont {C.}~\bibnamefont {Lupo}}, \bibinfo {author} {\bibfnamefont {P.~W.~R.}\ \bibnamefont {Connolly}}, \bibinfo {author} {\bibfnamefont {P.}~\bibnamefont {Kok}}, \bibinfo {author} {\bibfnamefont {G.~S.}\ \bibnamefont {Buller}},\ and\ \bibinfo {author} {\bibfnamefont {Z.}~\bibnamefont {Huang}},\ }\href {https://doi.org/10.1038/s41467-022-32977-8} {\bibfield  {journal} {\bibinfo  {journal} {Nature Communications}\ }\textbf {\bibinfo {volume} {13}},\ \bibinfo {pages} {5373} (\bibinfo {year} {2022})}\BibitemShut {NoStop}%
\bibitem [{\citenamefont {Ang}\ \emph {et~al.}(2017)\citenamefont {Ang}, \citenamefont {Nair},\ and\ \citenamefont {Tsang}}]{Ang2017}%
  \BibitemOpen
  \bibfield  {author} {\bibinfo {author} {\bibfnamefont {S.~Z.}\ \bibnamefont {Ang}}, \bibinfo {author} {\bibfnamefont {R.}~\bibnamefont {Nair}},\ and\ \bibinfo {author} {\bibfnamefont {M.}~\bibnamefont {Tsang}},\ }\href {https://doi.org/10.1103/PhysRevA.95.063847} {\bibfield  {journal} {\bibinfo  {journal} {Physical Review A}\ }\textbf {\bibinfo {volume} {95}},\ \bibinfo {pages} {063847} (\bibinfo {year} {2017})}\BibitemShut {NoStop}%
\bibitem [{\citenamefont {Napoli}\ \emph {et~al.}(2019)\citenamefont {Napoli}, \citenamefont {Piano}, \citenamefont {Leach}, \citenamefont {Adesso},\ and\ \citenamefont {Tufarelli}}]{Napoli2019}%
  \BibitemOpen
  \bibfield  {author} {\bibinfo {author} {\bibfnamefont {C.}~\bibnamefont {Napoli}}, \bibinfo {author} {\bibfnamefont {S.}~\bibnamefont {Piano}}, \bibinfo {author} {\bibfnamefont {R.}~\bibnamefont {Leach}}, \bibinfo {author} {\bibfnamefont {G.}~\bibnamefont {Adesso}},\ and\ \bibinfo {author} {\bibfnamefont {T.}~\bibnamefont {Tufarelli}},\ }\href {https://doi.org/10.1103/PhysRevLett.122.140505} {\bibfield  {journal} {\bibinfo  {journal} {Physical Review Letters}\ }\textbf {\bibinfo {volume} {122}},\ \bibinfo {pages} {140505} (\bibinfo {year} {2019})},\ \Eprint {https://arxiv.org/abs/1805.04116} {1805.04116} \BibitemShut {NoStop}%
\bibitem [{\citenamefont {Prasad}(2020{\natexlab{a}})}]{Prasad2020a}%
  \BibitemOpen
  \bibfield  {author} {\bibinfo {author} {\bibfnamefont {S.}~\bibnamefont {Prasad}},\ }\href {https://doi.org/10.1103/PhysRevA.102.033726} {\bibfield  {journal} {\bibinfo  {journal} {Physical Review A}\ }\textbf {\bibinfo {volume} {102}},\ \bibinfo {pages} {033726} (\bibinfo {year} {2020}{\natexlab{a}})}\BibitemShut {NoStop}%
\bibitem [{\citenamefont {Prasad}(2020{\natexlab{b}})}]{Prasad2020b}%
  \BibitemOpen
  \bibfield  {author} {\bibinfo {author} {\bibfnamefont {S.}~\bibnamefont {Prasad}},\ }\href {https://doi.org/10.1103/PhysRevA.102.063719} {\bibfield  {journal} {\bibinfo  {journal} {Physical Review A}\ }\textbf {\bibinfo {volume} {102}},\ \bibinfo {pages} {063719} (\bibinfo {year} {2020}{\natexlab{b}})},\ \Eprint {https://arxiv.org/abs/2008.09946} {2008.09946} \BibitemShut {NoStop}%
\bibitem [{\citenamefont {Yu}\ and\ \citenamefont {Prasad}(2018)}]{Yu2018}%
  \BibitemOpen
  \bibfield  {author} {\bibinfo {author} {\bibfnamefont {Z.}~\bibnamefont {Yu}}\ and\ \bibinfo {author} {\bibfnamefont {S.}~\bibnamefont {Prasad}},\ }\href {https://doi.org/10.1103/PhysRevLett.121.180504} {\bibfield  {journal} {\bibinfo  {journal} {Physical Review Letters}\ }\textbf {\bibinfo {volume} {121}},\ \bibinfo {pages} {180504} (\bibinfo {year} {2018})}\BibitemShut {NoStop}%
\bibitem [{\citenamefont {Prasad}(2020{\natexlab{c}})}]{Prasad2020}%
  \BibitemOpen
  \bibfield  {author} {\bibinfo {author} {\bibfnamefont {S.}~\bibnamefont {Prasad}},\ }\href {https://doi.org/10.1088/1402-4896/ab573d} {\bibfield  {journal} {\bibinfo  {journal} {Physica Scripta}\ }\textbf {\bibinfo {volume} {95}},\ \bibinfo {pages} {054004} (\bibinfo {year} {2020}{\natexlab{c}})}\BibitemShut {NoStop}%
\bibitem [{\citenamefont {Zhou}\ and\ \citenamefont {Jiang}(2019)}]{Zhou2019}%
  \BibitemOpen
  \bibfield  {author} {\bibinfo {author} {\bibfnamefont {S.}~\bibnamefont {Zhou}}\ and\ \bibinfo {author} {\bibfnamefont {L.}~\bibnamefont {Jiang}},\ }\href {https://doi.org/10.1103/PhysRevA.99.013808} {\bibfield  {journal} {\bibinfo  {journal} {Physical Review A}\ }\textbf {\bibinfo {volume} {99}},\ \bibinfo {pages} {013808} (\bibinfo {year} {2019})}\BibitemShut {NoStop}%
\bibitem [{\citenamefont {Bisketzi}\ \emph {et~al.}(2019)\citenamefont {Bisketzi}, \citenamefont {Branford},\ and\ \citenamefont {Datta}}]{Bisketzi2019}%
  \BibitemOpen
  \bibfield  {author} {\bibinfo {author} {\bibfnamefont {E.}~\bibnamefont {Bisketzi}}, \bibinfo {author} {\bibfnamefont {D.}~\bibnamefont {Branford}},\ and\ \bibinfo {author} {\bibfnamefont {A.}~\bibnamefont {Datta}},\ }\href {https://doi.org/10.1088/1367-2630/ab58a0} {\bibfield  {journal} {\bibinfo  {journal} {New Journal of Physics}\ }\textbf {\bibinfo {volume} {21}},\ \bibinfo {pages} {123032} (\bibinfo {year} {2019})}\BibitemShut {NoStop}%
\bibitem [{\citenamefont {Lupo}\ and\ \citenamefont {Pirandola}(2016)}]{Lupo2016}%
  \BibitemOpen
  \bibfield  {author} {\bibinfo {author} {\bibfnamefont {C.}~\bibnamefont {Lupo}}\ and\ \bibinfo {author} {\bibfnamefont {S.}~\bibnamefont {Pirandola}},\ }\href {https://doi.org/10.1103/PhysRevLett.117.190802} {\bibfield  {journal} {\bibinfo  {journal} {Physical Review Letters}\ }\textbf {\bibinfo {volume} {117}},\ \bibinfo {pages} {190802} (\bibinfo {year} {2016})}\BibitemShut {NoStop}%
\bibitem [{\citenamefont {Nair}\ and\ \citenamefont {Tsang}(2016)}]{Nair2016}%
  \BibitemOpen
  \bibfield  {author} {\bibinfo {author} {\bibfnamefont {R.}~\bibnamefont {Nair}}\ and\ \bibinfo {author} {\bibfnamefont {M.}~\bibnamefont {Tsang}},\ }\href {https://doi.org/10.1103/PhysRevLett.117.190801} {\bibfield  {journal} {\bibinfo  {journal} {Physical Review Letters}\ }\textbf {\bibinfo {volume} {117}},\ \bibinfo {pages} {190801} (\bibinfo {year} {2016})}\BibitemShut {NoStop}%
\bibitem [{\citenamefont {Wang}\ \emph {et~al.}(2021{\natexlab{a}})\citenamefont {Wang}, \citenamefont {Zhang},\ and\ \citenamefont {Lorenz}}]{Wang2021}%
  \BibitemOpen
  \bibfield  {author} {\bibinfo {author} {\bibfnamefont {Y.}~\bibnamefont {Wang}}, \bibinfo {author} {\bibfnamefont {Y.}~\bibnamefont {Zhang}},\ and\ \bibinfo {author} {\bibfnamefont {V.~O.}\ \bibnamefont {Lorenz}},\ }\href {https://doi.org/10.1103/PhysRevA.104.022613} {\bibfield  {journal} {\bibinfo  {journal} {Physical Review A}\ }\textbf {\bibinfo {volume} {104}},\ \bibinfo {pages} {022613} (\bibinfo {year} {2021}{\natexlab{a}})}\BibitemShut {NoStop}%
\bibitem [{\citenamefont {{Ben-Abdallah}}(2019)}]{Ben-Abdallah2019}%
  \BibitemOpen
  \bibfield  {author} {\bibinfo {author} {\bibfnamefont {P.}~\bibnamefont {{Ben-Abdallah}}},\ }\href {https://doi.org/10.1103/PhysRevLett.123.264301} {\bibfield  {journal} {\bibinfo  {journal} {Physical Review Letters}\ }\textbf {\bibinfo {volume} {123}},\ \bibinfo {pages} {264301} (\bibinfo {year} {2019})}\BibitemShut {NoStop}%
\bibitem [{\citenamefont {Pushkina}\ \emph {et~al.}(2021)\citenamefont {Pushkina}, \citenamefont {Maltese}, \citenamefont {{Costa-Filho}}, \citenamefont {Patel},\ and\ \citenamefont {Lvovsky}}]{Pushkina2021}%
  \BibitemOpen
  \bibfield  {author} {\bibinfo {author} {\bibfnamefont {A.~A.}\ \bibnamefont {Pushkina}}, \bibinfo {author} {\bibfnamefont {G.}~\bibnamefont {Maltese}}, \bibinfo {author} {\bibfnamefont {J.~I.}\ \bibnamefont {{Costa-Filho}}}, \bibinfo {author} {\bibfnamefont {P.}~\bibnamefont {Patel}},\ and\ \bibinfo {author} {\bibfnamefont {A.~I.}\ \bibnamefont {Lvovsky}},\ }\href {https://doi.org/10.1103/PhysRevLett.127.253602} {\bibfield  {journal} {\bibinfo  {journal} {Physical Review Letters}\ }\textbf {\bibinfo {volume} {127}},\ \bibinfo {pages} {253602} (\bibinfo {year} {2021})}\BibitemShut {NoStop}%
\bibitem [{\citenamefont {Matlin}\ and\ \citenamefont {Zipp}(2022)}]{Matlin2022}%
  \BibitemOpen
  \bibfield  {author} {\bibinfo {author} {\bibfnamefont {E.~F.}\ \bibnamefont {Matlin}}\ and\ \bibinfo {author} {\bibfnamefont {L.~J.}\ \bibnamefont {Zipp}},\ }\href {https://doi.org/10.1038/s41598-022-06644-3} {\bibfield  {journal} {\bibinfo  {journal} {Scientific Reports}\ }\textbf {\bibinfo {volume} {12}},\ \bibinfo {pages} {2810} (\bibinfo {year} {2022})}\BibitemShut {NoStop}%
\bibitem [{\citenamefont {Brennan}\ \emph {et~al.}(2022)\citenamefont {Brennan}, \citenamefont {Howard}, \citenamefont {Marzouk},\ and\ \citenamefont {{Dresselhaus-Marais}}}]{Brennan2022}%
  \BibitemOpen
  \bibfield  {author} {\bibinfo {author} {\bibfnamefont {M.~C.}\ \bibnamefont {Brennan}}, \bibinfo {author} {\bibfnamefont {M.}~\bibnamefont {Howard}}, \bibinfo {author} {\bibfnamefont {Y.}~\bibnamefont {Marzouk}},\ and\ \bibinfo {author} {\bibfnamefont {L.~E.}\ \bibnamefont {{Dresselhaus-Marais}}},\ }\href {https://doi.org/10.1007/s10853-022-07465-5} {\bibfield  {journal} {\bibinfo  {journal} {Journal of Materials Science}\ }\textbf {\bibinfo {volume} {57}},\ \bibinfo {pages} {14890} (\bibinfo {year} {2022})},\ \Eprint {https://arxiv.org/abs/2203.05671} {2203.05671} \BibitemShut {NoStop}%
\bibitem [{\citenamefont {Kolesov}\ \emph {et~al.}(2018)\citenamefont {Kolesov}, \citenamefont {Lasse}, \citenamefont {Rothfuchs}, \citenamefont {Wieck}, \citenamefont {Xia}, \citenamefont {Kornher},\ and\ \citenamefont {Wrachtrup}}]{Kolesov2018}%
  \BibitemOpen
  \bibfield  {author} {\bibinfo {author} {\bibfnamefont {R.}~\bibnamefont {Kolesov}}, \bibinfo {author} {\bibfnamefont {S.}~\bibnamefont {Lasse}}, \bibinfo {author} {\bibfnamefont {C.}~\bibnamefont {Rothfuchs}}, \bibinfo {author} {\bibfnamefont {A.~D.}\ \bibnamefont {Wieck}}, \bibinfo {author} {\bibfnamefont {K.}~\bibnamefont {Xia}}, \bibinfo {author} {\bibfnamefont {T.}~\bibnamefont {Kornher}},\ and\ \bibinfo {author} {\bibfnamefont {J.}~\bibnamefont {Wrachtrup}},\ }\href {https://doi.org/10.1103/PhysRevLett.120.033903} {\bibfield  {journal} {\bibinfo  {journal} {Physical Review Letters}\ }\textbf {\bibinfo {volume} {120}},\ \bibinfo {pages} {033903} (\bibinfo {year} {2018})}\BibitemShut {NoStop}%
\bibitem [{\citenamefont {Horodynski}\ \emph {et~al.}(2021)\citenamefont {Horodynski}, \citenamefont {Bouchet}, \citenamefont {K{\"u}hmayer},\ and\ \citenamefont {Rotter}}]{Horodynski2021}%
  \BibitemOpen
  \bibfield  {author} {\bibinfo {author} {\bibfnamefont {M.}~\bibnamefont {Horodynski}}, \bibinfo {author} {\bibfnamefont {D.}~\bibnamefont {Bouchet}}, \bibinfo {author} {\bibfnamefont {M.}~\bibnamefont {K{\"u}hmayer}},\ and\ \bibinfo {author} {\bibfnamefont {S.}~\bibnamefont {Rotter}},\ }\href {https://doi.org/10.1103/PhysRevLett.127.233201} {\bibfield  {journal} {\bibinfo  {journal} {Physical Review Letters}\ }\textbf {\bibinfo {volume} {127}},\ \bibinfo {pages} {233201} (\bibinfo {year} {2021})}\BibitemShut {NoStop}%
\bibitem [{\citenamefont {Xie}\ \emph {et~al.}(2022)\citenamefont {Xie}, \citenamefont {Xu},\ and\ \citenamefont {Wang}}]{Xie2022}%
  \BibitemOpen
  \bibfield  {author} {\bibinfo {author} {\bibfnamefont {D.}~\bibnamefont {Xie}}, \bibinfo {author} {\bibfnamefont {C.}~\bibnamefont {Xu}},\ and\ \bibinfo {author} {\bibfnamefont {A.~M.}\ \bibnamefont {Wang}},\ }\href {https://doi.org/10.1016/j.rinp.2022.105957} {\bibfield  {journal} {\bibinfo  {journal} {Results in Physics}\ }\textbf {\bibinfo {volume} {42}},\ \bibinfo {pages} {105957} (\bibinfo {year} {2022})}\BibitemShut {NoStop}%
\bibitem [{\citenamefont {Bai}\ \emph {et~al.}(2019)\citenamefont {Bai}, \citenamefont {Peng}, \citenamefont {Luo},\ and\ \citenamefont {An}}]{Bai2019}%
  \BibitemOpen
  \bibfield  {author} {\bibinfo {author} {\bibfnamefont {K.}~\bibnamefont {Bai}}, \bibinfo {author} {\bibfnamefont {Z.}~\bibnamefont {Peng}}, \bibinfo {author} {\bibfnamefont {H.-G.}\ \bibnamefont {Luo}},\ and\ \bibinfo {author} {\bibfnamefont {J.-H.}\ \bibnamefont {An}},\ }\href {https://doi.org/10.1103/PhysRevLett.123.040402} {\bibfield  {journal} {\bibinfo  {journal} {Physical Review Letters}\ }\textbf {\bibinfo {volume} {123}},\ \bibinfo {pages} {040402} (\bibinfo {year} {2019})}\BibitemShut {NoStop}%
\bibitem [{\citenamefont {Gessner}\ \emph {et~al.}(2020)\citenamefont {Gessner}, \citenamefont {Fabre},\ and\ \citenamefont {Treps}}]{Gessner2020}%
  \BibitemOpen
  \bibfield  {author} {\bibinfo {author} {\bibfnamefont {M.}~\bibnamefont {Gessner}}, \bibinfo {author} {\bibfnamefont {C.}~\bibnamefont {Fabre}},\ and\ \bibinfo {author} {\bibfnamefont {N.}~\bibnamefont {Treps}},\ }\href {https://doi.org/10.1103/PhysRevLett.125.100501} {\bibfield  {journal} {\bibinfo  {journal} {Physical Review Letters}\ }\textbf {\bibinfo {volume} {125}},\ \bibinfo {pages} {100501} (\bibinfo {year} {2020})}\BibitemShut {NoStop}%
\bibitem [{\citenamefont {Oh}\ \emph {et~al.}(2021)\citenamefont {Oh}, \citenamefont {Zhou}, \citenamefont {Wong},\ and\ \citenamefont {Jiang}}]{Oh2021}%
  \BibitemOpen
  \bibfield  {author} {\bibinfo {author} {\bibfnamefont {C.}~\bibnamefont {Oh}}, \bibinfo {author} {\bibfnamefont {S.}~\bibnamefont {Zhou}}, \bibinfo {author} {\bibfnamefont {Y.}~\bibnamefont {Wong}},\ and\ \bibinfo {author} {\bibfnamefont {L.}~\bibnamefont {Jiang}},\ }\href {https://doi.org/10.1103/PhysRevLett.126.120502} {\bibfield  {journal} {\bibinfo  {journal} {Physical Review Letters}\ }\textbf {\bibinfo {volume} {126}},\ \bibinfo {pages} {120502} (\bibinfo {year} {2021})}\BibitemShut {NoStop}%
\bibitem [{\citenamefont {Mikhalychev}\ \emph {et~al.}(2021)\citenamefont {Mikhalychev}, \citenamefont {Novik}, \citenamefont {Karuseichyk}, \citenamefont {Lyakhov}, \citenamefont {Michels},\ and\ \citenamefont {Mogilevtsev}}]{Mikhalychev2021}%
  \BibitemOpen
  \bibfield  {author} {\bibinfo {author} {\bibfnamefont {A.}~\bibnamefont {Mikhalychev}}, \bibinfo {author} {\bibfnamefont {P.}~\bibnamefont {Novik}}, \bibinfo {author} {\bibfnamefont {I.}~\bibnamefont {Karuseichyk}}, \bibinfo {author} {\bibfnamefont {D.~A.}\ \bibnamefont {Lyakhov}}, \bibinfo {author} {\bibfnamefont {D.~L.}\ \bibnamefont {Michels}},\ and\ \bibinfo {author} {\bibfnamefont {D.}~\bibnamefont {Mogilevtsev}},\ }\href {https://doi.org/10.1038/s41534-021-00465-4} {\bibfield  {journal} {\bibinfo  {journal} {npj Quantum Information}\ }\textbf {\bibinfo {volume} {7}},\ \bibinfo {pages} {125} (\bibinfo {year} {2021})},\ \Eprint {https://arxiv.org/abs/2101.05849} {2101.05849} \BibitemShut {NoStop}%
\bibitem [{\citenamefont {Tsang}(2023)}]{Tsang2023}%
  \BibitemOpen
  \bibfield  {author} {\bibinfo {author} {\bibfnamefont {M.}~\bibnamefont {Tsang}},\ }\href {https://doi.org/10.1103/PhysRevA.107.012611} {\bibfield  {journal} {\bibinfo  {journal} {Physical Review A}\ }\textbf {\bibinfo {volume} {107}},\ \bibinfo {pages} {012611} (\bibinfo {year} {2023})},\ \Eprint {https://arxiv.org/abs/2209.06104} {2209.06104} \BibitemShut {NoStop}%
\bibitem [{\citenamefont {G{\'o}recki}\ \emph {et~al.}(2022)\citenamefont {G{\'o}recki}, \citenamefont {Riccardi},\ and\ \citenamefont {Maccone}}]{Gorecki2022a}%
  \BibitemOpen
  \bibfield  {author} {\bibinfo {author} {\bibfnamefont {W.}~\bibnamefont {G{\'o}recki}}, \bibinfo {author} {\bibfnamefont {A.}~\bibnamefont {Riccardi}},\ and\ \bibinfo {author} {\bibfnamefont {L.}~\bibnamefont {Maccone}},\ }\href {https://doi.org/10.1103/PhysRevLett.129.240503} {\bibfield  {journal} {\bibinfo  {journal} {Physical Review Letters}\ }\textbf {\bibinfo {volume} {129}},\ \bibinfo {pages} {240503} (\bibinfo {year} {2022})}\BibitemShut {NoStop}%
\bibitem [{\citenamefont {Sajjad}\ \emph {et~al.}(2021)\citenamefont {Sajjad}, \citenamefont {Grace}, \citenamefont {Zhuang},\ and\ \citenamefont {Guha}}]{Sajjad2021}%
  \BibitemOpen
  \bibfield  {author} {\bibinfo {author} {\bibfnamefont {A.}~\bibnamefont {Sajjad}}, \bibinfo {author} {\bibfnamefont {M.~R.}\ \bibnamefont {Grace}}, \bibinfo {author} {\bibfnamefont {Q.}~\bibnamefont {Zhuang}},\ and\ \bibinfo {author} {\bibfnamefont {S.}~\bibnamefont {Guha}},\ }\href {https://doi.org/10.1103/PhysRevA.104.022410} {\bibfield  {journal} {\bibinfo  {journal} {Physical Review A}\ }\textbf {\bibinfo {volume} {104}},\ \bibinfo {pages} {022410} (\bibinfo {year} {2021})}\BibitemShut {NoStop}%
\bibitem [{\citenamefont {Wang}\ \emph {et~al.}(2021{\natexlab{b}})\citenamefont {Wang}, \citenamefont {Xu}, \citenamefont {Li},\ and\ \citenamefont {Zhang}}]{Wang2021a}%
  \BibitemOpen
  \bibfield  {author} {\bibinfo {author} {\bibfnamefont {B.}~\bibnamefont {Wang}}, \bibinfo {author} {\bibfnamefont {L.}~\bibnamefont {Xu}}, \bibinfo {author} {\bibfnamefont {J.-c.}\ \bibnamefont {Li}},\ and\ \bibinfo {author} {\bibfnamefont {L.}~\bibnamefont {Zhang}},\ }\href {https://doi.org/10.1364/PRJ.417613} {\bibfield  {journal} {\bibinfo  {journal} {Photonics Research}\ }\textbf {\bibinfo {volume} {9}},\ \bibinfo {pages} {1522} (\bibinfo {year} {2021}{\natexlab{b}})}\BibitemShut {NoStop}%
\bibitem [{\citenamefont {Prasad}\ and\ \citenamefont {Yu}(2019)}]{Prasad2019a}%
  \BibitemOpen
  \bibfield  {author} {\bibinfo {author} {\bibfnamefont {S.}~\bibnamefont {Prasad}}\ and\ \bibinfo {author} {\bibfnamefont {Z.}~\bibnamefont {Yu}},\ }\href {https://doi.org/10.1103/PhysRevA.99.022116} {\bibfield  {journal} {\bibinfo  {journal} {Physical Review A}\ }\textbf {\bibinfo {volume} {99}},\ \bibinfo {pages} {022116} (\bibinfo {year} {2019})}\BibitemShut {NoStop}%
\bibitem [{\citenamefont {Grace}\ and\ \citenamefont {Guha}(2022)}]{Grace2022}%
  \BibitemOpen
  \bibfield  {author} {\bibinfo {author} {\bibfnamefont {M.~R.}\ \bibnamefont {Grace}}\ and\ \bibinfo {author} {\bibfnamefont {S.}~\bibnamefont {Guha}},\ }\href {https://doi.org/10.1103/PhysRevLett.129.180502} {\bibfield  {journal} {\bibinfo  {journal} {Physical Review Letters}\ }\textbf {\bibinfo {volume} {129}},\ \bibinfo {pages} {180502} (\bibinfo {year} {2022})}\BibitemShut {NoStop}%
\bibitem [{\citenamefont {Fiderer}\ \emph {et~al.}(2021)\citenamefont {Fiderer}, \citenamefont {Tufarelli}, \citenamefont {Piano},\ and\ \citenamefont {Adesso}}]{Fiderer2021}%
  \BibitemOpen
  \bibfield  {author} {\bibinfo {author} {\bibfnamefont {L.~J.}\ \bibnamefont {Fiderer}}, \bibinfo {author} {\bibfnamefont {T.}~\bibnamefont {Tufarelli}}, \bibinfo {author} {\bibfnamefont {S.}~\bibnamefont {Piano}},\ and\ \bibinfo {author} {\bibfnamefont {G.}~\bibnamefont {Adesso}},\ }\href {https://doi.org/10.1103/PRXQuantum.2.020308} {\bibfield  {journal} {\bibinfo  {journal} {PRX Quantum}\ }\textbf {\bibinfo {volume} {2}},\ \bibinfo {pages} {020308} (\bibinfo {year} {2021})}\BibitemShut {NoStop}%
\bibitem [{\citenamefont {Mazelanik}\ \emph {et~al.}(2022)\citenamefont {Mazelanik}, \citenamefont {Leszczy{\'n}ski},\ and\ \citenamefont {Parniak}}]{Mazelanik2022}%
  \BibitemOpen
  \bibfield  {author} {\bibinfo {author} {\bibfnamefont {M.}~\bibnamefont {Mazelanik}}, \bibinfo {author} {\bibfnamefont {A.}~\bibnamefont {Leszczy{\'n}ski}},\ and\ \bibinfo {author} {\bibfnamefont {M.}~\bibnamefont {Parniak}},\ }\href {https://doi.org/10.1038/s41467-022-28066-5} {\bibfield  {journal} {\bibinfo  {journal} {Nature Communications}\ }\textbf {\bibinfo {volume} {13}},\ \bibinfo {pages} {691} (\bibinfo {year} {2022})}\BibitemShut {NoStop}%
\bibitem [{\citenamefont {Zheltikov}(2022)}]{Zheltikov2022}%
  \BibitemOpen
  \bibfield  {author} {\bibinfo {author} {\bibfnamefont {A.~M.}\ \bibnamefont {Zheltikov}},\ }\href {https://doi.org/10.1002/jrs.6345} {\bibfield  {journal} {\bibinfo  {journal} {Journal of Raman Spectroscopy}\ }\textbf {\bibinfo {volume} {53}},\ \bibinfo {pages} {1094} (\bibinfo {year} {2022})}\BibitemShut {NoStop}%
\bibitem [{\citenamefont {Zhao}\ \emph {et~al.}(2021)\citenamefont {Zhao}, \citenamefont {Zhang}, \citenamefont {Liu}, \citenamefont {Guan}, \citenamefont {Zhang}, \citenamefont {Li}, \citenamefont {Bai}, \citenamefont {Li}, \citenamefont {Liu}, \citenamefont {You}, \citenamefont {Zhang}, \citenamefont {Fan}, \citenamefont {Xu}, \citenamefont {Zhang},\ and\ \citenamefont {Pan}}]{Zhao2021}%
  \BibitemOpen
  \bibfield  {author} {\bibinfo {author} {\bibfnamefont {S.-R.}\ \bibnamefont {Zhao}}, \bibinfo {author} {\bibfnamefont {Y.-Z.}\ \bibnamefont {Zhang}}, \bibinfo {author} {\bibfnamefont {W.-Z.}\ \bibnamefont {Liu}}, \bibinfo {author} {\bibfnamefont {J.-Y.}\ \bibnamefont {Guan}}, \bibinfo {author} {\bibfnamefont {W.}~\bibnamefont {Zhang}}, \bibinfo {author} {\bibfnamefont {C.-L.}\ \bibnamefont {Li}}, \bibinfo {author} {\bibfnamefont {B.}~\bibnamefont {Bai}}, \bibinfo {author} {\bibfnamefont {M.-H.}\ \bibnamefont {Li}}, \bibinfo {author} {\bibfnamefont {Y.}~\bibnamefont {Liu}}, \bibinfo {author} {\bibfnamefont {L.}~\bibnamefont {You}}, \bibinfo {author} {\bibfnamefont {J.}~\bibnamefont {Zhang}}, \bibinfo {author} {\bibfnamefont {J.}~\bibnamefont {Fan}}, \bibinfo {author} {\bibfnamefont {F.}~\bibnamefont {Xu}}, \bibinfo {author} {\bibfnamefont {Q.}~\bibnamefont {Zhang}},\ and\ \bibinfo {author} {\bibfnamefont {J.-W.}\ \bibnamefont {Pan}},\ }\href {https://doi.org/10.1103/PhysRevX.11.031009} {\bibfield  {journal}
  {\bibinfo  {journal} {Physical Review X}\ }\textbf {\bibinfo {volume} {11}},\ \bibinfo {pages} {031009} (\bibinfo {year} {2021})}\BibitemShut {NoStop}%
\bibitem [{\citenamefont {K{\"o}se}\ and\ \citenamefont {Braun}(2023)}]{Kose2023}%
  \BibitemOpen
  \bibfield  {author} {\bibinfo {author} {\bibfnamefont {E.}~\bibnamefont {K{\"o}se}}\ and\ \bibinfo {author} {\bibfnamefont {D.}~\bibnamefont {Braun}},\ }\href {https://doi.org/10.1103/PhysRevA.107.032607} {\bibfield  {journal} {\bibinfo  {journal} {Physical Review A}\ }\textbf {\bibinfo {volume} {107}},\ \bibinfo {pages} {032607} (\bibinfo {year} {2023})}\BibitemShut {NoStop}%
\bibitem [{\citenamefont {Huszka}\ and\ \citenamefont {Gijs}(2019)}]{Huszka2019}%
  \BibitemOpen
  \bibfield  {author} {\bibinfo {author} {\bibfnamefont {G.}~\bibnamefont {Huszka}}\ and\ \bibinfo {author} {\bibfnamefont {M.~A.~M.}\ \bibnamefont {Gijs}},\ }\href {https://doi.org/10.1016/j.mne.2018.11.005} {\bibfield  {journal} {\bibinfo  {journal} {Micro and Nano Engineering}\ }\textbf {\bibinfo {volume} {2}},\ \bibinfo {pages} {7} (\bibinfo {year} {2019})}\BibitemShut {NoStop}%
\bibitem [{\citenamefont {Cremer}\ and\ \citenamefont {Masters}(2013)}]{Cremer2013}%
  \BibitemOpen
  \bibfield  {author} {\bibinfo {author} {\bibfnamefont {C.}~\bibnamefont {Cremer}}\ and\ \bibinfo {author} {\bibfnamefont {B.~R.}\ \bibnamefont {Masters}},\ }\href {https://doi.org/10.1140/epjh/e2012-20060-1} {\bibfield  {journal} {\bibinfo  {journal} {The European Physical Journal H}\ }\textbf {\bibinfo {volume} {38}},\ \bibinfo {pages} {281} (\bibinfo {year} {2013})}\BibitemShut {NoStop}%
\bibitem [{\citenamefont {Patterson}\ \emph {et~al.}(2010)\citenamefont {Patterson}, \citenamefont {Davidson}, \citenamefont {Manley},\ and\ \citenamefont {{Lippincott-Schwartz}}}]{Patterson2010}%
  \BibitemOpen
  \bibfield  {author} {\bibinfo {author} {\bibfnamefont {G.}~\bibnamefont {Patterson}}, \bibinfo {author} {\bibfnamefont {M.}~\bibnamefont {Davidson}}, \bibinfo {author} {\bibfnamefont {S.}~\bibnamefont {Manley}},\ and\ \bibinfo {author} {\bibfnamefont {J.}~\bibnamefont {{Lippincott-Schwartz}}},\ }\href {https://doi.org/10.1146/annurev.physchem.012809.103444} {\bibfield  {journal} {\bibinfo  {journal} {Annual Review of Physical Chemistry}\ }\textbf {\bibinfo {volume} {61}},\ \bibinfo {pages} {345} (\bibinfo {year} {2010})}\BibitemShut {NoStop}%
\bibitem [{\citenamefont {Sheng}\ \emph {et~al.}(2016)\citenamefont {Sheng}, \citenamefont {Durak},\ and\ \citenamefont {Ling}}]{Sheng2016}%
  \BibitemOpen
  \bibfield  {author} {\bibinfo {author} {\bibfnamefont {T.~Z.}\ \bibnamefont {Sheng}}, \bibinfo {author} {\bibfnamefont {K.}~\bibnamefont {Durak}},\ and\ \bibinfo {author} {\bibfnamefont {A.}~\bibnamefont {Ling}},\ }\href@noop {} {\bibinfo {title} {Fault-tolerant and finite-error localization for point emitters within the diffraction limit}} (\bibinfo {year} {2016}),\ \Eprint {https://arxiv.org/abs/arXiv:1605.07297} {arxiv:arXiv:1605.07297} \BibitemShut {NoStop}%
\bibitem [{\citenamefont {Zhou}\ \emph {et~al.}(2019)\citenamefont {Zhou}, \citenamefont {Yang}, \citenamefont {Hassett}, \citenamefont {Rafsanjani}, \citenamefont {Mirhosseini}, \citenamefont {Vamivakas}, \citenamefont {Jordan}, \citenamefont {Shi},\ and\ \citenamefont {Boyd}}]{Zhou2019a}%
  \BibitemOpen
  \bibfield  {author} {\bibinfo {author} {\bibfnamefont {Y.}~\bibnamefont {Zhou}}, \bibinfo {author} {\bibfnamefont {J.}~\bibnamefont {Yang}}, \bibinfo {author} {\bibfnamefont {J.~D.}\ \bibnamefont {Hassett}}, \bibinfo {author} {\bibfnamefont {S.~M.~H.}\ \bibnamefont {Rafsanjani}}, \bibinfo {author} {\bibfnamefont {M.}~\bibnamefont {Mirhosseini}}, \bibinfo {author} {\bibfnamefont {A.~N.}\ \bibnamefont {Vamivakas}}, \bibinfo {author} {\bibfnamefont {A.~N.}\ \bibnamefont {Jordan}}, \bibinfo {author} {\bibfnamefont {Z.}~\bibnamefont {Shi}},\ and\ \bibinfo {author} {\bibfnamefont {R.~W.}\ \bibnamefont {Boyd}},\ }\href {https://doi.org/10.1364/OPTICA.6.000534} {\bibfield  {journal} {\bibinfo  {journal} {Optica}\ }\textbf {\bibinfo {volume} {6}},\ \bibinfo {pages} {534} (\bibinfo {year} {2019})},\ \Eprint {https://arxiv.org/abs/1810.01027} {1810.01027} \BibitemShut {NoStop}%
\bibitem [{\citenamefont {Fabre}\ and\ \citenamefont {Treps}(2020)}]{Fabre2020}%
  \BibitemOpen
  \bibfield  {author} {\bibinfo {author} {\bibfnamefont {C.}~\bibnamefont {Fabre}}\ and\ \bibinfo {author} {\bibfnamefont {N.}~\bibnamefont {Treps}},\ }\href {https://doi.org/10.1103/RevModPhys.92.035005} {\bibfield  {journal} {\bibinfo  {journal} {Reviews of Modern Physics}\ }\textbf {\bibinfo {volume} {92}},\ \bibinfo {pages} {035005} (\bibinfo {year} {2020})},\ \Eprint {https://arxiv.org/abs/1912.09321} {1912.09321} \BibitemShut {NoStop}%
\bibitem [{\citenamefont {Lupo}\ \emph {et~al.}(2020)\citenamefont {Lupo}, \citenamefont {Huang},\ and\ \citenamefont {Kok}}]{Lupo2020}%
  \BibitemOpen
  \bibfield  {author} {\bibinfo {author} {\bibfnamefont {C.}~\bibnamefont {Lupo}}, \bibinfo {author} {\bibfnamefont {Z.}~\bibnamefont {Huang}},\ and\ \bibinfo {author} {\bibfnamefont {P.}~\bibnamefont {Kok}},\ }\href {https://doi.org/10.1103/PhysRevLett.124.080503} {\bibfield  {journal} {\bibinfo  {journal} {Physical Review Letters}\ }\textbf {\bibinfo {volume} {124}},\ \bibinfo {pages} {080503} (\bibinfo {year} {2020})}\BibitemShut {NoStop}%
\bibitem [{\citenamefont {Huang}\ and\ \citenamefont {Lupo}(2021)}]{Huang2021}%
  \BibitemOpen
  \bibfield  {author} {\bibinfo {author} {\bibfnamefont {Z.}~\bibnamefont {Huang}}\ and\ \bibinfo {author} {\bibfnamefont {C.}~\bibnamefont {Lupo}},\ }\href {https://doi.org/10.1103/PhysRevLett.127.130502} {\bibfield  {journal} {\bibinfo  {journal} {Physical Review Letters}\ }\textbf {\bibinfo {volume} {127}},\ \bibinfo {pages} {130502} (\bibinfo {year} {2021})}\BibitemShut {NoStop}%
\bibitem [{\citenamefont {Datta}\ \emph {et~al.}(2021)\citenamefont {Datta}, \citenamefont {Len}, \citenamefont {{\L}ukanowski}, \citenamefont {Banaszek},\ and\ \citenamefont {Jarzyna}}]{Datta2021a}%
  \BibitemOpen
  \bibfield  {author} {\bibinfo {author} {\bibfnamefont {C.}~\bibnamefont {Datta}}, \bibinfo {author} {\bibfnamefont {Y.~L.}\ \bibnamefont {Len}}, \bibinfo {author} {\bibfnamefont {K.}~\bibnamefont {{\L}ukanowski}}, \bibinfo {author} {\bibfnamefont {K.}~\bibnamefont {Banaszek}},\ and\ \bibinfo {author} {\bibfnamefont {M.}~\bibnamefont {Jarzyna}},\ }\href {https://doi.org/10.1364/OE.433990} {\bibfield  {journal} {\bibinfo  {journal} {Optics Express}\ }\textbf {\bibinfo {volume} {29}},\ \bibinfo {pages} {35592} (\bibinfo {year} {2021})}\BibitemShut {NoStop}%
\bibitem [{\citenamefont {G{\'o}recki}\ and\ \citenamefont {{Demkowicz-Dobrza{\'n}ski}}(2022)}]{Gorecki2022}%
  \BibitemOpen
  \bibfield  {author} {\bibinfo {author} {\bibfnamefont {W.}~\bibnamefont {G{\'o}recki}}\ and\ \bibinfo {author} {\bibfnamefont {R.}~\bibnamefont {{Demkowicz-Dobrza{\'n}ski}}},\ }\href {https://doi.org/10.1103/PhysRevLett.128.040504} {\bibfield  {journal} {\bibinfo  {journal} {Physical Review Letters}\ }\textbf {\bibinfo {volume} {128}},\ \bibinfo {pages} {040504} (\bibinfo {year} {2022})}\BibitemShut {NoStop}%
\bibitem [{\citenamefont {Nair.Ranjith}(2016)}]{Nair2016a}%
  \BibitemOpen
  \bibfield  {author} {\bibinfo {author} {\bibfnamefont {T.}~\bibnamefont {Nair.Ranjith}},\ }\href@noop {} {\bibfield  {journal} {\bibinfo  {journal} {Optics Express}\ }\textbf {\bibinfo {volume} {24}},\ \bibinfo {pages} {3684} (\bibinfo {year} {2016})}\BibitemShut {NoStop}%
\bibitem [{\citenamefont {Hassett}\ \emph {et~al.}(2018)\citenamefont {Hassett}, \citenamefont {Malhorta}, \citenamefont {Alonso}, \citenamefont {Boyd}, \citenamefont {Hashemi~Rafsanjani},\ and\ \citenamefont {Vamivakas}}]{Hassett2018}%
  \BibitemOpen
  \bibfield  {author} {\bibinfo {author} {\bibfnamefont {J.}~\bibnamefont {Hassett}}, \bibinfo {author} {\bibfnamefont {T.}~\bibnamefont {Malhorta}}, \bibinfo {author} {\bibfnamefont {M.~A.}\ \bibnamefont {Alonso}}, \bibinfo {author} {\bibfnamefont {R.~W.}\ \bibnamefont {Boyd}}, \bibinfo {author} {\bibfnamefont {S.~M.}\ \bibnamefont {Hashemi~Rafsanjani}},\ and\ \bibinfo {author} {\bibfnamefont {A.~N.}\ \bibnamefont {Vamivakas}},\ }in\ \href {https://doi.org/10.1364/FIO.2018.JW4A.124} {\emph {\bibinfo {booktitle} {Frontiers in {{Optics}} / {{Laser Science}}}}}\ (\bibinfo  {publisher} {{OSA}},\ \bibinfo {address} {{Washington, DC}},\ \bibinfo {year} {2018})\ p.\ \bibinfo {pages} {JW4A.124}\BibitemShut {NoStop}%
\bibitem [{\citenamefont {Tham}\ \emph {et~al.}(2017)\citenamefont {Tham}, \citenamefont {Ferretti},\ and\ \citenamefont {Steinberg}}]{Tham2017}%
  \BibitemOpen
  \bibfield  {author} {\bibinfo {author} {\bibfnamefont {W.-K.}\ \bibnamefont {Tham}}, \bibinfo {author} {\bibfnamefont {H.}~\bibnamefont {Ferretti}},\ and\ \bibinfo {author} {\bibfnamefont {A.~M.}\ \bibnamefont {Steinberg}},\ }\href {https://doi.org/10.1103/PhysRevLett.118.070801} {\bibfield  {journal} {\bibinfo  {journal} {Physical Review Letters}\ }\textbf {\bibinfo {volume} {118}},\ \bibinfo {pages} {070801} (\bibinfo {year} {2017})}\BibitemShut {NoStop}%
\bibitem [{\citenamefont {Pa{\'u}r}\ \emph {et~al.}(2018)\citenamefont {Pa{\'u}r}, \citenamefont {Stoklasa}, \citenamefont {Grover}, \citenamefont {Krzic}, \citenamefont {{S{\'a}nchez-Soto}}, \citenamefont {Hradil},\ and\ \citenamefont {{\v R}eh{\'a}{\v c}ek}}]{Paur2018}%
  \BibitemOpen
  \bibfield  {author} {\bibinfo {author} {\bibfnamefont {M.}~\bibnamefont {Pa{\'u}r}}, \bibinfo {author} {\bibfnamefont {B.}~\bibnamefont {Stoklasa}}, \bibinfo {author} {\bibfnamefont {J.}~\bibnamefont {Grover}}, \bibinfo {author} {\bibfnamefont {A.}~\bibnamefont {Krzic}}, \bibinfo {author} {\bibfnamefont {L.~L.}\ \bibnamefont {{S{\'a}nchez-Soto}}}, \bibinfo {author} {\bibfnamefont {Z.}~\bibnamefont {Hradil}},\ and\ \bibinfo {author} {\bibfnamefont {J.}~\bibnamefont {{\v R}eh{\'a}{\v c}ek}},\ }\href {https://doi.org/10.1364/OPTICA.5.001177} {\bibfield  {journal} {\bibinfo  {journal} {Optica}\ }\textbf {\bibinfo {volume} {5}},\ \bibinfo {pages} {1177} (\bibinfo {year} {2018})}\BibitemShut {NoStop}%
\bibitem [{\citenamefont {Yang}\ \emph {et~al.}(2016)\citenamefont {Yang}, \citenamefont {Tashchilina}, \citenamefont {Moiseev}, \citenamefont {Simon},\ and\ \citenamefont {Lvovsky}}]{Yang2016}%
  \BibitemOpen
  \bibfield  {author} {\bibinfo {author} {\bibfnamefont {F.}~\bibnamefont {Yang}}, \bibinfo {author} {\bibfnamefont {A.}~\bibnamefont {Tashchilina}}, \bibinfo {author} {\bibfnamefont {E.~S.}\ \bibnamefont {Moiseev}}, \bibinfo {author} {\bibfnamefont {C.}~\bibnamefont {Simon}},\ and\ \bibinfo {author} {\bibfnamefont {A.~I.}\ \bibnamefont {Lvovsky}},\ }\href {https://doi.org/10.1364/OPTICA.3.001148} {\bibfield  {journal} {\bibinfo  {journal} {Optica}\ }\textbf {\bibinfo {volume} {3}},\ \bibinfo {pages} {1148} (\bibinfo {year} {2016})}\BibitemShut {NoStop}%
\bibitem [{\citenamefont {Pa{\'u}r}\ \emph {et~al.}(2016)\citenamefont {Pa{\'u}r}, \citenamefont {Stoklasa}, \citenamefont {Hradil}, \citenamefont {{S{\'a}nchez-Soto}},\ and\ \citenamefont {Rehacek}}]{Paur2016}%
  \BibitemOpen
  \bibfield  {author} {\bibinfo {author} {\bibfnamefont {M.}~\bibnamefont {Pa{\'u}r}}, \bibinfo {author} {\bibfnamefont {B.}~\bibnamefont {Stoklasa}}, \bibinfo {author} {\bibfnamefont {Z.}~\bibnamefont {Hradil}}, \bibinfo {author} {\bibfnamefont {L.~L.}\ \bibnamefont {{S{\'a}nchez-Soto}}},\ and\ \bibinfo {author} {\bibfnamefont {J.}~\bibnamefont {Rehacek}},\ }\href {https://doi.org/10.1364/OPTICA.3.001144} {\bibfield  {journal} {\bibinfo  {journal} {Optica}\ }\textbf {\bibinfo {volume} {3}},\ \bibinfo {pages} {1144} (\bibinfo {year} {2016})}\BibitemShut {NoStop}%
\bibitem [{\citenamefont {Wadood}\ \emph {et~al.}(2021)\citenamefont {Wadood}, \citenamefont {Liang}, \citenamefont {Zhou}, \citenamefont {Yang}, \citenamefont {Alonso}, \citenamefont {Qian}, \citenamefont {Malhotra}, \citenamefont {Rafsanjani}, \citenamefont {Jordan}, \citenamefont {Boyd},\ and\ \citenamefont {Vamivakas}}]{Wadood2021}%
  \BibitemOpen
  \bibfield  {author} {\bibinfo {author} {\bibfnamefont {S.~A.}\ \bibnamefont {Wadood}}, \bibinfo {author} {\bibfnamefont {K.}~\bibnamefont {Liang}}, \bibinfo {author} {\bibfnamefont {Y.}~\bibnamefont {Zhou}}, \bibinfo {author} {\bibfnamefont {J.}~\bibnamefont {Yang}}, \bibinfo {author} {\bibfnamefont {M.~A.}\ \bibnamefont {Alonso}}, \bibinfo {author} {\bibfnamefont {X.-F.}\ \bibnamefont {Qian}}, \bibinfo {author} {\bibfnamefont {T.}~\bibnamefont {Malhotra}}, \bibinfo {author} {\bibfnamefont {S.~M.~H.}\ \bibnamefont {Rafsanjani}}, \bibinfo {author} {\bibfnamefont {A.~N.}\ \bibnamefont {Jordan}}, \bibinfo {author} {\bibfnamefont {R.~W.}\ \bibnamefont {Boyd}},\ and\ \bibinfo {author} {\bibfnamefont {A.~N.}\ \bibnamefont {Vamivakas}},\ }\href {https://doi.org/10.1364/OE.427734} {\bibfield  {journal} {\bibinfo  {journal} {Optics Express}\ }\textbf {\bibinfo {volume} {29}},\ \bibinfo {pages} {22034} (\bibinfo {year} {2021})},\ \Eprint {https://arxiv.org/abs/2102.01603} {2102.01603} \BibitemShut {NoStop}%
\bibitem [{\citenamefont {Donohue}\ \emph {et~al.}(2018)\citenamefont {Donohue}, \citenamefont {Ansari}, \citenamefont {{\v R}eh{\'a}{\v c}ek}, \citenamefont {Hradil}, \citenamefont {Stoklasa}, \citenamefont {Pa{\'u}r}, \citenamefont {{S{\'a}nchez-Soto}},\ and\ \citenamefont {Silberhorn}}]{Donohue2018}%
  \BibitemOpen
  \bibfield  {author} {\bibinfo {author} {\bibfnamefont {J.~M.}\ \bibnamefont {Donohue}}, \bibinfo {author} {\bibfnamefont {V.}~\bibnamefont {Ansari}}, \bibinfo {author} {\bibfnamefont {J.}~\bibnamefont {{\v R}eh{\'a}{\v c}ek}}, \bibinfo {author} {\bibfnamefont {Z.}~\bibnamefont {Hradil}}, \bibinfo {author} {\bibfnamefont {B.}~\bibnamefont {Stoklasa}}, \bibinfo {author} {\bibfnamefont {M.}~\bibnamefont {Pa{\'u}r}}, \bibinfo {author} {\bibfnamefont {L.~L.}\ \bibnamefont {{S{\'a}nchez-Soto}}},\ and\ \bibinfo {author} {\bibfnamefont {C.}~\bibnamefont {Silberhorn}},\ }\href {https://doi.org/10.1103/PhysRevLett.121.090501} {\bibfield  {journal} {\bibinfo  {journal} {Physical Review Letters}\ }\textbf {\bibinfo {volume} {121}},\ \bibinfo {pages} {090501} (\bibinfo {year} {2018})}\BibitemShut {NoStop}%
\bibitem [{\citenamefont {Shah}\ and\ \citenamefont {Fan}(2021)}]{Shah2021}%
  \BibitemOpen
  \bibfield  {author} {\bibinfo {author} {\bibfnamefont {M.}~\bibnamefont {Shah}}\ and\ \bibinfo {author} {\bibfnamefont {L.}~\bibnamefont {Fan}},\ }\href {https://doi.org/10.1103/PhysRevApplied.15.034071} {\bibfield  {journal} {\bibinfo  {journal} {Physical Review Applied}\ }\textbf {\bibinfo {volume} {15}},\ \bibinfo {pages} {034071} (\bibinfo {year} {2021})}\BibitemShut {NoStop}%
\bibitem [{\citenamefont {Tan}\ \emph {et~al.}(2017)\citenamefont {Tan}, \citenamefont {An},\ and\ \citenamefont {Ohl}}]{Tan2017}%
  \BibitemOpen
  \bibfield  {author} {\bibinfo {author} {\bibfnamefont {B.~H.}\ \bibnamefont {Tan}}, \bibinfo {author} {\bibfnamefont {H.}~\bibnamefont {An}},\ and\ \bibinfo {author} {\bibfnamefont {C.-D.}\ \bibnamefont {Ohl}},\ }\href {https://doi.org/10.1103/PhysRevLett.118.054501} {\bibfield  {journal} {\bibinfo  {journal} {Physical Review Letters}\ }\textbf {\bibinfo {volume} {118}},\ \bibinfo {pages} {054501} (\bibinfo {year} {2017})}\BibitemShut {NoStop}%
\bibitem [{\citenamefont {Backlund}\ \emph {et~al.}(2018)\citenamefont {Backlund}, \citenamefont {Shechtman},\ and\ \citenamefont {Walsworth}}]{Backlund2018}%
  \BibitemOpen
  \bibfield  {author} {\bibinfo {author} {\bibfnamefont {M.~P.}\ \bibnamefont {Backlund}}, \bibinfo {author} {\bibfnamefont {Y.}~\bibnamefont {Shechtman}},\ and\ \bibinfo {author} {\bibfnamefont {R.~L.}\ \bibnamefont {Walsworth}},\ }\href {https://doi.org/10.1103/PhysRevLett.121.023904} {\bibfield  {journal} {\bibinfo  {journal} {Physical Review Letters}\ }\textbf {\bibinfo {volume} {121}},\ \bibinfo {pages} {023904} (\bibinfo {year} {2018})}\BibitemShut {NoStop}%
\bibitem [{\citenamefont {{\v R}eh{\'a}{\v c}ek}\ \emph {et~al.}(2019)\citenamefont {{\v R}eh{\'a}{\v c}ek}, \citenamefont {Pa{\'u}r}, \citenamefont {Stoklasa}, \citenamefont {Koutn{\'y}}, \citenamefont {Hradil},\ and\ \citenamefont {{S{\'a}nchez-Soto}}}]{Rehacek2019}%
  \BibitemOpen
  \bibfield  {author} {\bibinfo {author} {\bibfnamefont {J.}~\bibnamefont {{\v R}eh{\'a}{\v c}ek}}, \bibinfo {author} {\bibfnamefont {M.}~\bibnamefont {Pa{\'u}r}}, \bibinfo {author} {\bibfnamefont {B.}~\bibnamefont {Stoklasa}}, \bibinfo {author} {\bibfnamefont {D.}~\bibnamefont {Koutn{\'y}}}, \bibinfo {author} {\bibfnamefont {Z.}~\bibnamefont {Hradil}},\ and\ \bibinfo {author} {\bibfnamefont {L.~L.}\ \bibnamefont {{S{\'a}nchez-Soto}}},\ }\href {https://doi.org/10.1103/PhysRevLett.123.193601} {\bibfield  {journal} {\bibinfo  {journal} {Physical Review Letters}\ }\textbf {\bibinfo {volume} {123}},\ \bibinfo {pages} {193601} (\bibinfo {year} {2019})}\BibitemShut {NoStop}%
\bibitem [{\citenamefont {Deist}\ \emph {et~al.}(2022)\citenamefont {Deist}, \citenamefont {Gerber}, \citenamefont {Lu}, \citenamefont {Zeiher},\ and\ \citenamefont {{Stamper-Kurn}}}]{Deist2022}%
  \BibitemOpen
  \bibfield  {author} {\bibinfo {author} {\bibfnamefont {E.}~\bibnamefont {Deist}}, \bibinfo {author} {\bibfnamefont {J.~A.}\ \bibnamefont {Gerber}}, \bibinfo {author} {\bibfnamefont {Y.-H.}\ \bibnamefont {Lu}}, \bibinfo {author} {\bibfnamefont {J.}~\bibnamefont {Zeiher}},\ and\ \bibinfo {author} {\bibfnamefont {D.~M.}\ \bibnamefont {{Stamper-Kurn}}},\ }\href {https://doi.org/10.1103/PhysRevLett.128.083201} {\bibfield  {journal} {\bibinfo  {journal} {Physical Review Letters}\ }\textbf {\bibinfo {volume} {128}},\ \bibinfo {pages} {083201} (\bibinfo {year} {2022})}\BibitemShut {NoStop}%
\bibitem [{\citenamefont {Cheng}\ \emph {et~al.}(2020)\citenamefont {Cheng}, \citenamefont {Chen},\ and\ \citenamefont {Li}}]{Cheng2020}%
  \BibitemOpen
  \bibfield  {author} {\bibinfo {author} {\bibfnamefont {T.}~\bibnamefont {Cheng}}, \bibinfo {author} {\bibfnamefont {D.}~\bibnamefont {Chen}},\ and\ \bibinfo {author} {\bibfnamefont {H.}~\bibnamefont {Li}},\ }\href {https://doi.org/10.1088/1742-6596/1651/1/012177} {\bibfield  {journal} {\bibinfo  {journal} {Journal of Physics: Conference Series}\ }\textbf {\bibinfo {volume} {1651}},\ \bibinfo {pages} {012177} (\bibinfo {year} {2020})}\BibitemShut {NoStop}%
\bibitem [{\citenamefont {Howard}\ \emph {et~al.}(2019)\citenamefont {Howard}, \citenamefont {Gillett}, \citenamefont {Pearce}, \citenamefont {Abrahao}, \citenamefont {Weinhold}, \citenamefont {Kok},\ and\ \citenamefont {White}}]{Howard2019}%
  \BibitemOpen
  \bibfield  {author} {\bibinfo {author} {\bibfnamefont {L.~A.}\ \bibnamefont {Howard}}, \bibinfo {author} {\bibfnamefont {G.~G.}\ \bibnamefont {Gillett}}, \bibinfo {author} {\bibfnamefont {M.~E.}\ \bibnamefont {Pearce}}, \bibinfo {author} {\bibfnamefont {R.~A.}\ \bibnamefont {Abrahao}}, \bibinfo {author} {\bibfnamefont {T.~J.}\ \bibnamefont {Weinhold}}, \bibinfo {author} {\bibfnamefont {P.}~\bibnamefont {Kok}},\ and\ \bibinfo {author} {\bibfnamefont {A.~G.}\ \bibnamefont {White}},\ }\href {https://doi.org/10.1103/PhysRevLett.123.143604} {\bibfield  {journal} {\bibinfo  {journal} {Physical Review Letters}\ }\textbf {\bibinfo {volume} {123}},\ \bibinfo {pages} {143604} (\bibinfo {year} {2019})}\BibitemShut {NoStop}%
\bibitem [{\citenamefont {Huang}\ \emph {et~al.}(2022)\citenamefont {Huang}, \citenamefont {Brennen},\ and\ \citenamefont {Ouyang}}]{Huang2022}%
  \BibitemOpen
  \bibfield  {author} {\bibinfo {author} {\bibfnamefont {Z.}~\bibnamefont {Huang}}, \bibinfo {author} {\bibfnamefont {G.~K.}\ \bibnamefont {Brennen}},\ and\ \bibinfo {author} {\bibfnamefont {Y.}~\bibnamefont {Ouyang}},\ }\href {https://doi.org/10.1103/PhysRevLett.129.210502} {\bibfield  {journal} {\bibinfo  {journal} {Physical Review Letters}\ }\textbf {\bibinfo {volume} {129}},\ \bibinfo {pages} {210502} (\bibinfo {year} {2022})}\BibitemShut {NoStop}%
\bibitem [{\citenamefont {{\v R}eha{\v c}ek}\ \emph {et~al.}(2017)\citenamefont {{\v R}eha{\v c}ek}, \citenamefont {Hradil}, \citenamefont {Stoklasa}, \citenamefont {Pa{\'u}r}, \citenamefont {Grover}, \citenamefont {Krzic},\ and\ \citenamefont {{S{\'a}nchez-Soto}}}]{Rehacek2017}%
  \BibitemOpen
  \bibfield  {author} {\bibinfo {author} {\bibfnamefont {J.}~\bibnamefont {{\v R}eha{\v c}ek}}, \bibinfo {author} {\bibfnamefont {Z.}~\bibnamefont {Hradil}}, \bibinfo {author} {\bibfnamefont {B.}~\bibnamefont {Stoklasa}}, \bibinfo {author} {\bibfnamefont {M.}~\bibnamefont {Pa{\'u}r}}, \bibinfo {author} {\bibfnamefont {J.}~\bibnamefont {Grover}}, \bibinfo {author} {\bibfnamefont {A.}~\bibnamefont {Krzic}},\ and\ \bibinfo {author} {\bibfnamefont {L.~L.}\ \bibnamefont {{S{\'a}nchez-Soto}}},\ }\href {https://doi.org/10.1103/PhysRevA.96.062107} {\bibfield  {journal} {\bibinfo  {journal} {Physical Review A}\ }\textbf {\bibinfo {volume} {96}},\ \bibinfo {pages} {062107} (\bibinfo {year} {2017})}\BibitemShut {NoStop}%
\bibitem [{\citenamefont {Larson}\ and\ \citenamefont {Saleh}(2018)}]{Larson2018}%
  \BibitemOpen
  \bibfield  {author} {\bibinfo {author} {\bibfnamefont {W.}~\bibnamefont {Larson}}\ and\ \bibinfo {author} {\bibfnamefont {B.~E.~A.}\ \bibnamefont {Saleh}},\ }\href {https://doi.org/10.1364/OPTICA.5.001382} {\bibfield  {journal} {\bibinfo  {journal} {Optica}\ }\textbf {\bibinfo {volume} {5}},\ \bibinfo {pages} {1382} (\bibinfo {year} {2018})}\BibitemShut {NoStop}%
\bibitem [{\citenamefont {Tsang}\ and\ \citenamefont {Nair}(2019)}]{Tsang2019c}%
  \BibitemOpen
  \bibfield  {author} {\bibinfo {author} {\bibfnamefont {M.}~\bibnamefont {Tsang}}\ and\ \bibinfo {author} {\bibfnamefont {R.}~\bibnamefont {Nair}},\ }\href {https://doi.org/10.1364/OPTICA.6.000400} {\bibfield  {journal} {\bibinfo  {journal} {Optica}\ }\textbf {\bibinfo {volume} {6}},\ \bibinfo {pages} {400} (\bibinfo {year} {2019})}\BibitemShut {NoStop}%
\bibitem [{\citenamefont {Larson}\ and\ \citenamefont {Saleh}(2019)}]{Larson2019a}%
  \BibitemOpen
  \bibfield  {author} {\bibinfo {author} {\bibfnamefont {W.}~\bibnamefont {Larson}}\ and\ \bibinfo {author} {\bibfnamefont {B.~E.~A.}\ \bibnamefont {Saleh}},\ }\href {https://doi.org/10.1364/OPTICA.6.000402} {\bibfield  {journal} {\bibinfo  {journal} {Optica}\ }\textbf {\bibinfo {volume} {6}},\ \bibinfo {pages} {402} (\bibinfo {year} {2019})}\BibitemShut {NoStop}%
\bibitem [{\citenamefont {Tsang}(2021)}]{Tsang2021a}%
  \BibitemOpen
  \bibfield  {author} {\bibinfo {author} {\bibfnamefont {M.}~\bibnamefont {Tsang}},\ }\href {https://doi.org/10.22331/q-2021-08-19-527} {\bibfield  {journal} {\bibinfo  {journal} {Quantum}\ }\textbf {\bibinfo {volume} {5}},\ \bibinfo {pages} {527} (\bibinfo {year} {2021})},\ \Eprint {https://arxiv.org/abs/2103.08532} {2103.08532} \BibitemShut {NoStop}%
\bibitem [{\citenamefont {De}\ \emph {et~al.}(2021)\citenamefont {De}, \citenamefont {{Gil-Lopez}}, \citenamefont {Brecht}, \citenamefont {Silberhorn}, \citenamefont {{S{\'a}nchez-Soto}}, \citenamefont {Hradil},\ and\ \citenamefont {{\v R}eh{\'a}{\v c}ek}}]{De2021}%
  \BibitemOpen
  \bibfield  {author} {\bibinfo {author} {\bibfnamefont {S.}~\bibnamefont {De}}, \bibinfo {author} {\bibfnamefont {J.}~\bibnamefont {{Gil-Lopez}}}, \bibinfo {author} {\bibfnamefont {B.}~\bibnamefont {Brecht}}, \bibinfo {author} {\bibfnamefont {C.}~\bibnamefont {Silberhorn}}, \bibinfo {author} {\bibfnamefont {L.~L.}\ \bibnamefont {{S{\'a}nchez-Soto}}}, \bibinfo {author} {\bibfnamefont {Z.}~\bibnamefont {Hradil}},\ and\ \bibinfo {author} {\bibfnamefont {J.}~\bibnamefont {{\v R}eh{\'a}{\v c}ek}},\ }\href {https://doi.org/10.1103/PhysRevResearch.3.033082} {\bibfield  {journal} {\bibinfo  {journal} {Physical Review Research}\ }\textbf {\bibinfo {volume} {3}},\ \bibinfo {pages} {033082} (\bibinfo {year} {2021})}\BibitemShut {NoStop}%
\bibitem [{\citenamefont {Karuseichyk}\ \emph {et~al.}(2022)\citenamefont {Karuseichyk}, \citenamefont {Sorelli}, \citenamefont {Walschaers}, \citenamefont {Treps},\ and\ \citenamefont {Gessner}}]{Karuseichyk2022}%
  \BibitemOpen
  \bibfield  {author} {\bibinfo {author} {\bibfnamefont {I.}~\bibnamefont {Karuseichyk}}, \bibinfo {author} {\bibfnamefont {G.}~\bibnamefont {Sorelli}}, \bibinfo {author} {\bibfnamefont {M.}~\bibnamefont {Walschaers}}, \bibinfo {author} {\bibfnamefont {N.}~\bibnamefont {Treps}},\ and\ \bibinfo {author} {\bibfnamefont {M.}~\bibnamefont {Gessner}},\ }\href {https://doi.org/10.1103/PhysRevResearch.4.043010} {\bibfield  {journal} {\bibinfo  {journal} {Physical Review Research}\ }\textbf {\bibinfo {volume} {4}},\ \bibinfo {pages} {043010} (\bibinfo {year} {2022})}\BibitemShut {NoStop}%
\bibitem [{\citenamefont {Liang}\ \emph {et~al.}(2023)\citenamefont {Liang}, \citenamefont {Wadood},\ and\ \citenamefont {Vamivakas}}]{Liang2023}%
  \BibitemOpen
  \bibfield  {author} {\bibinfo {author} {\bibfnamefont {K.}~\bibnamefont {Liang}}, \bibinfo {author} {\bibfnamefont {S.~A.}\ \bibnamefont {Wadood}},\ and\ \bibinfo {author} {\bibfnamefont {A.~N.}\ \bibnamefont {Vamivakas}},\ }\href {https://doi.org/10.1364/OE.474036} {\bibfield  {journal} {\bibinfo  {journal} {Optics Express}\ }\textbf {\bibinfo {volume} {31}},\ \bibinfo {pages} {2726} (\bibinfo {year} {2023})}\BibitemShut {NoStop}%
\bibitem [{\citenamefont {Sajia}\ and\ \citenamefont {Qian}(2022)}]{Sajia2022}%
  \BibitemOpen
  \bibfield  {author} {\bibinfo {author} {\bibfnamefont {A.}~\bibnamefont {Sajia}}\ and\ \bibinfo {author} {\bibfnamefont {X.-F.}\ \bibnamefont {Qian}},\ }\href {https://doi.org/10.1103/PhysRevResearch.4.033244} {\bibfield  {journal} {\bibinfo  {journal} {Physical Review Research}\ }\textbf {\bibinfo {volume} {4}},\ \bibinfo {pages} {033244} (\bibinfo {year} {2022})}\BibitemShut {NoStop}%
\bibitem [{\citenamefont {Sajia}\ and\ \citenamefont {Qian}(2023)}]{Sajia2023}%
  \BibitemOpen
  \bibfield  {author} {\bibinfo {author} {\bibfnamefont {A.}~\bibnamefont {Sajia}}\ and\ \bibinfo {author} {\bibfnamefont {X.-F.}\ \bibnamefont {Qian}},\ }\href@noop {} {\bibinfo {title} {Superresolution picks entanglement over coherence}} (\bibinfo {year} {2023}),\ \Eprint {https://arxiv.org/abs/arXiv:2302.04909} {arXiv:arXiv:2302.04909} \BibitemShut {NoStop}%
\bibitem [{\citenamefont {Chrostowski}\ \emph {et~al.}(2017)\citenamefont {Chrostowski}, \citenamefont {{Demkowicz-Dobrza{\'n}ski}}, \citenamefont {Jarzyna},\ and\ \citenamefont {Banaszek}}]{Chrostowski2017a}%
  \BibitemOpen
  \bibfield  {author} {\bibinfo {author} {\bibfnamefont {A.}~\bibnamefont {Chrostowski}}, \bibinfo {author} {\bibfnamefont {R.}~\bibnamefont {{Demkowicz-Dobrza{\'n}ski}}}, \bibinfo {author} {\bibfnamefont {M.}~\bibnamefont {Jarzyna}},\ and\ \bibinfo {author} {\bibfnamefont {K.}~\bibnamefont {Banaszek}},\ }\href {https://doi.org/10.1142/S0219749917400056} {\bibfield  {journal} {\bibinfo  {journal} {International Journal of Quantum Information}\ }\textbf {\bibinfo {volume} {15}},\ \bibinfo {pages} {1740005} (\bibinfo {year} {2017})}\BibitemShut {NoStop}%
\bibitem [{\citenamefont {Mandel}(9 01)}]{Mandel1959}%
  \BibitemOpen
  \bibfield  {author} {\bibinfo {author} {\bibfnamefont {L.}~\bibnamefont {Mandel}},\ }\href@noop {} {\bibfield  {journal} {\bibinfo  {journal} {Proc. Phys. Soc.}\ } (\bibinfo {year} {1959-09-01})}\BibitemShut {NoStop}%
\bibitem [{\citenamefont {Goodman}(1996)}]{Goodman1996}%
  \BibitemOpen
  \bibfield  {author} {\bibinfo {author} {\bibfnamefont {J.~W.}\ \bibnamefont {Goodman}},\ }\href@noop {} {\emph {\bibinfo {title} {Introduction to {{Fourier}} Optics}}},\ \bibinfo {edition} {2nd}\ ed.,\ {{McGraw-Hill}} Series in Electrical and Computer Engineering\ (\bibinfo  {publisher} {{McGraw-Hill}},\ \bibinfo {address} {{New York}},\ \bibinfo {year} {1996})\BibitemShut {NoStop}%
\bibitem [{\citenamefont {Van~Trees}(2001)}]{VanTrees2001}%
  \BibitemOpen
  \bibfield  {author} {\bibinfo {author} {\bibfnamefont {H.~L.}\ \bibnamefont {Van~Trees}},\ }\href {https://doi.org/10.1002/0471221082} {\emph {\bibinfo {title} {Detection, {{Estimation}}, and {{Modulation Theory}}, {{Part I}}}}}\ (\bibinfo  {publisher} {{John Wiley \& Sons, Inc.}},\ \bibinfo {address} {{New York, USA}},\ \bibinfo {year} {2001})\BibitemShut {NoStop}%
\bibitem [{\citenamefont {Nielsen}\ and\ \citenamefont {Chuang}(2012)}]{Nielsen2012}%
  \BibitemOpen
  \bibfield  {author} {\bibinfo {author} {\bibfnamefont {M.~A.}\ \bibnamefont {Nielsen}}\ and\ \bibinfo {author} {\bibfnamefont {I.~L.}\ \bibnamefont {Chuang}},\ }\href {https://doi.org/10.1017/CBO9780511976667} {\emph {\bibinfo {title} {Quantum {{Computation}} and {{Quantum Information}}: 10th {{Anniversary Edition}}}}},\ \bibinfo {edition} {1st}\ ed.\ (\bibinfo  {publisher} {{Cambridge University Press}},\ \bibinfo {year} {2012})\BibitemShut {NoStop}%
\bibitem [{\citenamefont {Ragy}\ \emph {et~al.}(2016)\citenamefont {Ragy}, \citenamefont {Jarzyna},\ and\ \citenamefont {{Demkowicz-Dobrza{\'n}ski}}}]{Ragy2016}%
  \BibitemOpen
  \bibfield  {author} {\bibinfo {author} {\bibfnamefont {S.}~\bibnamefont {Ragy}}, \bibinfo {author} {\bibfnamefont {M.}~\bibnamefont {Jarzyna}},\ and\ \bibinfo {author} {\bibfnamefont {R.}~\bibnamefont {{Demkowicz-Dobrza{\'n}ski}}},\ }\href {https://doi.org/10.1103/PhysRevA.94.052108} {\bibfield  {journal} {\bibinfo  {journal} {Phys. Rev. A}\ }\textbf {\bibinfo {volume} {94}},\ \bibinfo {pages} {052108} (\bibinfo {year} {2016})}\BibitemShut {NoStop}%
\bibitem [{\citenamefont {Suzuki}\ \emph {et~al.}(2020)\citenamefont {Suzuki}, \citenamefont {Yang},\ and\ \citenamefont {Hayashi}}]{Suzuki2020}%
  \BibitemOpen
  \bibfield  {author} {\bibinfo {author} {\bibfnamefont {J.}~\bibnamefont {Suzuki}}, \bibinfo {author} {\bibfnamefont {Y.}~\bibnamefont {Yang}},\ and\ \bibinfo {author} {\bibfnamefont {M.}~\bibnamefont {Hayashi}},\ }\href {https://doi.org/10.1088/1751-8121/ab8b78} {\bibfield  {journal} {\bibinfo  {journal} {Journal of Physics A: Mathematical and Theoretical}\ }\textbf {\bibinfo {volume} {53}},\ \bibinfo {pages} {453001} (\bibinfo {year} {2020})}\BibitemShut {NoStop}%
\bibitem [{\citenamefont {{\v R}eh{\'a}{\v c}ek}\ \emph {et~al.}(2018)\citenamefont {{\v R}eh{\'a}{\v c}ek}, \citenamefont {Hradil}, \citenamefont {Koutn{\'y}}, \citenamefont {Grover}, \citenamefont {Krzic},\ and\ \citenamefont {{S{\'a}nchez-Soto}}}]{Rehacek2018}%
  \BibitemOpen
  \bibfield  {author} {\bibinfo {author} {\bibfnamefont {J.}~\bibnamefont {{\v R}eh{\'a}{\v c}ek}}, \bibinfo {author} {\bibfnamefont {Z.}~\bibnamefont {Hradil}}, \bibinfo {author} {\bibfnamefont {D.}~\bibnamefont {Koutn{\'y}}}, \bibinfo {author} {\bibfnamefont {J.}~\bibnamefont {Grover}}, \bibinfo {author} {\bibfnamefont {A.}~\bibnamefont {Krzic}},\ and\ \bibinfo {author} {\bibfnamefont {L.~L.}\ \bibnamefont {{S{\'a}nchez-Soto}}},\ }\href {https://doi.org/10.1103/PhysRevA.98.012103} {\bibfield  {journal} {\bibinfo  {journal} {Physical Review A}\ }\textbf {\bibinfo {volume} {98}},\ \bibinfo {pages} {012103} (\bibinfo {year} {2018})}\BibitemShut {NoStop}%
\bibitem [{\citenamefont {Yang}\ \emph {et~al.}(2023)\citenamefont {Yang}, \citenamefont {Zhao}, \citenamefont {Chen}, \citenamefont {Du}, \citenamefont {Fan}, \citenamefont {Zhang},\ and\ \citenamefont {Zhao}}]{Yang2023}%
  \BibitemOpen
  \bibfield  {author} {\bibinfo {author} {\bibfnamefont {X.}~\bibnamefont {Yang}}, \bibinfo {author} {\bibfnamefont {R.}~\bibnamefont {Zhao}}, \bibinfo {author} {\bibfnamefont {H.}~\bibnamefont {Chen}}, \bibinfo {author} {\bibfnamefont {Y.}~\bibnamefont {Du}}, \bibinfo {author} {\bibfnamefont {C.}~\bibnamefont {Fan}}, \bibinfo {author} {\bibfnamefont {G.}~\bibnamefont {Zhang}},\ and\ \bibinfo {author} {\bibfnamefont {Z.}~\bibnamefont {Zhao}},\ }\href {https://doi.org/10.3390/photonics10121350} {\bibfield  {journal} {\bibinfo  {journal} {Photonics}\ }\textbf {\bibinfo {volume} {10}},\ \bibinfo {pages} {1350} (\bibinfo {year} {2023})}\BibitemShut {NoStop}%
\bibitem [{\citenamefont {Xiao}\ \emph {et~al.}(2024)\citenamefont {Xiao}, \citenamefont {Kedem~Orange}, \citenamefont {Opatovski}, \citenamefont {Parizat}, \citenamefont {Nehme}, \citenamefont {Alalouf},\ and\ \citenamefont {Shechtman}}]{Xiao2024}%
  \BibitemOpen
  \bibfield  {author} {\bibinfo {author} {\bibfnamefont {D.}~\bibnamefont {Xiao}}, \bibinfo {author} {\bibfnamefont {R.}~\bibnamefont {Kedem~Orange}}, \bibinfo {author} {\bibfnamefont {N.}~\bibnamefont {Opatovski}}, \bibinfo {author} {\bibfnamefont {A.}~\bibnamefont {Parizat}}, \bibinfo {author} {\bibfnamefont {E.}~\bibnamefont {Nehme}}, \bibinfo {author} {\bibfnamefont {O.}~\bibnamefont {Alalouf}},\ and\ \bibinfo {author} {\bibfnamefont {Y.}~\bibnamefont {Shechtman}},\ }\href {https://doi.org/10.1126/sciadv.adj3656} {\bibfield  {journal} {\bibinfo  {journal} {Science Advances}\ }\textbf {\bibinfo {volume} {10}},\ \bibinfo {pages} {eadj3656} (\bibinfo {year} {2024})}\BibitemShut {NoStop}%
\bibitem [{\citenamefont {Pawley}\ and\ \citenamefont {Masters}(2008)}]{Pawley2008}%
  \BibitemOpen
  \bibfield  {author} {\bibinfo {author} {\bibfnamefont {J.~B.}\ \bibnamefont {Pawley}}\ and\ \bibinfo {author} {\bibfnamefont {B.~R.}\ \bibnamefont {Masters}},\ }\href {https://doi.org/10.1117/1.2911629} {\bibfield  {journal} {\bibinfo  {journal} {Journal of Biomedical Optics}\ }\textbf {\bibinfo {volume} {13}},\ \bibinfo {pages} {029902} (\bibinfo {year} {2008})}\BibitemShut {NoStop}%
\bibitem [{\citenamefont {Shechtman}\ \emph {et~al.}(2014)\citenamefont {Shechtman}, \citenamefont {Sahl}, \citenamefont {Backer},\ and\ \citenamefont {Moerner}}]{Shechtman2014}%
  \BibitemOpen
  \bibfield  {author} {\bibinfo {author} {\bibfnamefont {Y.}~\bibnamefont {Shechtman}}, \bibinfo {author} {\bibfnamefont {S.~J.}\ \bibnamefont {Sahl}}, \bibinfo {author} {\bibfnamefont {A.~S.}\ \bibnamefont {Backer}},\ and\ \bibinfo {author} {\bibfnamefont {W.~E.}\ \bibnamefont {Moerner}},\ }\href {https://doi.org/10.1103/PhysRevLett.113.133902} {\bibfield  {journal} {\bibinfo  {journal} {Physical Review Letters}\ }\textbf {\bibinfo {volume} {113}},\ \bibinfo {pages} {133902} (\bibinfo {year} {2014})}\BibitemShut {NoStop}%
\bibitem [{\citenamefont {Li}\ \emph {et~al.}(2018)\citenamefont {Li}, \citenamefont {Xue}, \citenamefont {Qu}, \citenamefont {Ho},\ and\ \citenamefont {Blu}}]{Li2018}%
  \BibitemOpen
  \bibfield  {author} {\bibinfo {author} {\bibfnamefont {J.}~\bibnamefont {Li}}, \bibinfo {author} {\bibfnamefont {F.}~\bibnamefont {Xue}}, \bibinfo {author} {\bibfnamefont {F.}~\bibnamefont {Qu}}, \bibinfo {author} {\bibfnamefont {Y.-P.}\ \bibnamefont {Ho}},\ and\ \bibinfo {author} {\bibfnamefont {T.}~\bibnamefont {Blu}},\ }\href {https://doi.org/10.1364/OE.26.026120} {\bibfield  {journal} {\bibinfo  {journal} {Optics Express}\ }\textbf {\bibinfo {volume} {26}},\ \bibinfo {pages} {26120} (\bibinfo {year} {2018})}\BibitemShut {NoStop}%
\bibitem [{\citenamefont {Pawley}(2006)}]{Pawley2006}%
  \BibitemOpen
  \bibinfo {editor} {\bibfnamefont {J.~B.}\ \bibnamefont {Pawley}},\ ed.,\ \href@noop {} {\emph {\bibinfo {title} {Handbook of Biological Confocal Microscopy}}},\ \bibinfo {edition} {3rd}\ ed.\ (\bibinfo  {publisher} {{Springer}},\ \bibinfo {address} {{New York, NY}},\ \bibinfo {year} {2006})\BibitemShut {NoStop}%
\bibitem [{\citenamefont {Liu}\ \emph {et~al.}(2014)\citenamefont {Liu}, \citenamefont {Xiong}, \citenamefont {Song},\ and\ \citenamefont {Wang}}]{Liu2014}%
  \BibitemOpen
  \bibfield  {author} {\bibinfo {author} {\bibfnamefont {J.}~\bibnamefont {Liu}}, \bibinfo {author} {\bibfnamefont {H.-N.}\ \bibnamefont {Xiong}}, \bibinfo {author} {\bibfnamefont {F.}~\bibnamefont {Song}},\ and\ \bibinfo {author} {\bibfnamefont {X.}~\bibnamefont {Wang}},\ }\href {https://doi.org/10.1016/j.physa.2014.05.028} {\bibfield  {journal} {\bibinfo  {journal} {Physica A: Statistical Mechanics and its Applications}\ }\textbf {\bibinfo {volume} {410}},\ \bibinfo {pages} {167} (\bibinfo {year} {2014})}\BibitemShut {NoStop}%
\end{thebibliography}%

\appendix
\onecolumngrid

\section{Precision limit of direct imaging\protect\label{sec:Precision-limit-of}}

Suppose the probability distribution of a single photon of position
$x$ on the image plane is
\begin{equation}
\Lambda(x)=\frac{1-\dn}{2}|\psi_{1}(x)|^{2}+\frac{1+\dn}{2}|\psi_{2}(x)|^{2}.\label{eq:lamx}
\end{equation}
Here, $\psi_{1}(x)$ and $\psi_{2}(x)$ are the point-spread functions
of the two sources, and $\dn$ is the ratio between the difference
and the total of the photon numbers of the two sources,
\begin{equation}
\dn=\frac{\nb[2]-\nb[1]}{\nb[2]+\nb[1]}.
\end{equation}

In order to quantify the precision limit of estimating the separation
of two incoherent optical sources, the well-known Cramér-Rao bound
\cite{Helstrom1969a,Yang2019,Liu2020,Albarelli2020} from the classical
parameter estimation theory needs to be invoked, which demonstrates
that over all possible estimation strategies, the covariance matrix
$C$ for the estimates of the parameters $g_{1},\cdots,g_{n}$ is
lower bounded by 
\begin{equation}
C[g_{1},...,g_{n}]\geq M^{-1}\mathcal{J}^{-1}[g_{1},...,g_{n}].\label{eq:cramer-rao-1}
\end{equation}
The ``$\geq$'' sign represents semi-definite positivity of matrix,
$M$ is the number of samples, and $\mathcal{J}[g_{1},...,g_{n}]$
denotes the Fisher information matrix defined as
\begin{equation}
\mathcal{J}_{ij}=\int_{-\infty}^{\infty}\frac{1}{P(x)}\frac{\partial P(x)}{\partial g_{i}}\frac{\partial P(x)}{\partial g_{j}}dx,\label{eq:classical=000020fisher-1}
\end{equation}
where $P(x)$ is a probability distribution of random variable $x$
dependent on the parameters $g_{1},\cdots,g_{n}$. The benefit of
the Cramér-Rao inequality is that its lower bound can always be asymptotically
saturated by the maximum likelihood estimation strategy when $M$
is sufficiently large \cite{Helstrom1969a}.

For the task of resolving two point sources, the positions of the
two sources, $X_{1}$ and $X_{2}$, or equivalently the centroid $\bar{X}=\frac{X_{1}+X_{2}}{2}$
and the separation $d=X_{2}-X_{1}$ are unknown, so $\bar{X}$ and
$d$ are the parameters to estimate, and $P(x)$ is the position distribution
$\Lambda(x)$ \eqref{eq:lamx} of the photon from the incoherent sources.

For the direct imaging with an ideal continuum photodetector in the
image plane, if the intensities and the point-spread functions of
the two optical sources are equal, the position distribution \eqref{eq:lamx}
of the arrival single photon can be rewritten as
\begin{equation}
\Lambda(x)=\frac{1}{2}I(x-\bar{X}+d/2)+\frac{1}{2}I(x-\bar{X}-d/2),\label{eq:probability-1-1}
\end{equation}
where $I(x)$ is defined as
\begin{equation}
I(x)=|\psi(x)|^{2}.
\end{equation}

When the separation $d$ is small, one can expand $\Lambda(x)$ to
the second order of $d$,
\begin{equation}
\Lambda(x)=I(x-\bar{X})+\frac{1}{8}d^{2}I''(x-\bar{X}).
\end{equation}
Then according Eq.\eqref{eq:classical=000020fisher-1}, the classical
Fisher information matrix for the parameters $\bar{X}$, $d$ can
be obtained as
\begin{equation}
\mathcal{J}=\begin{bmatrix}\alpha_{1}-\frac{\alpha_{2}d^{2}}{8} & 0\\
0 & \frac{\alpha_{2}d^{2}}{16}
\end{bmatrix},
\end{equation}
where
\begin{equation}
\alpha_{i}=\int_{-\infty}^{\infty}dx\frac{1}{I(x-\bar{X})}\left[\partial_{x}^{i}I(x-\bar{X})\right]^{2}.
\end{equation}

If one is interested in the parameter $d$ only (though $\bar{X}$
is still unknown), a weight matrix $W={\rm Diag}\{0,1\}$ can be multiplied
to the both sides of the Cramér-Rao inequality \eqref{eq:cramer-rao-1},
and the variance of the estimate of $d$ is lower bounded by
\begin{align}
{\rm Var}(\hat{d}) & \geqslant{\rm Tr}(W\mathcal{J}^{-1}),\label{eq:crb-1-1-1}
\end{align}
where $\hat{d}$ denotes the estimate of the separation $d$. Considering
the variance may diverge when $d\rightarrow0$, we use the reciprocal
of the variance to quantify the estimation precision of $d$. Thus
the estimation precision of the separation $d$ by direct imaging
is
\begin{equation}
\mathcal{H}_{d}^{({\rm direct})}=\nt/(\mathcal{J}^{-1})_{22}=\frac{1}{8}d^{2}\alpha_{2}\text{\ensuremath{\nb}}.
\end{equation}
Obviously, when $d\rightarrow0$, the estimation separation precision
vanishes, so one can hardly obtain any information about the separation
by direct imaging when the two point sources are sufficiently close
to each other. This is called the Rayleigh's curse \cite{Tsang2016}
for direct imaging of two identical optical point sources.

The above result can be generalized to the case that the two point-spread
functions are the same but the intensities are unequal. In this case,
the position distribution of single photon can be written as
\begin{equation}
\Lambda(x)=\frac{1-\dn}{2}I(x-\bar{X}+d/2)+\frac{1+\dn}{2}I(x-\bar{X}-d/2).
\end{equation}
Expanding the $\Lambda(x)$ to the second order of $d$, one can obtain
\begin{align}
\Lambda(x) & =I(x-\bar{X})-\frac{1}{2}d\epsilon I^{\prime}(x-\bar{X})+\frac{1}{8}d^{2}I^{\prime\prime}(x-\bar{X}).
\end{align}
The Fisher information matrix $\mathcal{J}$ with respect to the parameters
$\bar{X}$, $d$ can be obtained, and the entries are
\begin{equation}
\mathcal{J}=\begin{bmatrix}\alpha_{1}+\frac{1}{8}d^{2}\left[2\beta\epsilon^{2}-\alpha_{2}\left(2\epsilon^{2}+1\right)\right] & \frac{\alpha_{1}\epsilon}{2}+\frac{\epsilon d^{2}}{16}\left[2\beta\epsilon^{2}-\alpha_{2}\left(2\epsilon^{2}+1\right)\right]\\
\frac{\epsilon\alpha_{1}}{2}+\frac{\epsilon d^{2}}{16}\left[2\beta\epsilon^{2}-\alpha_{2}\left(2\epsilon^{2}+1\right)\right] & \frac{\alpha_{1}\epsilon^{2}}{4}+\frac{d^{2}}{32}\left[\alpha_{2}\left(2-5\epsilon^{2}\right)+2\beta\epsilon^{4}\right]
\end{bmatrix},
\end{equation}
where $\beta=\int_{-\infty}^{\infty}dx\frac{1}{I(x-\bar{X})^{3}}\left[\partial_{x}I(x-\bar{X})\right]^{4}$.
Then the estimation precision of the separation $d$ by direct imaging
is
\begin{equation}
\mathcal{H}_{d}^{({\rm direct})}=\nt/(\mathcal{J}^{-1})_{22}=\frac{\nt\alpha_{2}}{16}d^{2}\left(1-\epsilon^{2}\right)^{2},
\end{equation}
which means that no information about the separation $d$ can be acquired
by direct imaging when $d\rightarrow0$ in this generalized case.

\section{\protect\label{sec:Derivation=000020f}Multiparameter quantum metrology
theory for superresolution}

The quantum parameter estimation theory considers the estimation precision
for unknown parameters in a parameter-dependent quantum state by quantum
measurements. The core of the quantum parameter estimation theory
is the quantum Cramér-Rao inequality. For the single-parameter case,
the quantum Cramér-Rao inequality tells that the variance of the estimate
of the unknown parameter is lower bounded by the inverse of the quantum
Fisher information which maximizes the classical Fisher information
over all possible quantum measurements on the parameter-dependent
quantum state. For the multi-parameter case, the variance and the
quantum Fisher information are generalized to the covariance matrix
and the quantum Fisher information matrix respectively between the
different parameters.

In detail, if one measures a quantum system with its density matrix
$\hro$ dependent on unknown parameters $g_{1},\cdots,g_{n}$ to determine
these parameters, the covariance matrix for the estimates of these
parameters is bounded by the following quantum Cramér-Rao's inequality,
\begin{equation}
C[g_{1},\cdots,g_{n}]\geq M^{-1}\mathcal{Q}^{-1}[g_{1},\cdots,g_{n}],\label{eq:mcr}
\end{equation}
where $M$ is the number of the quantum states, $C[g_{1},\cdots,g_{n}]$
is the covariance matrix for the estimates of $g_{1},\cdots,g_{n}$,
$\mathcal{Q}[g_{1},\cdots,g_{n}]$ is the quantum Fisher information
matrix about $g_{1},\cdots,g_{n}$ with entries
\begin{equation}
\mathcal{Q}_{ij}(\hat{\rho})=\frac{1}{2}{\rm Tr}[\hro\{\hl i,\hl j\}],\label{eq:fisher0-2}
\end{equation}
and $\hl i$, $\hl j$ are the symmetric logarithmic derivatives (SLD)
of the density operator $\hro$,
\begin{equation}
\frac{1}{2}(\hl i\hro+\hro\hl i)=\partial_{\gi}\hro.\label{eq:sld-2}
\end{equation}

To obtain the quantum Fisher information matrix for $\hro$ which
depends on the unknown parameters $g_{i}$, one can decompose the
density matrix $\hro$ as
\begin{equation}
\hro=\sum_{k}\lambda_{k}|\lambda_{k}\rangle\langle\lambda_{k}|,
\end{equation}
where $\lambda_{k}$ and $|\lambda_{k}\rangle$ are eigenvalue and
eigenstate of $\hro$. The matrix elements of both sides of Eq. \eqref{eq:sld-2}
with respect to the eigenbasis $|\lambda_{k}\rangle$ can be obtained
as
\begin{equation}
\langle\lambda_{k}|\partial_{g_{i}}\hat{\rho}|\lambda_{h}\rangle=\frac{1}{2}\left(\langle\lambda_{k}|\hat{\rho}\hl i|\lambda_{h}\rangle+\langle\lambda_{k}|\hl i\hat{\rho}|\lambda_{h}\rangle\right),\label{eq:fisher2}
\end{equation}
which leads to
\begin{equation}
\hl i=\sum_{\lambda_{k}+\lambda_{h}\neq0}\frac{2\langle\lambda_{k}|\partial_{g_{i}}\hat{\rho}|\lambda_{h}\rangle}{\lambda_{k}+\lambda_{h}}\ket{\lambda_{k}}\bra{\lambda_{h}}.
\end{equation}
Thus, the elements of the quantum Fisher information matrix can be
obtained as
\begin{equation}
\mathcal{Q}_{ij}(\hat{\rho})=2\sum_{\lambda_{k}+\lambda_{l}\neq0}\frac{\langle\lambda_{k}|\partial_{g_{i}}\hat{\rho}|\lambda_{l}\rangle\langle\lambda_{l}|\partial_{g_{j}}\hat{\rho}|\lambda_{k}\rangle}{\lambda_{k}+\lambda_{l}}.
\end{equation}
As the density matrix \eqref{eq:rho} of a single photon is low-rank,
it is more convenient to compute the quantum Fisher information matrix
on the support of $\hro$,
\begin{equation}
\begin{aligned}\mathcal{Q}_{ij}(\hat{\rho})= & \sum_{\lambda_{k}\neq0}\frac{4\langle\lambda_{k}|\partial_{g_{i}}\hat{\rho}\partial_{g_{j}}\hat{\rho}|\lambda_{k}\rangle}{\lambda_{k}}\\
 & +\sum_{\lambda_{k},\lambda_{h}\neq0}2\left(\frac{1}{\lambda_{k}+\lambda_{h}}-\frac{1}{\lambda_{k}}-\frac{1}{\lambda_{h}}\right)\\
 & \times\langle\lambda_{h}|\partial_{g_{i}}\hat{\rho}|\lambda_{k}\rangle\langle\lambda_{k}|\partial_{g_{j}}\hat{\rho}|\lambda_{h}\rangle,
\end{aligned}
\label{eq:fisher=000020exp}
\end{equation}
which is a variant of the result in Ref. \cite{Liu2014}.

For the superresolution of two point sources, the positions of the
two point sources, $\xk[1]$ and $\xk[2]$, or equivalently the centroid $\xb=\frac{\xk[1]+\xk[2]}{2}$
and the separation $d=\xk[2]-\xk[1]$, are the unknown parameters
to estimate, so the multi-parameter quantum Cramér-Rao bound \eqref{eq:mcr}
tells that 
\begin{equation}
C[\bar{X},\,d]\geq\nt[-1]\mathcal{Q}^{-1}[\bar{X},\,d],\label{eq:qcr}
\end{equation}
where $\nt$ is the total number of photons $\nb[1]+\nb[2]$, $C[\bar{X},\,d]$
is the covariance matrix for the estimates of $\bar{X},\,d$, $\mathcal{Q}[\bar{X},\,d]$
is the quantum Fisher information matrix about $\bar{X},\,d$ with
entries in Eq. \eqref{eq:fisher0-2}.

If one is interested in only one or a few unknown parameters, e.g.,
the separation as in the superresolution, a positive semidefinite
weight matrix $W$ can be imposed on both sides of the quantum Cramér-Rao
inequality \eqref{eq:qcr}, then the matrix quantum Cramér-Rao bound
turns to a scalar form,
\begin{equation}
{\rm Tr}(WC)\geq\nt[-1]{\rm Tr}(W\mathcal{Q}^{-1}).\label{eq:scalar=000020qcr}
\end{equation}

Note that for quantum multiparameter estimation, the optimal measurements
for different parameters are not always compatible, so the matrix
quantum Cramér-Rao bound is not always achievable. The saturation
condition given by Ragy et al. \cite{Ragy2016} for the above scalar
quantum Cramér-Rao bound is
\begin{equation}
{\rm Tr}(\hat{\rho}[\hl i,\hl j])={\rm Im[Tr}(\hat{\rho}\hl i\hl j)]=0,\label{eq:ragy}
\end{equation}
as the scalar quantum Cramér-Rao bound $N_{tot}^{-1}{\rm Tr}(W\mathcal{Q}^{-1})$
coincides with the Holevo bound in this case and the latter is always
asymptotically achievable when $\nt\rightarrow\infty$. As the point-spread
functions $\psi_{i}(x)$ are real in the superresolution problem,
the density matrix $\hro$ and the eigenvalues and eigenstates of
$\hro$ are also real in this case as will be shown in the following
sections, so the compatibility condition \eqref{eq:ragy} is saturated,
and the scalar quantum Cramér-Rao bound \eqref{eq:scalar=000020qcr}
can be achieved by the maximum-likelihood estimation strategy for
$\nt\rightarrow\infty$.

\section{\protect\label{sec:Quantum-fisher-information=000020of=000020two=000020identical=000020source}Precision
for superresolution of two identical optical sources}

For two incoherent optical sources if one photon is emitted, the quantum
density operator for the single photon of the optical field on the
image plane can be written as
\begin{equation}
\hro=\frac{1-\dn}{2}\nei{\xk[1]}\ksi\bsi\ei{\xk[1]}+\frac{1+\dn}{2}\nei{\xk[2]}\ksi\bsi\ei{\xk[2]},
\end{equation}
where we define $\hp=-i\partial_{x}$ for the sake of convenience.
The density matrix $\hro$ can be infinite-dimensional in principle,
but has only two nonzero eigenvalues which can be obtained as 
\begin{equation}
\lambda_{1,2}=\frac{1}{2}\left(1\mp\sqrt{\epsilon^{2}+\delta^{2}\left(1-\epsilon^{2}\right)}\right).\label{eq:ev-1-1-1}
\end{equation}
The associated eigenstates of $\hro$ are 
\begin{align}
|\lambda_{1,2}\rangle & =\frac{1}{\sqrt{\mathscr{Q}_{1,2}}}\times\{\nei{\xk[1]}\ksi\mp\frac{\left(\sqrt{\epsilon^{2}+\delta^{2}\left(1-\epsilon^{2}\right)}\mp\epsilon\right)}{\delta(1-\epsilon)}\nei{\xk[2]}\ksi\},\label{eq:es-1-1-1}
\end{align}
where $\mathscr{Q}_{1,2}$ are the normalization constants,
\begin{align}
\mathscr{Q}_{1,2} & =\left(1-\delta^{2}\right)\left(1+\frac{\delta^{2}\left(1-\delta^{2}\right)(1-\epsilon)^{2}}{\left(\sqrt{\delta^{2}+\epsilon^{2}-\delta^{2}\epsilon^{2}}\pm\delta^{2}(1-\epsilon)\pm\epsilon\right)^{2}}\right),
\end{align}
and because point-spread functions are assumed to be real, so it can
simplify to
\begin{equation}
\delta=\langle\psi|\e^{-id\hat{P}}|\psi\rangle=\avg[\cos(d\hp)].\label{eq:det}
\end{equation}

\subsection{With known arbitrary photon numbers\protect\label{subsec:With-known-arbitrary}}

When the photon numbers $\nb[1]$ and $\nb[2]$ of the two point sources
are known, one can derive the quantum Fisher information matrix $\mathcal{Q}$
for parameters $\bar{X},\,d$ by plugging the above eigenvalues and
eigenstates of $\hro$ into Eq. \eqref{eq:fisher=000020exp},
\begin{equation}
\mathcal{Q}=\begin{bmatrix}4\kappa-4\gamma^{2}\left(1-\epsilon^{2}\right) & 2\kappa\epsilon\\
2\kappa\epsilon & \kappa
\end{bmatrix}.
\end{equation}
Taking the inverse of the quantum Fisher information matrix \eqref{eq:qfims},
the estimation precision of the separation $d$ can be worked out
as
\begin{equation}
\mathcal{H}_{d}=\nt/(\mathcal{Q}^{-1})_{22}=\frac{\nt\kappa\left(1-\epsilon^{2}\right)\left(\kappa-\gamma^{2}\right)}{\kappa-\gamma^{2}\left(1-\epsilon^{2}\right)},\label{eq:snhd-1}
\end{equation}
where
\begin{equation}
\kappa=\langle\psi|\hat{P}^{2}|\psi\rangle=\avg,\;\gamma=i\langle\psi|\hat{P}\e^{-id\hat{P}}|\psi\rangle=\avg[\sin(d\hp)\hp].\label{eq:kgs}
\end{equation}
When the separation goes to zero, the estimation separation precision
$\mathcal{H}_{d}$ becomes
\begin{equation}
\dlm{\mathcal{H}_{d}}=\nt\left(1-\epsilon^{2}\right)\avg[\hp^{2}].\label{eq:hds}
\end{equation}
For two Gaussian point-spread functions, the estimation precision
for the separation $d$ is
\begin{equation}
\hd=\frac{\nt\left(1-\epsilon^{2}\right)\left(4\sigma^{2}\e^{\frac{d^{2}}{4\sigma^{2}}}-d^{2}\right)}{4\sigma^{2}\left(4\sigma^{2}\e^{\frac{d^{2}}{4\sigma^{2}}}-d^{2}\left(1-\epsilon^{2}\right)\right)}.\label{eq:ndhd}
\end{equation}
In the limit $d\rightarrow0$, the estimation precision of the separation
\eqref{eq:hds} can be simplified to
\begin{equation}
\dlm{\hd}=\frac{\nt\left(1-\epsilon^{2}\right)}{4\sigma^{2}},\label{eq:hdg}
\end{equation}
which shows $\mathcal{H}_{d}$ keeps nonzero when $d$ is zero.

When the photon numbers of the two optical sources are the same, $\epsilon=0$,
$\nb[1]=\nb[2]=\nb$ and $\nt=2\nb$, the general result \eqref{eq:hds}
for the estimation precision of $d$ in the limit $d\rightarrow0$
can be simplified to 
\begin{equation}
\mathcal{H}_{d}=2\nb\avg[\hp^{2}].
\end{equation}
And for the Gaussian case, the result becomes 
\begin{equation}
\mathcal{H}_{d}=\frac{\nb}{2\sigma^{2}}.
\end{equation}
This recovers the results for the original superresolution scheme
in the  paper by Tsang et al. \cite{Tsang2016}.

\subsection{Unknown arbitrary photon numbers\protect\label{subsec:Unknown-arbitrary-photon}}

For two incoherent optical sources with unknown photon numbers, according
to Eq. \eqref{eq:fisher=000020exp}, one can derive the quantum Fisher
information matrix $\mathcal{Q}$ for parameters $\bar{X},\,d,\,\epsilon$
as
\begin{equation}
\mathcal{Q}=\left[\begin{array}{ccc}
4\kappa-4\gamma^{2}\left(1-\epsilon^{2}\right) & 2\kappa\epsilon & 2\gamma\delta\\
2\kappa\epsilon & \kappa & 0\\
2\gamma\delta & 0 & \frac{1-\delta^{2}}{1-\epsilon^{2}}
\end{array}\right],\label{eq:qfims}
\end{equation}
where $\kappa$, $\gamma$ are defined in Eq. \eqref{eq:kgs} and
$\delta$ is defined in Eq. \eqref{eq:det}.Taking the inverse of
the quantum Fisher information matrix $\mathcal{Q}$ \eqref{eq:qfims},
the estimation precision of the separation $d$ can be worked out
as
\begin{equation}
\hd=\nt/(\mathcal{Q}^{-1})_{22}=\frac{\nt\kappa\left(1-\epsilon^{2}\right)\left[\left(1-\delta^{2}\right)\kappa-\gamma^{2}\right]}{(1-\delta^{2})\kappa-\gamma^{2}\left(1-\epsilon^{2}\right)}.\label{eq:snhd}
\end{equation}

If the two photon numbers $\nb[1]$, $\nb[2]$ are different and $\nb[2]-\nb[1]$
keeps finite, the precision of the separation \eqref{eq:snhd} can
be simplified to
\begin{equation}
\hd=\frac{\nt\left(1-\epsilon^{2}\right)\var(\hp^{2})d^{2}}{4\epsilon^{2}},\label{eq:hd0}
\end{equation}
which is of order $O(d^{2})$, where $\var(\hp^{2})=\avg[\hp^{4}]-\avg[][][2]$.
This indicates that estimation precision for the separation vanishes
when $d$ is small, in drastic constraint to the above results for
resolving two identical incoherent optical sources with known photon
numbers. So the superresolution scheme does not work and the Rayleigh's
curse takes effect again in this case, which is sometimes called the
resurgence of Rayleigh's curse. In order to study how to address this
issue, we need to take a deeper insight into the estimation precision
$\hd$ by expanding the numerator and denominator of $\hd$ \eqref{eq:snhd}
to the fourth order of $d$,
\begin{equation}
\mathcal{H}_{d}=\frac{A_{0}+A_{2}d^{2}+A_{4}d^{4}}{B_{0}+B_{2}d^{2}+B_{4}d^{4}},
\end{equation}
with the expansion coefficients
\begin{equation}
\begin{aligned}A_{0} & =0,\;B_{0}=0,\\
A_{2} & =0,\;B_{2}=12\dn^{2}\avg,\\
A_{4} & =3\nt(1-\dn^{2})\avg\var(\hp^{2}),\\
B_{4} & =3\var(\hp^{2})-4\dn^{2}\avg[\hp^{4}],
\end{aligned}
\end{equation}
which turns out to be
\begin{equation}
\mathcal{H}_{d}=\frac{3\nt d^{2}\left(1-\epsilon^{2}\right)\avg\var(\hp^{2})}{12\epsilon^{2}\avg+d^{2}\left[3\var(\hp^{2})-4\dn^{2}\avg[\hp^{4}]\right]}.
\end{equation}

Obviously, when the photon numbers of the two incoherent point sources
are the same, i.e., $\epsilon=0$, the leading constant term of the
denominator of $\mathcal{H}_{d}$ vanishes and both the numerator
and the denominator of $\mathcal{H}_{d}$ are of order $O(d^{2})$,
so $\mathcal{H}_{d}$ keeps nonzero and independent of $d$ when $d\rightarrow0$,
which is the quantum superresolution of two incoherent point sources.
In contrast, if the photon numbers of the two optical sources are
different, i.e., $\nb[1]\neq\nb[2]$, the magnitude of the numerator
of $\mathcal{H}_{d}$ is $O(d^{2})$ but the magnitude of the denominator
of $\mathcal{H}_{d}$ is $O(1)$ if the difference between the two
photon numbers keeps finite. In this case when $d\rightarrow0$ the
separation precision vanishes, which is the limitation imposed by
the Rayleigh's curse.

Take the Gaussian point-spread functions as an example. The estimation
precision $\mathcal{H}_{d}$ \eqref{eq:snhd} becomes
\begin{equation}
\hd=\frac{\nt\left(1-\epsilon^{2}\right)\left[d^{2}+4\sigma^{2}\left(1-\e^{\frac{d^{2}}{4\sigma^{2}}}\right)\right]}{4\sigma^{2}\left[4\sigma^{2}\left(1-\e^{\frac{d^{2}}{4\sigma^{2}}}\right)+d^{2}\left(1-\epsilon^{2}\right)\right]}.\label{eq:gauhd}
\end{equation}
If we further define a dimensionless separation between the two Gaussian
point sources as
\begin{equation}
\dt=\frac{d}{\sigma},\label{eq:dt11}
\end{equation}
the estimation precision $\hd$ \eqref{eq:gauhd} can be simplified
to
\begin{equation}
\hd=\frac{\left(4e^{\frac{\dt[2]}{4}}-\dt[2]-4\right)\left(1-\epsilon^{2}\right)}{4e^{\frac{\dt[2]}{4}}-\dt[2]\left(1-\epsilon^{2}\right)-4}\hdo,\label{eq:hdr}
\end{equation}
where $\hdo$ is the optimal estimation precision of $d$ when the
photon numbers of the two sources are the same, i.e., $\nb[1]=\nb[2]=\nb$,
\begin{equation}
\hdo=\frac{\nb}{2\sigma^{2}}.
\end{equation}
This implies that the ratio between the estimation precision $\hd$
to the optimal precision $\hdo$ is determined by the parameters $\epsilon$
and $r$ only.

By expanding both the numerator and denominator of Eq. \eqref{eq:gauhd}
to the fourth order of $d$, one can have
\begin{equation}
\hd=\frac{\nt\left(1-\epsilon^{2}\right)d^{2}}{32\sigma^{4}\epsilon^{2}+4d^{2}\sigma^{2}},
\end{equation}
where a common $d^{2}$ has been reduced from both the numerator and
denominator. It shows $\mathcal{H}_{d}$ keeps nonzero when $d$ is
small if $\epsilon=0$ which is the result of quantum superresolution
and vanishes if $\epsilon$ is nonzero and keeps finite which is the result
of Rayleigh's criterion.

The above analysis givens an important implication that the magnitude
of $\epsilon$ plays a critical role in determining the effect of
superresolution. If $\nb[1]$ and $\nb[2]$ are different but the
magnitude of $\epsilon$ is of order $O(d\sqrt{\var(\hp^{2})/\avg})$
or higher, the order of the denominator of $\mathcal{H}_{d}$ is still
$O(d^{2})$ then and $\mathcal{H}_{d}$ will not vanish in this case.
This immediately leads to the condition for the superresolution with
two identical incoherent optical sources in the main text.

\section{\protect\label{sec:Quantum-fisher-information=000020of=000020unequal=000020source}Precision
for resolving two arbitrary point sources}

\subsection{General point-spread functions}

For two arbitrary optical point sources, the density operator of a
single photon can be written as
\begin{equation}
\hro=\frac{1-\dn}{2}\nei{\xk[1]}\ksi[1]\bsi[1]\ei{\xk[1]}+\frac{1+\dn}{2}\nei{\xk[2]}\ksi[2]\bsi[2]\ei{\xk[2]}.
\end{equation}
To obtain the quantum Fisher information matrix, we need to work out
the eigenvalues and eigenstates of $\hro$. Note that $\hro$ can
be infinite-dimensional in principle, but has only two nonzero eigenvalues,
which turn out to be
\begin{equation}
\lambda_{1,2}=\frac{1}{2}\left(1\mp\sqrt{\epsilon^{2}+\delta^{2}\left(1-\epsilon^{2}\right)}\right),\label{eq:ev}
\end{equation}
and the associated eigenstates $|\lambda_{1,2}\rangle$ are
\begin{equation}
|\lambda_{1,2}\rangle=\frac{1}{\sqrt{\mathscr{Q}_{1,2}}}\times\{\nei{\xk[1]}\ksi[1]\mp\frac{\left(\sqrt{\epsilon^{2}+\delta^{2}\left(1-\epsilon^{2}\right)}\mp\epsilon\right)}{\delta(1-\epsilon)}\nei{\xk[2]}\ksi[2]\},\label{eq:es}
\end{equation}
where $\delta$ is the overlap between the point-spread functions
of the two sources,
\begin{equation}
\delta=\bra{\psi_{1}}\ei{\xk[1]}\nei{\xk[2]}\ket{\psi_{2}}=\avv{\e^{-id\hp}}=\avv{\cos(d\hp)},\label{eq:delta}
\end{equation}
with $\avg[\cdot][i]=\bsi[i]\cdot\ksi[i]$ and $\avv{\cdot}=\bsi[1]\cdot\ksi[2]$,
$\mathscr{Q}_{1,2}$ are the normalization constants,
\begin{equation}
\mathscr{Q}_{1,2}=\left(1-\delta^{2}\right)\left(1+\frac{\delta^{2}\left(1-\delta^{2}\right)(1-\epsilon)^{2}}{\left(\sqrt{\delta^{2}+\epsilon^{2}-\delta^{2}\epsilon^{2}}\pm\delta^{2}(1-\epsilon)\pm\epsilon\right)^{2}}\right).
\end{equation}

When the photon numbers of the two optical sources are unknown, one
needs to estimate the ratio $\dn$ of the difference between the two photon
numbers to the total photon number in addition to the centroid $\xb$
and the separation $d$ of the two sources. The quantum Fisher information
matrix $\mathcal{Q}$ for parameters $\bar{X},\,d,\,\epsilon$ can
be derived by Eq. \eqref{eq:fisher=000020exp} as
\begin{equation}
\mathcal{Q}=\left[\begin{array}{ccc}
2\kt(1+\chi\epsilon)-4\gamma^{2}\left(1-\epsilon^{2}\right) & \kt(\chi+\epsilon) & 2\gamma\delta\\
\kt(\chi+\epsilon) & \frac{1}{2}\kt(1+\chi\epsilon) & 0\\
2\gamma\delta & 0 & \frac{1-\delta^{2}}{1-\epsilon^{2}}
\end{array}\right],\label{eq:qfim}
\end{equation}
where
\begin{equation}
\begin{aligned}\kappa_{i}= & \langle\partial_{X_{i}}\psi_{i}(x-X_{i})|\partial_{X_{i}}\psi_{i}(x-X_{i})\rangle=\langle\psi_{i}|P^{2}|\psi_{i}\rangle=\avg[][i],\;i=1,2,\\
\gamma= & \langle\partial_{X_{1}}\psi_{1}(x-X_{1})|\psi_{2}(x-X_{2})\rangle=i\langle\psi_{1}|P\e^{-idP}|\psi_{2}\rangle=\avv{\sin(d\hp)\hp},\\
\kt= & \avg[][1]+\avg[][2],\\
\chi= & \frac{\avg[][2]-\avg[][1]}{\avg[][1]+\avg[][2]}.
\end{aligned}
\label{eq:gamkap}
\end{equation}
By taking the inverse of the matrix $\mathcal{Q}$, the quantum Cramér-Rao's
bound for estimation precision of the separation $d$ can be derived
as
\begin{equation}
\hd=\nt/(Q^{-1})_{22}=\frac{\nt\kt\left(1-\epsilon^{2}\right)\left[\left(1-\delta^{2}\right)\left(1-\chi^{2}\right)\kt-2\gamma^{2}(1+\chi\epsilon)\right]}{2\kt\left(1-\delta^{2}\right)(\chi\epsilon+1)-4\gamma^{2}\left(1-\epsilon^{2}\right)}.\label{eq:difh-1}
\end{equation}

Now, one can expand $\delta$ and $\gamma$ to the fourth order of
$d$ (note that $\kappa_{1}$, $\kappa_{2}$ are independent of $d$
as can be seen from Eq. \eqref{eq:gamkap}),
\begin{equation}
\begin{aligned}\delta & =\qo-\frac{1}{2}\avg[][12]d^{2}+\frac{1}{24}\avg[\hp^{4}][12]d^{4},\\
\gamma & =d\avg[][12]-\frac{1}{6}d^{3}\avg[\hp^{4}][12],
\end{aligned}
\label{eq:dr}
\end{equation}
Plugging Eq. \eqref{eq:dr} into $\mathcal{H}_{d}$ \eqref{eq:difh-1},
the numerator and the denominator of $\mathcal{H}_{d}$ can be expanded
to the fourth order of $d$ respectively, which gives
\begin{equation}
\mathcal{H}_{d}=\frac{A_{0}+A_{2}d^{2}+A_{4}d^{4}}{B_{0}+B_{2}d^{2}+B_{4}d^{4}},\label{eq:hdexpand-1}
\end{equation}
where
\begin{equation}
\begin{aligned}A_{0} & =-12\left(1-\chi^{2}\right)\left(1-\epsilon^{2}\right)\left(1-\norm{\qo}\right)\nt\kt,\\
B_{0} & =-24(1+\chi\epsilon)\left(1-\norm{\qo}\right),\\
A_{2} & =12\nt\kt[2]\xi\left(1-\epsilon^{2}\right)\left[2\xi-\qo+\chi^{2}\qo+2\xi\chi\epsilon\right],\\
B_{2} & =24\xi\kt\left[\left(2\xi-\qo\right)-\chi\epsilon\qo-2\xi\epsilon^{2}\right],\\
A_{4} & =\left(1-\epsilon^{2}\right)\nt\kt\left[3\xi^{2}\kt[2]-8\xi\avg[\hp^{4}][12]+\qo\avg[\hp^{4}][12]-8\xi\chi\epsilon\avg[\hp^{4}][12]-\chi^{2}\left(\qo\avg[\hp^{4}][12]+3\xi^{2}\kt[2]\right)\right],\\
B_{4} & =2\left[3\xi^{2}\kt[2]-8\xi\avg[\hp^{4}][12]+\qo\avg[\hp^{4}][12]+8\xi\epsilon^{2}\avg[\hp^{4}][12]+\chi\epsilon\left(\qo\avg[\hp^{4}][12]+3\xi^{2}\kt[2]\right)\right],
\end{aligned}
\label{eq:a0b0a2b2a4b4}
\end{equation}
and
\begin{equation}
\xi=\frac{\avg[][12]}{\avg[][2]+\avg[][1]}.
\end{equation}

If $\qo\neq1$ and $1-\norm{\qo}$ keeps nonzero and finite, when
$d\rightarrow0$, $\mathcal{H}_{d}$ can be simplified to
\begin{equation}
\dlm{\hd}=\frac{\nt\kt\left(1-\chi^{2}\right)\left(1-\epsilon^{2}\right)}{2(1+\chi\epsilon)}.
\end{equation}
For this case, the Rayleigh's criterion is violated and the superresolution
works. Actually in this case, only $A_{0}$ and $B_{0}$ are involved.

When the difference between the two point-spread functions and the
difference between the two photon numbers are comparable to the magnitude
of the separation $d$, the higher-order terms in Eq. \eqref{eq:hdexpand-1}
will also contribute to $\mathcal{H}_{d}$ in the limit $d\rightarrow0$
. And the results are summarized in Table \ref{tab:con} of the main
text.

\subsection{Gaussian point-spread functions}

For two Gaussian point-spread functions with positions $X_{1},X_{2}$
and widths $\sigma_{1},\sigma_{2}$, the parameters $\delta,\gamma,\kappa_{1},\kappa_{2}$
can be obtained as
\begin{equation}
\begin{aligned}\delta & =\sqrt{\frac{2\sigma_{1}\sigma_{2}}{\sigma_{1}^{2}+\sigma_{2}^{2}}}\exp\left[-\frac{d^{2}}{4\left(\sigma_{1}^{2}+\sigma_{2}^{2}\right)}\right],\\
\gamma & =d\sqrt{\frac{\sigma_{1}\sigma_{2}}{2(\sigma_{1}^{2}+\sigma_{2}^{2})^{3}}}\exp\left[-\frac{d^{2}}{4\left(\sigma_{1}^{2}+\sigma_{2}^{2}\right)}\right],\\
\kappa_{1} & =\frac{1}{4\sigma_{1}^{2}},\;\kappa_{2}=\frac{1}{4\sigma_{2}^{2}},
\end{aligned}
\end{equation}
where $\st=\sigma_{1}+\sigma_{2}$. The estimation precision of the
separation $d$ turns to be
\begin{equation}
\hd=\frac{\nt\left(1-\epsilon^{2}\right)\left\{ \left(\sigma_{1}^{2}+\sigma_{2}^{2}\right){}^{2}\left[\left(\sigma_{1}^{2}+\sigma_{2}^{2}\right)\e^{\frac{d^{2}}{2\left(\sigma_{1}^{2}+\sigma_{2}^{2}\right)}}-2\sigma_{2}\sigma_{1}\right]-d^{2}\left[\sigma_{1}^{3}\sigma_{2}(1+\epsilon)+\sigma_{1}\sigma_{2}^{3}(1-\epsilon)\right]\right\} }{2\left(\sigma_{1}^{2}+\sigma_{2}^{2}\right){}^{2}\left[\left(\sigma_{1}^{2}+\sigma_{2}^{2}\right)\e^{\frac{d^{2}}{2\left(\sigma_{1}^{2}+\sigma_{2}^{2}\right)}}-2\sigma_{1}\sigma_{2}\right]\left[\sigma_{1}^{2}(\epsilon+1)+\sigma_{2}^{2}(1-\epsilon)\right]-8\left(1-\epsilon^{2}\right)\sigma_{1}^{3}\sigma_{2}^{3}d^{2}}.\label{eq:hdg-1}
\end{equation}
For two Gaussian point-spread functions with different widths $\sigma_{1}$
and $\sigma_{2}$, the coefficients in Eq. \eqref{eq:hdexpand-1}
can be obtained as
\begin{equation}
\begin{aligned}A_{0}= & 8\nt\left(1-\epsilon^{2}\right)\left(\sigma_{1}-\sigma_{2}\right){}^{2}\left(\sigma_{1}^{2}+\sigma_{2}^{2}\right){}^{2},\\
B_{0}= & 16\left(\sigma_{1}-\sigma_{2}\right){}^{2}\left(\sigma_{1}^{2}+\sigma_{2}^{2}\right){}^{2}\left[\sigma_{1}^{2}(1+\epsilon)+\sigma_{2}^{2}(1-\epsilon)\right],\\
A_{2}= & 4\nt\left(1-\epsilon^{2}\right)\left\{ \left(\sigma_{1}^{2}+\sigma_{2}^{2}\right){}^{2}-2\sigma_{2}\sigma_{1}\left[\sigma_{1}^{2}(1+\epsilon)+\sigma_{2}^{2}(1-\epsilon)\right]\right\} ,\\
B_{2}= & 8\left\{ \left(\sigma_{1}^{2}+\sigma_{2}^{2}\right){}^{2}\left[\sigma_{1}^{2}(1+\epsilon)+\sigma_{2}^{2}(1-\epsilon)\right]-8\sigma_{1}^{3}\sigma_{2}^{3}\left(1-\epsilon^{2}\right)\right\} ,\\
A_{4}= & \nt\left(\sigma_{1}^{2}+\sigma_{2}^{2}\right)\left(1-\epsilon^{2}\right),\\
B_{4}= & 2\left(\sigma_{1}^{2}+\sigma_{2}^{2}\right)\left[\sigma_{1}^{2}(1+\epsilon)+\sigma_{2}^{2}(1-\epsilon)\right].
\end{aligned}
\end{equation}
When $d\rightarrow0$, it can be simplified to
\begin{equation}
\dlm{\hd}=\frac{\nt\left(1-\epsilon^{2}\right)}{2\left[\sigma_{1}^{2}(1+\epsilon)+\sigma_{2}^{2}(1-\epsilon)\right]},
\end{equation}
which is Eq. \eqref{eq:op} in the main text. The results of the estimation
precision $\hd$ in different regimes of the photon numbers and the
widths of the two point sources are summarized in Table \ref{tab:The-detailed-results-1}
of the main text.

If we denote
\begin{equation}
\eta=\frac{\sigma_{2}-\sigma_{1}}{\sigma_{1}+\sigma_{2}},\quad\dt=\frac{d}{(\sigma_{1}+\sigma_{2})/2},
\end{equation}
where $\dt$ is the dimensionless separation extended from the case
of two identical point-spread functions \eqref{eq:dt11}, the estimation
precision $\hd$ \eqref{eq:hdg-1} can be simplified as
\begin{equation}
\hd=\frac{\left(1-\epsilon^{2}\right)\left\{ 4\left(\eta^{2}+1\right)^{3}e^{\frac{\dt[2]}{4\eta^{2}+4}}-\left(1-\eta^{2}\right)\left[4\left(\eta^{2}+1\right)^{2}+\dt[2]\left(\eta^{2}-2\eta\epsilon+1\right)\right]\right\} }{4\left(\eta^{2}+1\right)^{2}\left(\eta^{2}-2\eta\epsilon+1\right)\left[\left(\eta^{2}+1\right)e^{\frac{\dt[2]}{4\eta^{2}+4}}-\left(1-\eta^{2}\right)\right]-\dt[2]\left(1-\eta^{2}\right)^{3}\left(1-\epsilon^{2}\right)}\hdo,\label{eq:hdeta}
\end{equation}
and $\hdo$ is the optimal estimation precision of $d$ when the photon
numbers and the widths of the two Gaussian sources are the same, i.e.,
$\nb[1]=\nb[2]=\nb$ and $\sigma_{1}=\sigma_{2}=\sigma$,
\begin{equation}
\hdo=\frac{\nb}{2\sigma^{2}}.
\end{equation}
This implies that the ratio between the precision $\hd$ and the optimal
precision $\hdo$ is determined by the parameters $\dn$, $\eta$
and $\dt$ only.

\section{\protect\label{sec:para}Competing quantities in the validity of
superresolution}

\subsection{Critical quantities\protect\label{subsec:Critical-quantities}}

It can be seen from the results for two identical point sources that
the validity of superresolution is highly sensitive to the magnitude
of the leading term in the denominator of $\mathcal{H}_{d}$ that
may be comparable to $d$ and it turns out that the ratio of the difference
between the photon numbers of the two optical sources to the total
photon number,
\begin{equation}
\dn=\frac{\nb[2]-\nb[1]}{\nb[2]+\nb[1]}.
\end{equation}
determines the magnitude of that term which leads to quantum superresolution
if it is of order $O(\dt)$ or even equal to zero. So, $\epsilon$
is a critical quantity that determines the validity of superresolution
for two identical point-spread functions.

When the two point-spread functions differ slightly from each other,
the difference between the two point-spread functions may change the
magnitude of the leading terms in the numerator and denominator of
$\mathcal{H}_{d}$ and render the additional terms in $\hd$ \eqref{eq:hdexpand-1}
comparable to $\dt$ which will compete with each other as well as
with the ratio $\dn$ between the photon number difference and the
total photon number in determining the validity of superresolution,
so the difference of the point-spread functions can have significant
effect on $\mathcal{H}_{d}$ and we need to investigate this effect
in detail.

As a first step to analyze the effect of the difference between the
point-spread functions, we figure out what quantities in Eq. \eqref{eq:hdexpand-1}
are determined by the differences between the two point-spread functions
in the following.

For the zeroth-order terms $A_{0}$ and $B_{0}$ of the numerator
and denominator of $\mathcal{H}_{d}$ \eqref{eq:hdexpand-1}, Eq.
\eqref{eq:a0b0a2b2a4b4} tells that
\begin{equation}
\begin{aligned}A_{0} & =-12\nt\kt\left(1-\chi^{2}\right)\left(1-\epsilon^{2}\right)\left(\underline{1-\norm{\qo}}\right),\\
B_{0} & =-24(1+\chi\epsilon)\left(\underline{1-\norm{\qo}}\right),
\end{aligned}
\end{equation}
where $\nt$ is the total photon number,
\begin{equation}
\nt=\nb[1]+\nb[2].
\end{equation}
It can be seen that $A_{0}$ and $B_{0}$ vanish when the two point-spread
functions are close to each other, i.e., $\psi_{1}(x)-\psi_{2}(x)\rightarrow0$
or equivalently $\qo\rightarrow1$, so $1-\norm{\qo}$ is also critical
to the estimation precision $\mathcal{H}_{d}$. As $1-\norm{\qo}=\left(1-\qo\right)\left(1+\qo\right)$
and $1+\qo\neq0$ when $\psi_{1}(x)$ and $\psi_{2}(x)$ are close
to each other, so for simplicity we choose $1-\qo$ to be the quantity
that is determined by the difference between the two point-spread
functions and critical to the estimation precision $\mathcal{H}_{d}$.

For the second-order terms $A_{2}d^{2}$ and $B_{2}d^{2}$ in the
numerator and denominator of $\mathcal{H}_{d}$ \eqref{eq:hdexpand-1},
the coefficients $A_{2}$ and $B_{2}$ can be rewritten as
\begin{equation}
\begin{aligned}A_{2}= & 12\nt\kt[2]\xi\left(1-\epsilon^{2}\right)\left[\underline{2\xi-\qo}+\underline{\chi^{2}}\qo+2\xi\underline{\chi}\epsilon\right],\\
B_{2}= & 24\xi\kt\left[\underline{2\xi-\qo}-\underline{\chi}\epsilon\qo-2\xi\epsilon^{2}\right],
\end{aligned}
\end{equation}
where
\begin{equation}
\text{\ensuremath{\kt}}=\kappa_{1}+\kappa_{2}=\avg[][2]+\avg[][1].\label{eq:kappabar}
\end{equation}
When the two point-spread functions are close to each other, $\qo\rightarrow1$,
$\chi\rightarrow0$, $2\xi-\qo=2\xi-1\rightarrow0$ , which leads
to $A_{2}\rightarrow0$, $B_{2}\rightarrow0$, so one can immediately
find that $\chi$ and $\frac{1}{2}-\xi$ are the two additional quantities
that are determined by the difference between the two point-spread
functions and can change the estimation precision of the separation
$d$.

For the fourth-order terms $A_{4}d^{4}$ and $B_{4}d^{4}$ in the
numerator and denominator of $\mathcal{H}_{d}$ \eqref{eq:hdexpand-1},
the coefficients $A_{4}$ and $B_{4}$ can be rewritten as 
\begin{equation}
\begin{aligned}A_{4}= & \left(1-\epsilon^{2}\right)\nt\kt\bigg[3\xi^{2}\kt[2]-8\xi\avg[\hp^{4}][12]+\qo\avg[\hp^{4}][12]-8\xi\chi\epsilon\avg[\hp^{4}][12]-\chi^{2}\left(\qo\avg[\hp^{4}][12]+3\xi^{2}\kt[2]\right)\bigg],\\
B_{4}= & 2\left[3\xi^{2}\kt[2]-8\xi\avg[\hp^{4}][12]+\qo\avg[\hp^{4}][12]+8\xi\epsilon^{2}\avg[\hp^{4}][12]+\chi\epsilon\left(\qo\avg[\hp^{4}][12]+3\xi^{2}\kt[2]\right)\right].
\end{aligned}
\end{equation}

When the two point-spread functions are close to each other, $\qo\rightarrow1$
and $\chi\rightarrow0$, $\xi\rightarrow\frac{1}{2}$, $\epsilon\rightarrow0$,
leading to 
\begin{equation}
\begin{aligned}A_{4}\rightarrow & \nt\kt\left(3\xi^{2}\kt[2]-8\xi\avg[\hp^{4}][12]+\qo\avg[\hp^{4}][12]\right),\\
B_{4}\rightarrow & 2\left(3\xi^{2}\kt[2]-8\xi\avg[\hp^{4}][12]+\qo\avg[\hp^{4}][12]\right).
\end{aligned}
\end{equation}
In this case,
\begin{equation}
3\xi^{2}\kt[2]-8\xi\avg[\hp^{4}][12]+\qo\avg[\hp^{4}][12]\rightarrow3\avg[][][2]-3\avg[\hp^{4}]=3\var(\hp^{2}).
\end{equation}
$\var(\hp^{2})$ is generally nonzero unless the point-spread function
$\psi(x)$ is approximately an eigenstate of $\hp$ or equivalently
an uniformly-distributed point-spread function over the position space.
But this is determined by the property of the point-spread function
of either optical source, not by the difference between the two point-spread
functions, so there is no new critical quantity determined by the
difference of the two point-spread functions from the fourth-order
terms in $\hd$.

In summary, we can identify $1-\qo$, $\chi$ and $\frac{1}{2}-\xi$
as the critical quantities that are determined by the difference between
the point-spread functions and will compete with each other and with
the ratio $\dn$ between the difference and the total of the photon
number in determining the validity and the estimation precision of
superresolution.

\subsection{\protect\label{sec:Derivation=000020hfe}Relation between critical
quantities}

When considering the competition between the critical quantities obtained
above in determining the validity of superresolution, it should be
noted that those quantities are not independent of each other but
closely related, and we need to find out the relation between those
quantities which will impose constraints on the range of estimation
precision for the separation.

Assume that the difference between $|\psi_{1}\rangle$ and $|\psi_{2}\rangle$
is small. To facilitate the following computation, we first construct
an orthonormal basis for the subspace spanned by $|\psi_{1}\rangle$
and $|\psi_{2}\rangle$,
\begin{equation}
|\psi_{1}\rangle,\;|\psi_{1}^{\bot}\rangle=\frac{|\psi_{2}\rangle-\delta|\psi_{1}\rangle}{\sqrt{1-\delta^{2}}}.
\end{equation}
$|\psi_{2}\rangle$ can be decomposed onto the basis states $|\psi_{1}\rangle$
and $|\psi_{1}^{\bot}\rangle$ as
\begin{equation}
|\psi_{2}\rangle=\alpha|\psi_{1}\rangle+\beta|\psi_{1}^{\bot}\rangle.\label{eq:psi2}
\end{equation}
It is straightforward to verify that
\begin{equation}
\alpha=1-u,\;\beta=\sqrt{2u-u^{2}},
\end{equation}
where 
\begin{equation}
u=1-\qo.
\end{equation}
Now, substituting $|\psi_{1}\rangle$ and $|\psi_{2}\rangle$ \eqref{eq:psi2}
into $\avg[][2]-\avg[][1]$, one can have
\begin{equation}
\begin{aligned}\chi= & \frac{\avg[][2]-\avg[][1]}{\avg[][2]+\avg[][1]},\\
= & \frac{(u^{2}-2u)\langle\psi_{1}|\hp^{2}|\psi_{1}\rangle+2(1-u)\sqrt{2u-u^{2}}\langle\psi_{1}|\hp^{2}|\psi_{1}^{\bot}\rangle+(2u-u^{2})\langle\psi_{1}^{\bot}|\hp^{2}|\psi_{1}^{\bot}\rangle}{(u^{2}-2u+2)\langle\psi_{1}|\hp^{2}|\psi_{1}\rangle+2(1-u)\sqrt{2u-u^{2}}\langle\psi_{1}|\hp^{2}|\psi_{1}^{\bot}\rangle+(2u-u^{2})\langle\psi_{1}^{\bot}|\hp^{2}|\psi_{1}^{\bot}\rangle}.
\end{aligned}
\end{equation}
In the above equation, $u$ is a small parameter when the two point-spread
functions are close to each other. The numerator ranges between $O(u^{\frac{1}{2}})$
and $O(u)$, and the denominator ranges between $O(1)$ and $O(u)$,
so the order of $\chi$ is between $O(u^{1/2})$ and $O(u)$.Similarly,
for $\frac{1}{2}-\xi$, we note that 
\begin{alignat}{1}
\avg[][12] & =\langle\psi_{1}|\hp^{2}|\psi_{2}\rangle,\nonumber \\
 & =(1-u)\langle\psi_{1}|\hp^{2}|\psi_{1}\rangle+\sqrt{2u-u^{2}}\langle\psi_{1}|\hp^{2}|\psi_{1}^{\bot}\rangle,
\end{alignat}
so,
\begin{equation}
\begin{aligned}\frac{1}{2}-\xi= & \frac{\frac{1}{2}\left(\avg[][2]+\avg[][1]\right)-\avg[][12]}{\avg[][2]+\avg[][1]},\\
= & \frac{\frac{1}{2}u^{2}\langle\psi_{1}|\hp^{2}|\psi_{1}\rangle-u\sqrt{2u-u^{2}}\langle\psi_{1}|\hp^{2}|\psi_{1}^{\bot}\rangle+\frac{1}{2}(2u-u^{2})\langle\psi_{1}^{\bot}|\hp^{2}|\psi_{1}^{\bot}\rangle}{(u^{2}-2u+2)\langle\psi_{1}|\hp^{2}|\psi_{1}\rangle+2(1-u)\sqrt{2u-u^{2}}\langle\psi_{1}|\hp^{2}|\psi_{1}^{\bot}\rangle+(2u-u^{2})\langle\psi_{1}^{\bot}|\hp^{2}|\psi_{1}^{\bot}\rangle}.
\end{aligned}
\end{equation}
The numerator ranges between $O(u)$ and $O(u^{2})$, and the denominator
ranges between $O(1)$ and $O(u)$, implying that $\frac{1}{2}-\xi$
scales between $O(u)$ and $O(u^{2})$ when the two point-spread functions
are close to each other.

In order to make the critical quantities comparable, we denote
\begin{equation}
u=1-\qo=b\dt[h],\;\frac{1}{2}-\xi=a\dt[f],\;\chi=c\dt[e].
\end{equation}
According to the above ranges of $\chi$ and $\frac{1}{2}-\xi$ in
terms of $u$, we can immediately obtain the constraint condition
on the exponents $h$, $f$ and $e$,
\begin{equation}
h/2\leq e\leq h\leq f\leq2h.\label{eq:expconstraint}
\end{equation}

\end{document}